\RequirePackage{fix-cm}
\documentclass[twocolumn,envcountsame]{svljour3}  
\renewcommand*\descriptionlabel[1]{\hspace\labelsep
  \normalfont\itshape#1}
\newcommand{\myitem}[1][]{\item{}\hspace{-\labelsep}{\itshape #1}\hspace{1.75pt}}
  
\smartqed  

\usepackage{graphicx}

\usepackage{aodv}
\renewcommand{\comma}{\mathbin{\text{\hskip-0.1em$,$\hskip-0.1em}}} 
\usepackage{amsfonts,amsmath,amssymb}
\usepackage{algorithm,algorithmic}
\makeatletter
\def\theHALC@line{\thealgorithm-\theALC@line}
\def\theHALC@rem{\thealgorithm-\theALC@rem}
\makeatother
\floatname{algorithm}{Process}

\usepackage{xcolor}
\usepackage{enumerate}
\usepackage{xspace}
\usepackage{xifthen}
\usepackage{longtable}
\usepackage{wrapfig}
\usepackage{hyperref,breakurl}
\usepackage{etex,etoolbox}

\newcounter{Hequation}

\makeatletter
\g@addto@macro\equation{\stepcounter{Hequation}}
\makeatother

\newcommand{\awn}{AWN\xspace}

\newcommand{\plat}[1]{\raisebox{0pt}[0pt][0pt]{#1}} 
\newcommand{\NN}{
    \ensuremath{%
        \mathop{\rm I\mkern-2.5mu N}%
        \nolimits%
    }%
}
\providecommand{\doi}[1]{doi:\href{http://dx.doi.org/#1}{#1}}

\newenvironment{myquote}
               {\list{}{\rightmargin=0.13in\leftmargin=0.13in}%
                \item\relax}
               {\endlist}
\newcommand{\undefined}{\multicolumn{2}{l}{\mbox{undefined otherwise\,.}}}
\newcommand{\ow}[2][.]{#2\ &\mbox{otherwise}\,#1}
\newcommand{\ifs}[2]{#1\ &\mbox{if }#2}
\newcommand{\nifs}[1]{&\phantom{\mbox{if }}#1}

\newcommand{\ifsnewline}[2]{%
	\multicolumn{2}{l}{#1}\\%
	\multicolumn{2}{l}{\hspace{1em}\mbox{if }#2}%
}
\makeatletter
\def\comesfrom{\@transition\leftarrowfill}
\def\goesto{\@transition\rightarrowfill}
\def\ngoesto{\@transition\nrightarrowfill}
\def\Goesto{\@transition\Rightarrowfill}
\def\nGoesto{\@transition\nRightarrowfill}
\def\xmapsto{\@transition\mapstofill}
\def\nxmapsto{\@transition\nmapstofill}
\def\@transition#1{\@@transition{#1}}
\newbox\@transbox
\newbox\@arrowbox
\newbox\@downbox
\def\@@transition#1#2%
   {\setbox\@transbox\hbox
      {\vrule height 1.5ex depth .9ex width 0ex\hskip0.25em$\scriptstyle#2$\hskip0.25em}
   \ifdim\wd\@transbox<1.5em
      \setbox\@transbox\hbox to 1.5em{\hfil\box\@transbox\hfil}\fi
   \setbox\@arrowbox\hbox to \wd\@transbox{#1}
   \ht\@arrowbox\z@\dp\@arrowbox\z@
   \setbox\@transbox\hbox{$\mathop{\box\@arrowbox}\limits^{\box\@transbox}$}
   \dp\@transbox\z@\ht\@transbox 10pt
   \mathrel{\box\@transbox}}
\def\nrightarrowfill{$\m@th\mathord-\mkern-6mu%
  \cleaders\hbox{$\mkern-2mu\mathord-\mkern-2mu$}\hfill
  \mkern-6mu\mathord\not\mkern-2mu\mathord\rightarrow$}
\def\Rightarrowfill{$\m@th\mathord=\mkern-6mu%
  \cleaders\hbox{$\mkern-2mu\mathord=\mkern-2mu$}\hfill
  \mkern-6mu\mathord\Rightarrow$}
\def\nRightarrowfill{$\m@th\mathord=\mkern-6mu%
  \cleaders\hbox{$\mkern-2mu\mathord=\mkern-2mu$}\hfill
  \mkern-6mu\mathord\not\mathord\Rightarrow$}
\def\mapstofill{$\m@th\mathord\mapstochar\mathord-\mkern-6mu%
  \cleaders\hbox{$\mkern-2mu\mathord-\mkern-2mu$}\hfill
  \mkern-6mu\mathord\rightarrow$}
\def\nmapstofill{$\m@th\mathord\mapstochar\mathord-\mkern-6mu%
  \cleaders\hbox{$\mkern-2mu\mathord-\mkern-2mu$}\hfill
  \mkern-6mu\mathord\not\mkern-2mu\mathord\rightarrow$}
\makeatother 
\newcommand{\ar}[1]{\mathrel{\goesto{#1}}}            

\makeatletter
\providecommand{\@thirdoffive}[5]{#3}

\def\fixstatement#1{%
  \AtEndEnvironment{#1}{%
    \xdef\pat@label{\expandafter\expandafter\expandafter
      \@thirdoffive\csname#1\endcsname\space\@currentlabel}}}
      
\globtoksblk\prooftoks{1000}
\newcounter{proofcount}

\newboolean{qedatend}
\newcommand{\prf}{\setboolean{qedatend}{true}\begin{proof}}
\newcommand{\prfnobox}{\setboolean{qedatend}{false}\begin{proof}}
\newcommand{\eprf}{\ifqedatend\qed\fi\end{proof}}

\long\def\prfatend#1\eprf{%
  \edef\next{\noexpand\begin{theopargself}\noexpand\prf[\noexpand\hspace{-1.5pt}of \pat@label\noexpand\hspace{1.75pt}]}%
  \toks\numexpr\prooftoks+\value{proofcount}\relax=\expandafter{\next#1\eprf\end{theopargself}}
  \stepcounter{proofcount}}
  
  \long\def\prfnoboxatend#1\eprf{%
  \edef\next{\noexpand\begin{theopargself}\noexpand\prfnobox[\noexpand\hspace{-1.5pt}of \pat@label\noexpand\hspace{1.5pt}]}%
  \toks\numexpr\prooftoks+\value{proofcount}\relax=\expandafter{\next#1\eprf\end{theopargself}}
  \stepcounter{proofcount}}

\def\printproofs{%
  \count@=\z@
  \loop
    \the\toks\numexpr\prooftoks+\count@\relax
    \ifnum\count@<\value{proofcount}%
    \advance\count@\@ne
  \repeat}
\makeatother

\fixstatement{lemma}
\fixstatement{theorem}
\fixstatement{corollary}
\fixstatement{proposition}

\renewcommand{\quote}[1]{``\emph{#1\/}''}

\journalname{Journal of Distributed Computing}

\begin{document}

\title{Modelling and Verifying the AODV Routing Protocol\thanks{NICTA is funded by the Australian Government through the
    Department of Communications and the Australian Research Council
    through the ICT Centre of Excellence Program.}
}

\author{
    Rob van Glabbeek \and 
    Peter H\"ofner \and 
    Marius Portmann \and 
    Wee Lum Tan
}

\authorrunning{R.\ van Glabbeek, P.\ H\"ofner, M.\ Portmann, W.L.\ Tan} 

\institute{
    R.\,J.\ van Glabbeek \at
	    NICTA and UNSW.
	    \email{rvg@cs.stanford.edu} \and
    P.\ H\"ofner \at
	    NICTA and UNSW.
		\email{Peter.Hoefner@nicta.com.au} \and
    M.\ Portmann \at
	    The University of Queensland.
		\email{marius@itee.uq.edu.au} \and 
    W.\,L.\ Tan \at
	    Griffith University.
		\email{w.tan@griffith.edu.au}
}

\date{\mbox{}}

\maketitle

\begin{abstract}
This paper presents a formal specification of the Ad hoc On-Demand
Distance Vector (AODV) routing protocol using {\awn} (Algebra for
Wireless Networks), a recent process algebra which has been
tailored for the modelling of Mobile Ad Hoc Networks and Wireless Mesh Network
protocols.  Our formalisation models the exact details of the core
functionality of AODV, such as route discovery, route maintenance and
error handling.  We demonstrate how {\awn} can be used to reason about
critical protocol properties by providing detailed proofs of loop
freedom and route correctness.

\end{abstract}

\section{Introduction}\label{sec:intro}
Routing protocols are crucial to the dissemination of data packets between nodes in 
Wireless Mesh Networks (WMNs) and Mobile Ad Hoc Networks (MANETs). 
One of the most popular protocols that is widely used in WMNs is the Ad hoc On-Demand Distance
Vector (AODV) routing protocol~\cite{rfc3561}. It is one of the four protocols
standardised by the IETF MANET working group, and it also forms the basis of new WMN
routing protocols, including 
the Hybrid Wireless Mesh Protocol (HWMP) in the  IEEE 802.11s wireless mesh network
standard~\cite{IEEE80211s}. The details of the AODV protocol are standardised in IETF RFC 3561~\cite{rfc3561}.
However, due to the use of English prose,
this specification contains ambiguities and contradictions. This can lead to significantly different
implementations of the AODV routing protocol, depending on the developer's understanding and reading
of the AODV RFC\@. In the worst case scenario, an AODV implementation may contain serious flaws,
such as routing loops~\cite{AODVloop}.

Traditional approaches to the analysis of AODV and many other AODV-based protocols~\cite{AODVv2,IEEE80211s,AODV-ST,SBM06,PPI08}
are simulation and test-bed experiments. While such methods are important and valid for protocol
evaluation, in particular for quantitative performance evaluation, they have limitations in regards
to the evaluation of basic protocol correctness properties. Experimental evaluation is resource
intensive and time consuming, and, even after a very long time of evaluation, only a finite set of
network scenarios can be considered---no general guarantee can be given about correct protocol
behaviour for a wide range of unpredictable deployment scenarios~\cite{Verisim}.  This problem is
illustrated by recent discoveries of limitations in AODV-like protocols that have been under intense
scrutiny over many years \cite{MK10}.

We believe that formal methods can help in this regard; they complement simulation and test-bed
experiments as methods for protocol evaluation and verification, and provide stronger and more
general assurances about protocol properties and behaviour.
The overall goal is to reduce
the ``time-to-market'' for better (new or modified) WMN protocols, and
to increase the reliability and performance of the corresponding
networks.

In this paper we provide a complete and accurate formal specification of the core functionality of
AODV using the specification language \awn (\hspace{-1pt}Algebra of Wireless Networks) \cite{ESOP12}.
\awn provides the right level of abstraction to model key features such as 
unicast and broadcast, while abstracting from implementation-re\-la\-ted details. As its semantics is completely unambiguous, specifying a protocol in such a framework enforces total precision and the removal of any ambiguities. A key contribution is to demonstrate how {\awn} can be used to
support reasoning about protocol behaviour and to provide rigorous proofs of key protocol
properties, using the examples of loop freedom and route correctness.
In contrast to what can be achieved by model checking and test-bed experiments, our proofs apply to all conceivable dynamic network topologies.

Route correctness is a minimal sanity requirement for a routing protocol; it is the property
that the routing table entries stored at a node are entirely based on information on routes to other
nodes that either is currently valid or was valid at some point in the past.
Loop freedom is a critical property for any routing protocol, but it is particularly relevant and challenging for WMNs.
Descriptions as in~\cite{Garcia-Luna-Aceves89} capture the common understanding of loop freedom:
\quote{A routing-table loop is a path specified in the nodes' routing tables at a particular point in time that visits the same node more than once before reaching the intended destination.}
Packets caught in a routing loop,  until they are discarded by the IP Time-To-Live (TTL) mechanism, can quickly saturate the links and have a detrimental impact on network performance. It is therefore critical to ensure that protocols prevent routing loops.
We show that loop freedom can be guaranteed only if sequence numbers are used in a careful way, considering further rules and assumptions on the behaviour of the protocol.
The problem is, as shown in the case of AODV, that these additional rules and assumptions are not explicitly stated in the RFC, 
and that the RFC has significant ambiguities in regards to this.
To the best of our knowledge we are the first to give a complete and detailed proof of loop freedom.%
\footnote{Loop freedom of AODV has been ``proven'' at least thrice~\cite{AODV99,BOG02,ZYZW09}, but
 the proofs in \cite{AODV99} and \cite{BOG02} are not correct, and the one in \cite{ZYZW09} is based on a simple
   subset of AODV only, not including the ``intermediate route reply'' feature---a most likely source of loops.
   We elaborate on this in \Sect{relatedwork}.}
\footnote{In this paper, we abstract from timing issues by postulating that routing table entries never expire.
   Consequently, we can make no claim on routing loops resulting from premature expiration of
   routing tables entries.  This will be the subject of a forthcoming paper \cite{tawn}.}

The rigorous protocol analysis discussed in this paper has the
potential to save a significant amount of time in the development and
evaluation of new network protocols, can provide increased levels of
assurance of protocol correctness, and complements simulation and
other experimental protocol evaluation approaches.

The remainder of this paper is organised as follows.
\Sect{aodv} gives an informal introduction to AODV\@.
We briefly recapitulate {\awn} in \Sect{awn}.
\Sect{modelling_AODV} provides a detailed formal specification of AODV\@ in AWN\@.%
\footnote{Parts of
the specification have been published before in ``A Process Algebra for Wireless Mesh Networks''~\cite{ESOP12},
in ``Automated Analysis of AODV using UPPAAL''~\cite{TACAS12} and in ``A Rigorous Analysis of AODV and its
Variants''~\cite{MSWIM12}.} 
To achieve this, we present the basic data structure needed in \Sect{types}.
In \Sect{invariants} we formally prove some properties of AODV that can be expressed as invariants,
in  particular loop freedom and route correctness.\footnote{A sketch of the loop freedom proof is given in \cite{ESOP12} and in~\cite{MSWIM12}.}
\Sect{relatedwork} describes related work, and in \Sect{conclusion} we summarise our findings and
point at work that is yet to be done.

\section{The AODV Routing Protocol}\label{sec:aodv}
The Ad hoc On-Demand
Distance Vector \hspace{-0.35pt}(AODV)\hspace{-0.3pt} routing protocol \cite{rfc3561} is a widely-used routing protocol designed for
MANETs, and is one of the four protocols currently standardised by the
IETF MANET working
group\footnote{\url{http://datatracker.ietf.org/wg/manet/charter/}}.
It also forms the basis of new WMN routing protocols, including
the Hybrid Wireless Mesh Protocol (HWMP) in the IEEE 802.11s wireless mesh network standard~\cite{IEEE80211s}.

\begin{figure}[t]
\centering
   \begin{tabular}[b]{@{}r@{}l@{\hspace{7mm}}r@{}l@{}}
   (a)&
   \includegraphics[scale=1.35]{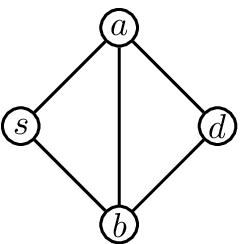}&
   (b)&
   \includegraphics[scale=1.35]{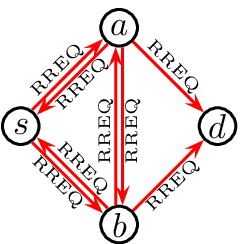}\\[3ex]
   \multicolumn{4}{c}{
   (c)
   \includegraphics[scale=1.35]{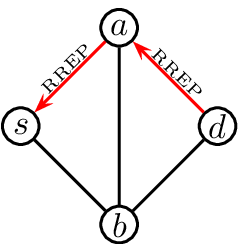}
   }
   \end{tabular}
   \caption{Example network topology}
   \label{fig:topology}
\end{figure}

AODV is a reactive protocol: routes are established only on demand. A
route from a source node $s$ to a destination node $d$ is a sequence
of nodes $[s,n_1,\dots,n_k,d]$, where $n_1$, $\dots$, $n_k$ are
intermediate nodes located on the path from $s$ to $d$.  Its basic
operation can best be explained using a simple example topology shown
in Figure~\ref{fig:topology}(a), where edges connect nodes within
transmission range. We assume node $s$ wants to send a data packet to
node~$d$, but $s$ does not have a valid routing table entry for
$d$. Node $s$ initiates a route discovery mechanism by broadcasting a
route request (RREQ) message, which is received by $s$'s immediate
neighbours $a$ and $b$. We assume that neither $a$ nor $b$ knows a
route to the destination node $d$.\footnote{In case an intermediate
  node knows a route to $d$, it directly sends a route reply back.}
Therefore, they simply re-broadcast the message, as shown in
Figure~\ref{fig:topology}(b). Each RREQ message has a unique identifier
which allows nodes to ignore duplicate RREQ messages that they have
handled before.

When forwarding the RREQ message, each intermediate node updates its
routing table and adds a ``reverse route'' entry to $s$, indicating
via which next hop the node $s$ can be reached, and the distance in
number of hops. Once the first RREQ message is received by the
destination node $d$ (we assume via $a$), $d$ also adds a reverse
route entry in its routing table, saying that node $s$ can be reached
via node $a$, at a distance of $2$ hops.

Node $d$ then responds by generating
 a route reply (RREP) message and sending it  back to
node $s$, as shown in Figure~\ref{fig:topology}(c). In contrast to the
RREQ message, the RREP is unicast, i.e., it is sent to an individual
next hop node only. The RREP is sent from $d$ to $a$, and then to $s$,
using the reverse routing table entries created during the forwarding
of the RREQ message. When processing the RREP message, a node creates
a ``forward route'' entry into its routing table. For example, upon
receiving the RREP via $a$, node $s$ creates an entry saying that $d$
can be reached via $a$, at a distance of $2$ hops. At the completion
of the route discovery process, a route has been established from $s$
to $d$, and data packets can start to flow.

In the event of link and route breaks, AODV uses route error (RERR)
messages to inform affected nodes. 

Sequence numbers, another important aspect of\linebreak AODV, indicate the freshness of routing information.
AODV \quote{uses destination sequence numbers to ensure loop freedom at all
 times (even in the face of anomalous delivery of routing control messages), ...}~\cite{rfc3561}.
A proof of loop freedom of AODV has been sketched in~\cite{AODV99}.  
Nodes maintain their own sequence number as well as a destination sequence number for each route discovered. 
This use of sequence numbers can be an efficient approach to address the problem of routing loops,
but has to be taken with caution, since loop freedom cannot be guaranteed a priori~\cite{AODVloop}.

\section{The Specification Language AWN}\label{sec:awn}
Ideally, any specification is free of ambiguities and contradictions. Using English prose only, this is nearly impossible to achieve.
Hence every specification should be equipped with a formal specification. 
The choice of an appropriate specification language is often secondary, although it has high impact on the analysis. 
The use of {\em any} formal specification language helps to avoid ambiguities and to precisely describe the intended behaviour. 
Examples of modelling languages are 
(i) the Alloy language, used by Zave to model Chord \cite{Zave12};
(ii) timed automata, which are the input language for the {\sc Uppaal} model checker and used by Chiyangwa, Kwiat\-kowska~\cite{CK05} and others~\cite{TACAS12} to reason about AODV;
(iii) routing algebra as introduced by Griffin and Sobrinho \cite{GS05}, 
or (iv) AWN, a process algebra particularly tailored for (wireless mesh) routing protocols~\cite{ESOP12,MSWIM12}.

For this paper we choose the modelling language \awn: 
on the one hand it is tailored for wireless protocols and therefore offers primitives such as {\bf broadcast}; 
on the other hand, it defines the protocol in a pseudo-code  that is easily readable.
(The language itself is implementation independent). 
\awn is a variant of standard
process algebras \cite{Mi89,Ho85,BK85,BB87}, extended with a local broadcast
mechanism and a novel
\emph{conditional unicast} op\-erator---allowing error handling in
response to failed communications while abstracting from link layer
implementations of the communication handling---and incorporating
data structures with assignments. It also
describes the interaction between nodes in a network with a dynamic
network topology.
Process algebras such as {\awn} are equipped with an operational semantics \cite{ESOP12,TR13}:
once a model has been described, its behaviour is governed by
the transitions allowed by the algebra's semantics.
This can significantly reduce the burden of proofs.
In this paper we abstain from a formal definition of the operational semantics.%
\footnote{Thereby we also abstain from explaining the modelling of
    the dynamic network topology in the semantics, i.e., the mechanism by
    which links between nodes break. This matter is explained in \cite{ESOP12,TR13},
    and completely orthogonal to the formal specification of the AODV
    protocol and the correctness properties that are the focus of this paper.
    In particular, the correctness properties 
    hold independently of the number of link breaks or link occurrences.}
Instead, we employ a correspondence between the transitions of
\awn processes and the execution of \emph{actions}---subexpressions as
occur in Entries 3--10 of  Table~\ref{tb:procexpr}---identified by
line numbers in protocol specifications in \awn. 

We use an underlying data structure (described in detail in \Sect{types}) with several types, variables
ranging over these types, operators and predicates. First
order predicate logic yields terms (or \emph{data expressions}) and
formulas to denote data values and statements about them.
The {\awn} data structure
always contains the types \tDATA, \tMSG, {\tIP} and $\pow(\tIP)$ of
\emph{application layer data}, \emph{messages}, \emph{IP addresses}---or any
other node iden\-tifiers---and \emph{sets of IP addresses}.
The messages comprise \emph{data packets}, containing application layer
data, and \emph{control messages}.
The rest of the data structure is customisable for any application of \awn.

In \awn, a WMN is modelled as an encapsulated parallel composition of network nodes.
On each node several processes may be running in parallel.
Network nodes communicate with their direct neighbours---those nodes
that are currently in transmission range---using either broadcast,
unicast, or an iterative unicast/multi\-cast (here called \emph{groupcast}).
The \emph{process expressions} are given
in Table~\ref{tb:procexpr}.
 \begin{table}[t]
\caption{process expressions}
 \centering
{\small
  \setlength{\tabcolsep}{2.6pt}
 \begin{tabular}{|l|p{4.55cm}|}
\hline
\rule[6.5pt]{0pt}{1pt}%
$X(\dexp{exp}_1,\ldots,\dexp{exp}_n)$& process name with arguments\\
$\p+\q$ & choice between proc.\ $\p$ and $\q$\\
$\cond{\varphi}\p$&conditional process\\
$\assignment{\keyw{var}:=\dexp{exp}}\p$&assignment followed by process $\p$\\
$\broadcastP{\dexp{ms}}.\p $&broadcast \dexp{ms} followed by $\p$\\
$\groupcastP{\dexp{dests}}{\dexp{ms}}.\p$&iterative unicast or
  multicast to all destinations \dexp{dests}\\
$\unicast{\dexp{dest}}{\dexp{ms}}.\p \prio \q$& unicast $\dexp{ms}$ to $\dexp{dest}$; if successful proceed with $\p$; otherwise with $\q\hspace{-2.5pt}$\\
$\send{\dexp{ms}}.\p$&synchronously transmit \dexp{ms} to\newline
     parallel process on same node\\
$\deliver{\dexp{data}}.\p$&deliver data to application layer\\
$\receive{\msg}.\p$&receive a message\\
\hline
\rule[6.5pt]{0pt}{1pt}%
$\xi,\p$         &process with valuation\\
$P\parl Q$	&parallel procs.\ on the same node\\
\hline
\rule[6.5pt]{0pt}{1pt}%
$a\mathop{:}P\mathop{:}R$  & node $a$ running $P$ with range $R$\\
$N\|M$		&parallel composition of nodes\\
$[N]$		&encapsulation\\
\hline
\end{tabular}}
\label{tb:procexpr}
\end{table}
A process name $X$ comes with a \emph{defining equation}\vspace{-1ex}
\[X(\keyw{var}_1,\ldots,\keyw{var}_n) \stackrel{{\it def}}{=} \p\,,\]
where $\p$ is a process expression, and the $\keyw{var}_i$ are data
variables maintained by process $X$. A named process is like a
\emph{procedure}; when it is called, data expressions $\dexp{exp}_i$
of the appropriate type are filled in for the variables $\keyw{var}_i$.
Furthermore, $\varphi$ is a condition,
$\keyw{var}\mathop{:=}\dexp{exp}$ an assignment of a data expression
\dexp{exp} to a variable \keyw{var} of the same type, \dexp{dest},
\dexp{dests}, \dexp{data} and \dexp{ms} data expressions of types
{\tIP}, $\pow(\tIP)$, {\tDATA} and {\tMSG}, respectively, and $\msg$ a
data variable of type \tMSG.

Given a valuation of the data variables by concrete data values, the
 process $\cond{\varphi}\p$ acts as $\p$ if $\varphi$
evaluates to {\tt true}, and deadlocks if $\varphi$ evaluates to
{\tt false}.\footnote{As
    \label{fn:undefvalues}%
    operators we also allow \emph{partial} functions with the
    convention that any atomic formula containing an undefined subterm
    evaluates to {\tt false}.}
     In case $\varphi$ contains free variables that are not
yet interpreted as data values, values are assigned to these variables
in any way that satisfies $\varphi$, if possible.
The  process $\assignment{\keyw{var}\mathop{:=}\dexp{exp}}\p$
acts as $\p$, but under an updated valuation of the data variables.
The  process $\p+\q$ may act either as $\p$ or as
$\q$, depending on which of the two is able to act at all.  In a
context where both are able to act, it is not specified how the choice
is made. The process
$\broadcastP{\dexp{ms}}.\p$ broadcasts (the data value bound to the
expression) $\dexp{ms}$ to the other network nodes within range,
and subsequently acts as $\p$, whereas the process $\unicast{\dexp{dest}}{\dexp{ms}}.\p \prio \q$
tries to unicast the message $\dexp{ms}$ to the destination
\dexp{dest}; if successful it continues to act as $\p$ and otherwise
as $\q$.\footnote{The unicast is unsuccessful if the destination
    \dval{dest} is out of transmission range of the node $\dval{ip}$ performing
    the unicast, i.e., if in the dynamic network topology there is
    currently no link between $\dval{ip}$ and $\dval{dest}$.}
The latter models an abstraction
of an acknowledgment-of-receipt mechanism that is typical for unicast
communication but absent in broadcast communication, as implemented by
the link layer of wireless standards such as IEEE 802.11.
The process $\groupcastP{\dexp{dests}}{\dexp{ms}}.\p$ tries
to transmit \dexp{ms} to all destinations $\dexp{dests}$, and proceeds
as $\p$ regardless of whether any of the transmissions is successful.
The process $\send{\dexp{ms}}.\p$ synchronously transmits a message to another
process running on the same network node; this action can occur only
when the other process is able to receive the message.
The process $\receive{\msg}.\p$ receives any message $m$ (a
data value of type \tMSG) either from another node, from another
 process running on the same node or from the application layer process
on the local node.  It then proceeds as $\p$, but with the data
variable $\msg$ bound to the value $m$.  In particular,
$\receive{\newpkt{\dexp{data}}{\dexp{dip}}}$ models the injection of
a data from the application layer, where the function $\newpktID$ generates
a message containing the application layer
$\dexp{data}$ and the intended
destination address $\dexp{dip}$.
Data is delivered to the
application layer by \deliver{\dexp{data}}.

A (state of a) \emph{valuated process} $P$ is given as a pair $(\xi, \p)$ 
of an expression $\p$ built from the above syntax, together with a
(partial) \emph{valuation} function $\xi$ that specifies values of the data variables maintained by $\p$.
Finally, $P\parl Q$ denotes a parallel composition of processes $P$ and $Q$,
with information piped from right to left; in our application $Q$ will
be a message queue.

In the full process algebra \cite{ESOP12}, \emph{node expressions} $a\mathop{:}P\mathop{:}R$ are
given by process expressions $P$, annotated with an \emph{address}~$a$ and
a set of nodes $R$ that are within \emph{transmission range} of~$a$.
A partial network is then modelled as a parallel composition of node expressions, using the operator $\|$,
and a complete network is obtained by placing this composition in the
scope of an encapsulation operator $[\_\!\_\,]$. The main purpose of the encapsulation
operator is to prevent the receipt of messages that have never been
sent by other nodes in the network---with the exception of messages
$\newpkt{\dexp{data}}{\dexp{dip}}$ stemming from the application layer
of a node.
More details on the language \awn can be found in~\cite{TR13}.

To illustrate the use of {\awn} we consider a network
of two nodes on which the same process is running.
One node broadcasts an integer value. A received broadcast message will be delivered to
the application layer if its value is $1$. Otherwise the node
decrements its value and broadcasts the new value. The behaviour of
each node can be modelled by:

\newcommand{\XP}{\keyw{X}}%
\newcommand{\YP}{\keyw{Y}}%
  \begin{algorithmic}%
	\item[\XP(\keyw{n})]\hspace{-\labelsep}\ $\stackrel{{\it def}}{=}$
 	\textbf{broadcast}(\keyw{n}).\YP()
	\item[\YP()]\hspace{-\labelsep}\hspace{1.7mm} $\stackrel{{\it def}}{=}$
 	\textbf{receive}(\keyw{m}).([$\keyw{m} \mathord= 1$] \textbf{deliver}(\keyw{m}).\YP()\,.\\
	\hspace{78.5pt}
	+ [$\keyw{m} \mathord{\not=} 1$] \XP($\keyw{m}\mathord-1$))
  \end{algorithmic}
\vspace{2pt}
If a node is in a state $\XP(\keyw{n})$ it will broadcast $\keyw{n}$
and continue in state $\YP()$. If a node is in state $\YP()$, and it receives $\keyw{m}$, it has two
ways to continue. Process $[\keyw{m} \mathord= 1]$
$\textbf{deliver}(\keyw{m}).\YP()$ is enabled if $\keyw{m} \mathord= 1$. In that case $\keyw{m}$
will be delivered to the application layer, and the process returns to $\YP()$. Alternatively, if
$\keyw{m} \mathop{\not=} 1$, the process continues as $\XP(\keyw{m}\mathord-1)$. Note that calls to
processes use expressions as parameters, in this case $\keyw{m}\mathord-1$.

Let us have a look at two network topologies.
First, assume that the nodes $a$ and $b$ are within transmission range of each other; node $a$ in state
$\XP(2)$, and node $b$ in $\YP()$. In {\awn} this is formally
expressed as
$[\colonact{a}{\XP(2)}\mathord{:}\{b\}\,\|\,\colonact{b}{\YP()}\mathord{:}\{a\}]$,
although below we simply write $\XP(2)\,\|\,\YP()$.
Then, node $a$ broadcasts $2$ and
continues as $\YP()$. Node $b$ receives $2$, and continues as $\XP(1)$. Next $b$
broadcasts $1$, and continues as $\YP()$, while node $a$ receives $1$, and, since the condition
$\keyw{m}\mathord= 1$ is satisfied, \textbf{deliver}s $1$ and continues as $\YP()$. 
This gives rise to transitions from one state to the other:
\newcommand{\sm}[1]{\mbox{$\scriptstyle #1$}}
\[\begin{array}{l}
\XP(2)\,\|\,\YP() \ar{{\sm{a}:\textbf{broadcast}\sm{(2)}}} \YP()\,\|\,\XP(1)
\ar{{\sm{b}:\textbf{broadcast}\sm{(1)}}}\\[2mm]
\phantom{\XP(2)\,\|\,\YP() }\ar{{\sm{a}:\textbf{deliver}\sm{(1)}}} \YP()\,\|\,\YP()\,.
\end{array}\]
In state $\YP()\,\|\,\YP()$ no further activity is possible; the network has reached a \emph{deadlock}.

Second, assume that the nodes are not within transmission range; formally
$[\colonact{a}{\XP(2)}\mathop{:}\emptyset\,\|\,\colonact{b}{\YP()}\mathop{:}\emptyset]$.
Again $a$ is in state $\XP(2)$, and $b$ in $\YP()$.
As before, node $a$ broadcasts $2$ and continues as $\YP()$; but this
time the message is not received by any node; hence
no message is forwarded or delivered and both nodes end up in state~$\YP()$.

For the last scenario, we assume that $a$ and $b$ are
within transmission range and that
both nodes have the same initial state $\XP(1)$.
Assuming that no packet collisions occur, and node $a$ sends first:
\[\begin{array}{l}
\XP(1)\,\|\,\XP(1) \ar{{\sm{a}:\textbf{broadcast}\sm{(1)}}} \YP()\,\|\,\XP(1)
\ar{{\sm{b}:\textbf{broadcast}\sm{(1)}}}\\[2mm]
\phantom{\XP(1)\,\|\,\XP(1) }\ar{{\sm{a}:\textbf{deliver}\sm{(1)}}} \YP()\,\|\,\YP()\,.
\end{array}\]
Unfortunately, node $b$ is in a state where it cannot receive a message,
so $a$'s message ``remains unheard'' and $b$ will never deliver that message.
To avoid this behaviour, and ensure that both messages get delivered,
as happens in real WMNs, a message queue can be
introduced (see Section~\ref{ssec:message_queue}).

\section{Data Structure for AODV}\label{sec:types}
In this section we present the data structure needed for the
detailed formal specification of AODV.  As well as describing
\emph{types} for the information handled at the nodes during the
execution of the protocol we also define functions which will be used
to describe the precise intention---and overall effect---of the
various update mechanisms in an AODV implementation. The definitions
are grouped roughly according to the various ``aspects" of AODV and
the host network.

Many of the presented type and function definitions are straightforward;
so in principle this section can be skipped or be used as reference material.

\subsection{Mandatory Types}\label{ssec:ip}
As stated in the previous section, the data structure always consists
of application layer data, messages, IP addresses and sets of IP addresses.

\begin{enumerate}[(a)]
\item\label{tDATA} The ultimate purpose of AODV is to deliver
  \emph{application layer data}.  The type $\tDATA$ describes a
  set of application layer data items. An item of data is thus a particular element of that
  set, denoted by the variable $\data\in\tDATA$.
\item \emph{Messages} are used to send information via the network. In
  our specification we use the variable $\msg$ of the type $\tMSG$.
  We distinguish AODV control messages (route request, route
  reply, and route error) as well as \emph{data packets}: messages for sending
  application layer data (see \SSect{messages}).
\item
  The type $\tIP$ describes a set of IP addresses or, more generally, a
  \emph{set of node identifiers}.  In the RFC 3561~\cite{rfc3561},
  $\tIP$ is defined as the set of all IP addresses.  
  We assume that
  each node has a unique identifier $\dval{ip}\in\tIP$.
  Moreover, in our  model, each node \dval{ip} maintains a variable {\ip}
  which always has the value \dval{ip}.
  In any AODV control message, the variable {\sip} holds the IP
  address of the sender, and if the message is part of the \emph{route
  discovery process}---a route request or route reply message---we use
  {\oip} and {\dip} for the origin and destination of the route
  sought.  Furthermore, {\rip} denotes an unreachable destination 
  (a destination to which a route was established earlier, but this route is now broken) and
  {\nhip} the next hop on some route.
\end{enumerate}

\subsection{Sequence Numbers}\label{ssec:sequence numbers}

As explained in \Sect{aodv}, any node maintains its own \emph{sequence number}---%
the value of the variable {\sn}---and
a routing table whose entries describe routes to other nodes. The value of {\sn} increases over time.
In AODV each routing table entry  is equipped with a
sequence number to constitute a measure approximating the
relative freshness of the information held---a smaller number denotes
older information.  All sequence numbers of routes to
$\dval{dip}\in \tIP$ stored in routing tables are ultimately derived from
\dval{dip}'s own sequence number at the time such a route was discovered.

We denote the set of sequence numbers by $\tSQN$ and assume it to be
totally ordered.  By default we take $\tSQN$ to be $\NN$, and use
standard functions such as $\max$.  The initial sequence number of any
node is $1$.  We reserve a special element $0\in\tSQN$ to be used 
for the sequence number of a route, whose semantics is that no
sequence number for that route is known.  Sequence numbers are
incremented by the function\pagebreak[2]
\hypertarget{inc}{
\[\begin{array}{r@{\hspace{0.5em}}c@{\hspace{0.5em}}l}
\fninc:\tSQN&\to&\tSQN\\
\inc{\dval{sn}}&=&
\left\{\begin{array}{ll}
  \ifs{\dval{sn}+1}{\dval{sn}\not=0}\\
  \ow{\dval{sn}}
\end{array}\right.
\vspace{-1.5ex}
\end{array}\]
}
The variables $\osn$, $\dsn$ and $\rsn$ of type $\tSQN$ are used to denote the
sequence numbers of routes leading to the nodes $\oip$, $\dip$ and
$\rip$.

AODV tags sequence numbers of routes as ``known'' or ``unknown''. This
indicates whether the value of the sequence number can be trusted. The
sequence-number-status flag is set to unknown (\unkno) when a routing table entry is
updated with information that is not equipped with a sequence number
itself.  In such a case the old sequence number of the entry is
maintained; hence the value {\unkno} does not indicate
that no sequence number for the entry is known.
Here we use the set $\tSQNK=\{\kno,\unkno\}$ for the possible values
of the sequence-number-status flag; we use the variable $\keyw{dsk}$
to range over type $\tSQNK$.

\subsection{Modelling Routes}

In a network, pairs $(\dval{ip}_0, \dval{ip}_k) \in \tIP \times \tIP$
of nodes are considered to be ``connected" if $\dval{ip}_0$ can send
to $\dval{ip}_k$ directly, i.e., $\dval{ip}_0$ is in transmission range of
$\dval{ip}_k$ and vice versa. We say that such nodes are connected by
a single \emph{hop}. When $\dval{ip}_0$ is not connected to $\dval{ip}_k$
then messages from $\dval{ip}_0$ directed to $\dval{ip}_k$ need to be
``routed" through intermediate nodes. We say that a \emph{route}
(from $\dval{ip}_0$ to $\dval{ip}_k$) is made up of a sequence
$[\dval{ip}_0,\dval{ip}_1,\dval{ip}_2,\dots,\dval{ip}_{k-1},\dval{ip}_k]$,
where $(\dval{ip}_{i}, \dval{ip}_{i+1})$, $i=0,\dots, k\mathord-1$, are
connected pairs; the \emph{length} or \emph{hop count} of the route is
the number of single hops, and any node $\dval{ip}_i$ needs only to
know the ``next hop" address $\dval{ip}_{i+1}$ in order to be able to
route messages intended for the final destination $\dval{ip}_{k}$.

In operation, information about routes to certain destinations is
stored in \emph{routing tables} maintained at each node. This information sometimes needs
to be re-evaluated in regard to
its validity. Routes may become \emph{invalid} if one of the pairs
$(\dval{ip}_i, \dval{ip}_{i+1})$ in the hop-to-hop sequence gets
disconnected. Then AODV may be reinvoked, as the need arises, to
discover alternative routes. 

In addition to the next hop and hop count, AODV also ``tags" a route
with its validity, sequence number and sequence-number status.
Information about invalid routes is preserved until fresh
information is received that establishes a valid replacement route.
The purpose of this is to compare the sequence number and hop
  count of the replacement route with that of the invalid one,
  to check that the information is indeed fresher (or equally fresh
  while the replacement route is shorter).
For every route, a node moreover stores a list of \emph{precursors},
modelled as a set of $\tIP$ addresses. This set collects all nodes
which are currently potential users of the route, and are located one
hop further ``upstream''.  When the interest of other
nodes emerges, these nodes are added to the precursor
list;\footnote{The RFC does not mention a situation where nodes are
dropped from the list, which seems curious.} the main purpose of
recording this information is to inform those nodes when the route
becomes invalid.

In summary, following the RFC, a routing table entry (or entry for short) is given by $7$ components:

\begin{enumerate}[(a)]
\item The destination IP address---an element of $\tIP$;
\item The destination sequence number---an element of $\tSQN$;
\item The sequence-number-status flag---an element of\linebreak the set $\tSQNK=\{\kno,\unkno\}$;
\item A flag tagging the route as being valid or invalid---an element
  of the set $\tFLAG= \{\val,\inval\}$. We use the variable $\flag$
  to range over type $\tFLAG$;
\item The hop count, which is an element of $\NN$.  The variable
  $\hops$ ranges over the type $\NN$ and we make use of the standard
  function $+1$;
\item The next hop, which is again an element of $\tIP$; and
\item A precursor list, which is modelled as an element of $\pow(\tIP)$.%
\footnote{The word ``precursor list'' is used in the RFC, but no properties of lists are used.}
  The variable $\pre$ ranges over $\pow(\tIP)$.
\end{enumerate}
We denote the type of routing table entries by $\tROUTE$, and use
the variable \keyw{r}.
A tuple
\[
(\dval{dip}\comma\dval{dsn}\comma\dval{dsk}\comma\dval{flag}\comma\dval{hops}\comma\dval{nhip}\comma\dval{pre})
\]
describes a route to $\dval{dip}$ of length $\dval{hops}$ and validity
$\dval{flag}$; the very next node on this route is $\dval{nhip}$; the
last time the entry was updated the destination sequence number was
$\dval{dsn}$; \dval{dsk} denotes whether the sequence number
is ``outdated'' or can be used to reason about freshness of the route.
Finally, $\dval{pre}$ is a set of all neighbours who are
``interested'' in the route to $\dval{dip}$.  A node being
``interested" in the route is somewhat sketchily defined as one which
has previously used the current node to route messages to
$\dval{dip}$. Interested nodes are recorded in case the route to
$\dval{dip}$ should ever become invalid, so that they may subsequently
be informed.  We use projections $\pi_{1},\dots\pi_{7}$ to select the
corresponding component from the $7$-tuple: For example,
$\pi_6:\tROUTE\to\tIP$ determines the next hop.

\subsection{Routing Tables}\label{ssec:rt}

{
\renewcommand{\ip}{\dval{ip}}
\renewcommand{\dip}{\dval{dip}}
\renewcommand{\oip}{\dval{oip}}
\renewcommand{\sip}{\dval{sip}}
\renewcommand{\rip}{\dval{rip}}
\renewcommand{\rt}{\dval{rt}}
  \newcommand{\nrt}{\dval{nrt}}
\renewcommand{\route}{\dval{r}}
  \newcommand{\s}{\dval{s}}
  \newcommand{\nr}{\dval{nr}}
  \newcommand{\ns}{\dval{ns}}
\renewcommand{\osn}{\dval{osn}}
\renewcommand{\dsn}{\dval{dsn}}
\renewcommand{\rsn}{\dval{rsn}}
\renewcommand{\flag}{\dval{flag}}
\renewcommand{\hops}{\dval{hops}}
\renewcommand{\nhip}{\dval{nhip}}
\renewcommand{\pre}{\dval{pre}}
  \newcommand{\npre}{\dval{npre}}
\renewcommand{\dests}{\dval{dests}}
\renewcommand{\rreqid}{\dval{rreqid}}
\renewcommand{\rreqs}{\dval{rreqs}}

Nodes store all their information about routes in their \emph{routing
tables}; a node \dval{ip}'s routing table consists of a set of routing
table entries, exactly one for each known destination.  Thus, a
routing table is defined as a set of entries, with the restriction
that each has a different destination $\dip$, i.e., the first
component of each entry in a routing table is unique.\footnote{As an
alternative to restricting the set, we could have defined routing
tables as partial functions from $\tIP$ to $\tROUTE$, in which case it
makes more sense to define an entry as a $6$-tuple, not including the
the destination IP as the first component.}  Formally, we define the
type $\tRT$ of routing tables by
\[\begin{array}{r@{\hspace{0.5em}}c@{\hspace{0.5em}}l}	
\tRT &:=& \{\rt \in\!\pow(\tROUTE)\mid \forall \route,\s\mathbin\in\rt:\route\not=\s\Rightarrow\pi_{1}(\route)\not=\pi_{1}(\s)\}\,.
\end{array}\]
AODV chooses between alternative routes if necessary to ensure
that only one route per destination ends up in a given node's routing table.
In our model, each node \dval{ip} maintains a variable \keyw{rt},
whose value is the current routing table of the node. 

In the formal model (and indeed in any AODV implementation) we need to
extract the components of the entry for any given destination from a
routing table. To this end, we define the following partial
functions---they are partial because the routing table need not have
an entry for the given destination.
We begin by selecting the entry in a routing table corresponding to a given destination $\dip$:
\hypertarget{selroute}{
\[\begin{array}{l}
		\fnselroute:\tRT\times\tIP\rightharpoonup\tROUTE\\
		\selr{\rt}{\dip}:=
		\left\{
		  \begin{array}{ll}
		   \ifs{\route}{\route\in\rt\wedge \pi_{1}(\route)=\dip}\\
		   \undefined
		\end{array}\right.\!\!\!\!
\end{array}\]
}
Through
the projections $\pi_{1},\dots,\pi_{7}$, defined above, we can now
select the components of a selected entry:
\begin{enumerate}[(a)]	
\item The \emph{destination sequence number} relative to $\dip$:
\hypertarget{sqn}{
\[\begin{array}{l}
		    	  \fnsqn : \tRT\times\tIP\to \tSQN\\
	    \sqn{\rt}{\dip}:=
	    \left\{
	    \begin{array}{ll}
	       \ifsnewline{\pi_{2}(\selr{\rt}{\dip})}{\selr{\rt}{\dip}\mbox{ is defined}}\\
	        \ow{0}
	    \end{array}\right.
	     \end{array}\]
}
\item The \emph{``known'' status} of a route's sequence number:
\hypertarget{sqnf}{
\[\begin{array}{l}
		    	  \fnsqnf : \tRT\times\tIP\to \tSQNK\\
	    \sqnf{\rt}{\dip}:=
	    \left\{
	    \begin{array}{ll}
	        \ifsnewline{\pi_{3}(\selr{\rt}{\dip})}{\selr{\rt}{\dip}\mbox{ is defined}}\\
	        \ow\unkno
	    \end{array}\right.
	     \end{array}\]
}
\item The \emph{validity status} of a recorded route:
\hypertarget{status}{
\[\begin{array}{l}
              \fnstatus :\ \tRT\times\tIP\rightharpoonup \tFLAG\\
	     \status{\rt}{\dip}:= \pi_{4}(\selr{\rt}{\dip})\,.
	\end{array}
\]	
}
\item The \emph{hop count} of the route from the current node (hosting $\rt$) to $\dip$:
\hypertarget{dhops}{
\[\begin{array}{l}
	       \fndhops :\tRT\times\tIP\rightharpoonup \NN\\
	     \dhops{\rt}{\dip}:=  \pi_{5}(\selr{\rt}{\dip})\,.
  \end{array}\]
}
\item The \emph{identity of the next node on the route to} $\dip$ (if such a route is known):
\hypertarget{nhop}{
\[\begin{array}{l}
 	\fnnhop :\tRT\times\tIP\rightharpoonup \tIP\\
         \nhop{\rt}{\dip}:= \pi_{6}(\selr{\rt}{\dip})\,.
\end{array}\]
}
\item The set of \emph{precursors} or neighbours interested in
  using the route from $\ip$ to $\dip$:
\hypertarget{precs}{
\[\begin{array}{l}
	    \fnprecs : \tRT\times\tIP\rightharpoonup\pow(\tIP)\\
	     \precs{\rt}{\dip}:= \pi_{7}(\selr{\rt}{\dip})\,.
  \end{array}\]
}
\end{enumerate}
The domain of these partial functions changes during the operation of
AODV as more routes are discovered and recorded in the routing table
$\rt$.  The first two functions are extended to be total functions:
whenever there is no route to $\dip$ inside the routing table under
consideration, the sequence number is set to ``unknown''  $(0)$ and 
the sequence-number-status flag is set to ``unknown'' $(\unkno)$, respectively. In the same
style each partial function could be turned into a total one. However,
in the specification we use these functions only when they are defined.

We are not only interested in information about a single route, but
also in information on a routing table:
\begin{enumerate}[(a)]	
\item The set of destination IP addresses for \emph{valid} routes in $\rt$ is given by
\[\begin{array}{l}
	      \fnakD :\tRT\to \pow(\tIP)\\
	    \akD{\rt}:= \{\dip\mid (\dip\comma*\comma*\comma\val\comma*\comma*\comma*)\in\rt\}\,.\footnotemark
\end{array} \]
\footnotetext{We use ``$*$'' as a wildcard.}
	
\item The set of destination IP addresses for \emph{invalid}\linebreak routes in $\rt$ is
\[\begin{array}{l}
	      \fnikD :\tRT\to \pow(\tIP)\\
	      \ikD{\rt}:= \{\dip\mid (\dip\comma*\comma*\comma\inval\comma*\comma*\comma*)\in\rt\}\,.
  \end{array} \]

\item Last, we define the set of destination IP addresses for \emph{known} routes by
\[\begin{array}{l}
	      \fnkD :\tRT\to \pow(\tIP)\\
	      \kD{\rt}:=\akD{\rt}\cup\ikD{\rt} \\
	       \phantom{\kD{\rt}\mbox{:}}= \{\dip\mid (\dip\comma*\comma*\comma*\comma*\comma*\comma*)\in\rt\}\,.
\end{array}\]
\end{enumerate}
The partial functions $\fnselroute$, {\fnstatus},
{\fndhops}, {\fnnhop} and {\fnprecs} are defined for {\rt} and {\dip}
iff $\dip \in \kD{\rt}$.

\subsection{Updating Routing Tables}

Routing tables can be updated for three principal reasons. The first
is when a node needs to adjust its list of precursors relative to a
given destination; the second is when a received request or response
carries information about network connectivity; and the last when
information is received to the effect that a previously valid route
should now be considered invalid. We define an update function for
each case.

\subsubsection{Updating Precursor Lists}

Recall that the precursors of a given node $\ip$ relative to a
particular destination $\dip$ are the nodes that are ``interested" in a route to {\dip}
via {\ip}.  The function $\fnaddprec$ takes a routing table entry and a set of IP addresses
$\npre$ and updates the entry by adding $\npre$ to the list of precursors
already present:
\[\begin{array}{l}
\fnaddprec : \tROUTE\times \pow(\tIP) \to \tROUTE\\
\addprec{(\dip\comma\dsn\comma\dval{dsk}\comma\flag\comma\hops\comma\nhip\comma\pre)}{\npre}:=\\
\qquad(\dip\comma\dsn\comma\dval{dsk}\comma\flag\comma\hops\comma\nhip\comma\pre\cup\npre)\,.
\end{array}\]

Often it is necessary to add  precursors to an entry of a given
routing table. For that, we define the function $\fnaddprecrt$, which takes a routing table \rt, a destination {\dip} and a set of IP addresses
$\npre$ and updates the entry with destination $\dip$ by adding $\npre$ to the list of precursors
already present. It is only 
defined if an entry for destination $\dip$ exists.
\hypertarget{addprert}{
\[\begin{array}{l}
\fnaddprecrt : \tRT\times \tIP\times \pow(\tIP) \rightharpoonup \tRT\\
\addprecrt{\rt}{\dip}{\npre} :=
(\rt - \{\selr{\rt}{\dip}\})\\
\qquad{}\cup\{\addprec{\selr{\rt}{\dip}}{\npre}\}
\end{array}
\]
}
Formally, we remove the entry with destination $\dip$ from the routing table and 
insert a new entry for that destination. This new entry is the same as before---only the precursors have been added.

\subsubsection{Inserting New Information in Routing Tables}\label{sssec:update}

If a node gathers new information about a route to a
destination \dip, then it updates its routing table depending on
its existing information on a route to \dip.
If no route to {\dip} was known at all, it inserts a new entry
in its routing table  recording the information received.
If it already has some (partial) information then it may update
this information, depending on whether the new route is fresher or
shorter than the one it has already.
We define an update function $\upd{\rt}{\route}$ of a routing table 
$\rt$ with an entry $\route$ only when
 $\route$ is valid, i.e., $\pi_{4}(\route)=\val$,
$\pi_{2}(\route)=0\Leftrightarrow\pi_{3}(\route)=\unkno$,
and $\pi_{3}(\dval{r})=\unkno\Rightarrow\pi_{5}(\dval{r})=1$.
After we have introduced our formal specification for AODV
 in Section~\ref{sec:modelling_AODV}, we will show that we only 
 use the function \fnupd\ if this condition is satisfied (\Prop{upd_well_defined});
hence this definition is sufficient.
\hypertarget{update}{
\[\begin{array}{l}
\fnupd : \tRT\times\tROUTE\ \ \rightharpoonup\ \ \tRT\label{df:update}\\
\upd{\rt}{\hspace{-1pt}\route}:=\\
\qquad \left\{
\begin{array}{@{\,}ll@{}}
\ifs		{\rt\cup\{\route\}}	{\pi_{1}(\route)\not\in\kD{\rt}}\\[1mm]
\ifs		{\nrt\cup\{\nr\}}		{\pi_{1}(\route)\in\kD{\rt}}\\
	\nifs	{\wedge\sqn{\rt}{\pi_{1}(\route)}<\pi_{2}(\route)}\\[1mm]
\ifs		{\nrt\cup\{\nr\}}		{\pi_{1}(\route)\in\kD{\rt}}\\
	\nifs	{\wedge\sqn{\rt}{\pi_{1}(\route)}=\pi_{2}(\route)}\\
	\nifs	{\wedge\dhops{\rt}{\pi_{1}(\route)}>\pi_{5}(\route)}\\[1mm]
\ifs		{\nrt\cup\{\nr\}}		{\pi_{1}(\route)\in\kD{\rt}}\\
	\nifs	{\wedge \sqn{\rt}{\pi_{1}(\route)}=\pi_{2}(\route)}\\
	\nifs	{\wedge \status{\rt}{\pi_{1}(\route)}=\inval}\\[1mm]
\ifs		{\nrt\cup\{\nr'\}}		{\pi_{1}(\route)\in\kD{\rt}}\\
	\nifs	{\wedge  \pi_3(\route)=\unkno}\\[1mm]
\ow[,]		{\nrt\cup\{\ns\}}
\end{array}
\right.
\end{array}\]
}
where $\s:=\selr{\rt}{\pi_{1}(\route)}$ is the current entry in the
routing table for the destination of $\route$ (if it exists), and
$\nrt := \rt -\{\s\}$ is the routing table without that entry.
The entry $\nr:=\addprec{\route}{\pi_{7}(\s)}$ is identical to~$\route$ except
that the precursors from $\s$ are added and $\ns:=\addprec{\s}{\pi_{7}(\route)}$
is generated from $\s$ by adding the precursors from $\route$. 
Lastly, 
$\nr'$ is identical to $\nr$ except that the sequence number is replaced by the one from 
the route $s$. More precisely,
$\nr':=(\dip_{\nr}\comma\pi_{2}(\s)\comma\dval{dsk}_{\nr}\comma\flag_{\nr}\comma\hops_{\nr}\comma\nhip_{\nr}\comma\pre_{\nr})$ if
$\nr=(\dip_{\nr}\comma*\comma\dval{dsk}_{\nr}\comma\flag_{\nr}\comma\hops_{\nr}\comma\nhip_{\nr}\comma\pre_{\nr})$.
In the situation where $\sqn{\rt}{\pi_{1}(\route)}=\pi_{2}(\route)$ both routes $\nr$ and $\nr'$ are equal.
Therefore, though the cases of the above definition are not
mutually exclusive, the function is well defined.

The first case describes the situation where the routing table does not contain any
information on a route to $\dip$. 
The second case models the situation where the new route has a greater
sequence number. As a consequence all the information from the incoming information 
is copied into the routing table. In the third and fourth case the sequence numbers 
are the same and cannot be used to identify better information. 
Hence other measures are used. The route inside the routing table is only replaced if 
either the new hop count is strictly smaller---a shorter route has been found---or if the route inside 
the routing table is marked as invalid.  The fifth case deals with the situation where a
new route to a known destination has been found without any
  information on its sequence number ($\pi_{2}(\route)=0\wedge\pi_{3}(\route)=\unkno$).
In that case the routing table entry to that destination is always
updated, but the existing sequence number is maintained, and marked as ``unknown''.

Note that we do not update if we receive a new entry where the sequence number and
the hop count are identical to the current entry in the routing table. Following the
RFC, the time period (till the valid route becomes invalid) should be reset; however
at the moment we do not model timing aspects.

\subsubsection{Invalidating Routes}\label{sssec:invalidate}

Invalidating routes is a main feature of AODV; if a route is not valid
any longer its validity flag has to be set to invalid. By doing
this, the stored information about the route, such as the sequence number or the hop count,
remains accessible. The process of invalidating a routing table entry follows four rules:
(a) any sequence number is incremented by $1$, except
(b) the truly unknown sequence number ($\dval{sqn}=0$, which will only
occur if $\dval{dsk}=\unkno$) is not incremented,
(c) the validity flag of the entry is set to \inval, and
(d) an invalid entry cannot be invalidated again.
However, in exception to (a) and (b), when the invalidation is in
response to an error message, this message also contains a new (and
already incremented) sequence number for each destination to be invalidated.

The function for invalidating routing table entries takes as arguments a routing
table and a set of destinations $\dests\in\pow(\tIP\times\tSQN)$.
Elements of this set are $(\rip,\rsn)$-pairs that not only identify an
unreachable destination $\rip$, but also a sequence number that
describes the freshness of the faulty route.  As for routing tables,
we restrict ourselves to sets that have at most one entry for each
destination; this time we formally define {\dests} as a \emph{partial
function} from $\tIP$ to $\tSQN$, i.e.\ a subset of $\tIP\times\tSQN$
satisfying 
\[(\rip\comma\rsn),(\rip\comma\rsn')\in \dests \ \Rightarrow\  \rsn=\rsn'\,.
\]
We use the variable \keyw{dests} to range over such sets.
When invoking {\fninv} we either distil {\dests} from an error
message, or determine {\dests} as a set of pairs $(\rip,\inc{\sqn{\rt}{\rip}}$,
where the operator {\fninc} (from \SSect{sequence numbers}) takes care of (a) and (b).
Moreover, we will distil or construct {\dests} in such a way that it only lists
destinations for which there is a valid entry in the routing table---this takes care
of (d).
\hypertarget{invalidate}{
\[\begin{array}{l}
\fninv : \tRT\times(\tIP\rightharpoonup\tSQN) \to\tRT\\
\inv{\rt}{\dests}:= \{\route\in\rt\,|\, (\pi_{1}(\route),*)\not\in\dests\}\\
\qquad\cup\{(\pi_{1}(\route)\comma\rsn\comma\pi_{3}(\route)\comma\inval\comma\pi_{5}(\route)\comma\pi_{6}(\route)\comma\pi_{7}(\route))\mid\\
\qquad\phantom{\cup\{}\route\in\rt\wedge(\pi_{1}(r),\rsn)\in\dests\}
\end{array}\]
}

\noindent
All entries in the routing table for a destination $\rip$ in $\dests$
are modified.
The modification replaces the value {\val} by {\inval} and
the sequence number in the entry by the corresponding sequence number from $\dests$.

Copying the sequence number from $\dests$ leaves the possibility
that the destination sequence number of an entry is decreased, which would violate one
of the fundamental assumption of AODV and may yield unexpected behaviour.
However, we will show that a decrease of a destination sequence number does not occur in our model of AODV.

\subsection{Route Requests}\label{ssec:rreqs}

A route request---RREQ---for a destination $\dip$ is initiated by a
node (with routing table $\rt$) if this node wants to transmit a data
packet to $\dip$ but there is no valid entry for $\dip$ in the routing
table, i.e.\ $\dip \mathbin{\not\in} \akD{\rt}$. When a new route request is sent
out it contains the identity of the originating node $\oip$, and a
\emph{route request identifier} (RREQ ID); the type of all such identifiers
 is denoted by $\tRREQID$, and the variable \keyw{rreqid}
ranges over this type. This information does not change, even when
the request is re-broadcast by any receiving node that does not
already know a route to the requested destination. In this way any
request still circulating through the network can be uniquely
identified by the pair $(\oip, \rreqid)\in\tIP\times\tRREQID$. For our
specification we set $\tRREQID=\NN$. In our model, each node
maintains a variable \keyw{rreqs} of type
		$\pow(\tIP\times\tRREQID)$
of sets of such pairs to store the sets of route requests seen
by the node so far. Within this set, the node records the requests
it has previously initiated itself.
\hypertarget{nrreqid}{To ensure a fresh {\rreqid} for each new RREQ it generates,
the node {\ip} applies the following function:
\[\begin{array}{l}
	     	  \fnnrreqid:\pow(\tIP\times\tRREQID)\times\tIP\to \tRREQID\\
		\nrreqid{\rreqs}{\ip}:=\max\{n\mid(\ip,n)\in\rreqs\}+1\,,
\end{array}\]
where we take the maximum of the empty set to be $0$.}

\subsection{Queued Packets}\label{ssec:store}
Strictly speaking the task of sending data packets is not regarded as
part of the AODV protocol---however, failure to send a packet because
either a route to the destination is unknown, or a previously known
route has become invalid, prompts AODV to be activated. In our
modelling we describe this interaction between packet sending and
AODV, providing the minimal infrastructure for our specification.

If a new packet is submitted by a client of AODV to a node, it may
need to be stored until a route to the packet's destination has been
found and the node is not busy carrying out other AODV tasks. We use a
queue-style data structure for modelling the store of packets at a
node, noting that at each node there may be many data queues, one for
each destination. In general, we denote queues of type $\tTYPE$ by
$[\tTYPE]$, denote the empty queue by $[\,]$, and make use of the
standard (partial) functions
$\hypertarget{head}{\fnhead}:[\tTYPE]\rightharpoonup\tTYPE$,
$\fntail:[\tTYPE]\rightharpoonup[\tTYPE]$ and
$\fnappend:\tTYPE\times[\tTYPE]\rightarrow[\tTYPE]$ that return the
``oldest'' element in the queue, remove the ``oldest'' element, and
add a packet to the queue, respectively. 

The data type\vspace{-1ex}
\[\begin{array}{l}
\tQUEUES := 
\big\{\dval{store} \in\pow(\tIP\times\tPendingRREQ\times[\tDATA])\mid\\
\qquad \big((\dip\comma p\comma q),(\dip\comma p'\!\comma q')\in\dval{store} \Rightarrow p\mathop=p' \wedge q\mathop=q'\big)\big\}
\end{array}\]
describes stores of enqueued data
packets for various destinations, where $\tPendingRREQ:=\{\pen,\nonpen\}$. An element $(\dip, p, q) \in \tIP\times\tPendingRREQ\times[\tDATA]$
denotes the queue $q$ of packets destined for $\dip$; the \penFlag $p$ is 
{\nonpen} if a new route discovery process for $\dip$ still needs to be initiated, i.e., a route request message needs to be sent.
The value {\pen} indicates that such a RREQ message has
been sent already, and either the reply is still pending or a route to $\dip$ has been established.
The flag is set to {\nonpen} when a routing table entry is invalidated. 

As for routing tables, we require that there is at most one entry for
every IP address.
In our model, each node maintains a variable {\queues} of type
{\tQUEUES} to record its current store of data packets.

\renewcommand{\queues}{\dval{store}} 
\renewcommand{\data}{\dval{d}}

We define some functions for inspecting a store:
\begin{enumerate}[(a)]
\item Similar to $\fnselroute$, we need a function that is able to extract the queue for a given destination:
\[\begin{array}{l}
\fnselqueue:\tQUEUES\times\tIP \to  [\tDATA]\\
\selq{\queues}{\dip}:=\left\{
\begin{array}{ll}
  \ifs{q}{(\dip\comma *\comma q)\in\queues}\\
  \ow{[\,]}
\end{array}\right.
\end{array}\]
\item We define a function $\fnqD$ to extract the destinations for which there are unsent packets:
\[\begin{array}{l}
\fnqD:\tQUEUES\to\pow(\tIP)\\
	\qD{\queues}:=
	\{\dip\mid(\dip\comma *\comma *)\in\queues\}\,.
\end{array}\]
\end{enumerate}
Next, we define operations for adding and removing data packets from a store.
\begin{enumerate}[(a)]
\setcounter{enumi}{2}
\item \hypertarget{add}{
Adding a data packet for a particular destination to a store is defined by:
\[\begin{array}{l@{}}
\fnadd:\tDATA\times\tIP\times\tQUEUES \to\tQUEUES\\
\add{\data}{\dip}{\queues}:=\\
\qquad\left\{
\begin{array}{ll@{}}
  \ifsnewline	{\queues\cup\{(\dip\comma\nonpen\comma\append{\data}{[\,]})\}}
  					{(\dval{dip}\comma*\comma*)\notin\queues}\\
  \ifsnewline	{\queues-\{(\dip,p,q)\}\cup\{(\dip\comma p\comma\append{\data}{q})\}}
  					{(\dval{dip}\comma p\comma q)\in\queues\,.}
\end{array}\right.\!\!\!
\end{array}\]}
Informally, the process selects 
the entry ${(\dip, p\comma q)}\in\queues\in\tQUEUES$,
where $\dip$ is the destination of the application layer data
$\data$,
and appends {\data} to queue $q$ of {\dip} in that triple;
the \penFlag $p$ remains unchanged.
In case there is no entry for {\dip} in \queues,
the process creates a new queue $[\data]$ of stored packets that only contains the
data packet under consideration and inserts it---together with $\dip$---into the store;
the \penFlag is set to \nonpen, since a route request needs to be sent.

\item \hypertarget{drop}{
To delete the oldest packet for a particular destination from a store, we define:
\hypertarget{drop}{
\[\begin{array}{l}
\fndrop:\tIP\times\tQUEUES \rightharpoonup\tQUEUES\\
\drop{\dip}{\queues}:=\\
\quad\left\{
\begin{array}{ll}
  \ifsnewline	{\queues-\{(\dip\comma*\comma q)\}}{\tail{q}=[\,]}\\
  \multicolumn{2}{l}{{\queues-\{(\dip\comma p\comma q)\}\cup\{(\dip\comma p\comma\tail{q})\}}}\\
  \nifs{\mbox{otherwise}\,,}
\end{array}\right.
\end{array}\]
}
\noindent where $q=\selq{\queues}{\dip}$ is the selected queue for destination $\dip$.
If $\dip\not\in\qD{\queues}$ then $q={[\,]}$. Therefore $\tail{q}$ and hence also $\drop{\dip}{\queues}$ is undefined.
Note that if $\data$ is the last queued packet for a specific
destination, the whole entry for the destination is removed from $\queues$.}
\end{enumerate}
\noindent In our model of AODV we use only {\fnadd} and {\fndrop} to update a store. 
This ensures that the store will never contain a triple $(\dip\comma*\comma [\,])$ with an empty data queue, 
that is
\begin{equation}\label{non-empty}
\dip\in \qD{\queues} \Rightarrow \selq{\queues}{\dip}\not={[\,]}\,.
\end{equation}
Finally, we define operations for reading and manipulating the \penFlag of a queue.
\begin{enumerate}[(a)]
\setcounter{enumi}{4}
\item We define a partial function $\fnfD$ to extract the flag for a destination for which there are unsent packets:
\hypertarget{qflag}{
\[\begin{array}{l}
\fnfD:\tQUEUES\times\tIP\rightharpoonup\tPendingRREQ\\
\fD{\queues}{\dip}:=\left\{
\begin{array}{ll}
  \ifs{p}{(\dip,p,*)\in\queues}\\
  \undefined
\end{array}\right.
\end{array}\]}

\item To change the status of the \penFlag, we define functions $\fnsetrrf$ and $\fnunsetrrf$.
After a route request for destination $\dip$ has been initiated, the \penFlag for $\dip$ has to be set to $\pen$.
\hypertarget{setrrf}{
\[\begin{array}{l}
\fnunsetrrf:\tQUEUES\times\tIP\to\tQUEUES\\
\unsetrrf{\queues}{\dip}:=\\
\qquad\left\{
\begin{array}{ll}
  \ifsnewline {\queues-\{(\dip,*,q)\}\cup\{(\dip,\pen,q)\}}{\{(\dip,*,q)\}\in\queues}\\
  \ow\queues
\end{array}\right.
\end{array}\]}
In case that 
there is no queued data for destination {\dip}, the {\queues} remains unchanged.

Whenever a route is invalidated the corresponding \penFlag has to be set to \nonpen;
this indicates that the protocol might need to initiate a new route
discovery process.
Since the function \hyperlink{invalidate}{$\fninv$} invalidates sets of routing table entries, we 
define a function with a set of destinations
$\dests\in\pow(\tIP\times\tSQN)$ as one of its arguments (annotated with sequence numbers, which are not used here).
\hypertarget{setrrf}{
\[\begin{array}{l}
\fnsetrrf:\tQUEUES\times(\tIP\rightharpoonup\tSQN)\to\tQUEUES\\
\setrrf{\queues}{\dests}:=\\
\qquad  \{(\dip\comma p\comma q) \in \queues\mid (\dip\comma *) \notin\dests\}\\
\qquad {}\cup\{(\dip\comma \nonpen\comma q) \mid (\dip\comma p\comma q) \in \queues\\
\qquad\phantom{{}\cup\{(\dip\comma \nonpen\comma q) \mid{}}
	 {}\wedge(\dip\comma *) \in\dests\}\;.
\end{array}\]}

\end{enumerate}

\subsection{Messages and Message Queues}\label{ssec:messages}
Messages are the main ingredient of any routing protocol.
The message types used in the AODV protocol are route request,
route reply, and route error. To generate theses
messages, we use functions
\vspace{-0.6ex}
\[\begin{array}{l}
\rreqID:\NN \times \tRREQID \times \tIP \times \tSQN \times \tSQNK\times \tIP \times \tSQN \times \tIP\\
\phantom{\rreqID:}\rightarrow \tMSG\\
\rrepID:\NN \times \tIP \times \tSQN \times \tIP \times \tIP \rightarrow \tMSG\\
\rerrID:(\tIP\rightharpoonup\tSQN) \times \tIP \rightarrow \tMSG\,.\footnotemark
\end{array}\vspace{-0.6ex}\]
\footnotetext{The ordering of the arguments follows the RFC.}%
The function $\rreq{\hops}{\rreqid}{\dip}{\dsn}{\dval{dsk}}{\oip}{\osn}{\sip}$
generates a route request. Here, $\hops$ indicates the hop count from
the originator $\oip$---that, at the time of sending, had the sequence
number $\osn$---to the sender of the message $\sip$; $\rreqid$
uniquely identifies the route request;  $\dsn$ is the least level
of freshness of a route to dip that is acceptable to \oip---it has been
obtained by incrementing the latest sequence number received in the
past by {\oip} for a route towards $\dip$; and \dval{dsk}
  indicates whether we can trust that number. In case no 
sequence number is known, $\dsn$ is set to $0$ and $\dval{dsk}$ to $\unkno$.
By $\rrep{\hops}{\dip}{\dsn}{\oip}{\sip}$ a route reply message is obtained.
Originally, it was generated by
 $\dip$---where $\dsn$ denotes the
sequence number of $\dip$ at the time of sending---and  is destined
for $\oip$; the last sender of the message was the node with IP
address $\sip$ and the distance between $\dip$ and $\sip$ is given by $\hops$.
The error message is generated by $\rerr{\dests}{\sip}$, where
$\dests:\tIP\rightharpoonup\tSQN$ is the list of unreachable
destinations and $\sip$ denotes the sender.
Every unreachable destination $\rip$ comes together with the incremented last-known sequence number $\rsn$.

Next to these AODV control messages, we use for our specification
also data packets: messages that carry application layer data.
\[\begin{array}{l}
\newpktID:\tDATA \times \tIP \rightarrow \tMSG\\
\pktID:\tDATA \times \tIP \times \tIP \rightarrow \tMSG
\end{array}\]
Although these messages are not part of the protocol itself, they are
necessary to initiate error messages, and to trigger the route discovery process.
$\newpkt{\data}{\dip}$ generates a message containing new application layer data
$\data$ destined for a particular destination $\dip$. Such a
message is submitted to a node by a client of the AODV protocol
hooked up to that node. The function \linebreak[4]$\pkt{\data}{\dip}{\sip}$
generates a message containing application layer data {\data},
that is sent by the sender $\sip$
to the next hop on the route towards $\dip$.

All messages received by a particular node are first stored in a queue
(see Section~\ref{ssec:message_queue} for a detailed description). To
model this behaviour we use a message queue, denoted by the variable
$\msgs$ of type $[\tMSG]$.  As for every other queue, we will freely
use the functions $\fnhead$, $\fntail$ and $\fnappend$.
}

Table~\ref{tab:types} provides a summary of the entire data structure we use.
\begin{table*}[p]
\caption{Data structure
\label{tab:types}}
\begin{tabular}{@{}|l|l|l|@{}}
\hline
\textbf{Basic Type} & \textbf{Variables} & \textbf{Description}\\
\hline
 \tIP			&\ip, \dip, \oip, \rip, \sip, \nhip	&node identifiers\\
 \tSQN		&\dsn, \osn, \rsn, \sn			&sequence numbers\\
 \tSQNK		&\keyw{dsk}        			&sequence-number-status flag\\
 \tFLAG		&\flag					&route validity\\
 \NN			&\hops					&hop counts\\
 \tROUTE  	& \keyw{r} 						&routing table entries\\
 \tRT			&\rt						&routing tables\\		
 \tRREQID		&\rreqid					&request identifiers\\
 \tPendingRREQ&		                                &\penFlag\\
 \tDATA 		&\data               				&application layer data\\
 \tQUEUES       	&\queues					&store of queued data packets\\
 \tMSG		&\msg					&messages\\
\hline\hline
\textbf{Complex Type} & \textbf{Variables} & \textbf{Description}\\
\hline
${[\tTYPE]}$						&			&queues with elements of type \tTYPE\\
\quad[\tMSG]						&\msgs		&message queues\\
$\pow(\tTYPE)$			        		&			&sets consisting of elements of type \tTYPE\\
\quad$\pow(\tIP)$					&$\pre$     	&sets of identifiers (precursors, destinations, \dots)\\
\quad$\pow(\tIP\times\tRREQID)$	        	&\rreqs		&sets of request identifiers  with originator IP\\
$\tTYPE_1\rightharpoonup\tTYPE_2$       &              	 	&partial functions from $\tTYPE_1$ to $\tTYPE_2$\\
\quad$\tIP\rightharpoonup\tSQN$         	&\dests         	&sets of destinations with sequence numbers\\
\hline
\hline
\multicolumn{2}{|l|}{\textbf{Constant/Predicate}}& \textbf{Description}\\
\hline
\multicolumn{2}{|l|}{$0:\tSQN,~1:\tSQN$}&
	unknown, smallest sequence number\\
\multicolumn{2}{|l|}{$\mathord{<} \subseteq \tSQN\times\tSQN$}&
	strict order on sequence numbers\\
\multicolumn{2}{|l|}{$\kno,\unkno:\tSQNK$}&
	constants to distinguish  known and unknown sqns\\
\multicolumn{2}{|l|}{$\val,\inval:\tFLAG$}&
	constants to distinguish  valid and invalid routes\\
\multicolumn{2}{|l|}{$\pen,\nonpen:\tPendingRREQ$}&
	constants indicating whether a RREQ is required\\
\multicolumn{2}{|l|}{$0:\NN,~1:\NN,~\mathord{<} \subseteq \NN\times\NN$}&
	standard constants/predicates of natural numbers\\
\multicolumn{2}{|l|}{${[\,]}:{[\tTYPE]},~\emptyset:\pow(\tTYPE)$}&
	empty queue, empty set\\
\multicolumn{2}{|l|}{$\mathord{\in}\subseteq\tTYPE\times\pow(\tTYPE)$}&
	membership, standard set theory\\
\hline
\hline
\multicolumn{2}{|l|}{\textbf{Function}} & \textbf{Description}\\
\hline
\multicolumn{2}{|l|}{$\fnhead:[\tTYPE]\rightharpoonup\tTYPE$}&
	returns the ``oldest'' element in the queue\\
\multicolumn{2}{|l|}{$\fntail:[\tTYPE]\rightharpoonup[\tTYPE]$}&
	removes the ``oldest'' element in the queue\\
\multicolumn{2}{|l|}{$\fnappend:\tTYPE\times[\tTYPE]\rightarrow[\tTYPE]$}&
	inserts a new element into the queue\\
\multicolumn{2}{|l|}{$\fndrop:\tIP\times\tQUEUES \rightharpoonup\tQUEUES$}&
	deletes a packet from the queued data packets\\
\multicolumn{2}{|l|}{$\fnadd:\tDATA\times\tIP\times\tQUEUES \to\tQUEUES$}&
	adds a packet to the queued data packets\\
\multicolumn{2}{|l|}{$\fnunsetrrf:\tQUEUES\times\tIP\to\tQUEUES$}&
	set the \penFlag to \pen\\
\multicolumn{2}{|l|}{$\fnsetrrf:\tQUEUES\times(\tIP\rightharpoonup\tSQN)\to\tQUEUES$}&
	set the \penFlag to \nonpen\\
\multicolumn{2}{|l|}{$\fnselqueue:\tQUEUES\times\tIP \rightarrow [\tDATA]$}&
	selects the data queue for a particular destination\\
\multicolumn{2}{|l|}{$\fnfD:\tQUEUES\times\tIP\rightharpoonup\tPendingRREQ$}&
	selects the flag for a destination from the store\\
\multicolumn{2}{|l|}{$\fnselroute:\tRT\times\tIP \rightharpoonup \tROUTE$}&
	selects the route for a particular destination\\
\multicolumn{2}{|l|}{$(\_\comma\_\comma\_\comma\_\comma\_\comma\_\comma\_\,)\!: \tIP
        \mathord\times \tSQN \mathord\times\tSQNK \mathord\times\tFLAG \mathord\times \NN
        \mathord\times \tIP \mathord\times \pow(\tIP) \mathop\rightarrow \tROUTE$}&
	generates a routing table entry\\
\multicolumn{2}{|l|}{$\fninc:\tSQN \rightarrow \tSQN$}&
	increments the sequence number\\
\multicolumn{2}{|l|}{$\max:\tSQN\times\tSQN \to\tSQN$}&
	returns the larger sequence number\\	
\multicolumn{2}{|l|}{$\fnsqn:\tRT \times \tIP \to \tSQN$}&
	returns the sequence number of a particular route\\	
\multicolumn{2}{|l|}{$\fnsqnf:\tRT \times \tIP \to \tSQNK$}&
	determines whether the sequence number is known\\
\multicolumn{2}{|l|}{$\fnstatus:\tRT\times\tIP\rightharpoonup\tFLAG$}&
	returns the validity of a particular route\\
\multicolumn{2}{|l|}{$+1:\NN \rightarrow \NN$}&
	increments the hop count\\
\multicolumn{2}{|l|}{$\fndhops:\tRT \times \tIP \rightharpoonup \NN$}&
	returns the hop count of a particular route\\
\multicolumn{2}{|l|}{$\fnnhop:\tRT \times \tIP \rightharpoonup \tIP$}&
	returns the next hop of a particular route\\
\multicolumn{2}{|l|}{$\fnprecs:\tRT \times \tIP \rightharpoonup \pow(\tIP)$}&
	returns the set of precursors of a particular route\\
\multicolumn{2}{|l|}{$\fnakD, \fnikD, \fnkD:\tRT \rightarrow\pow(\tIP)$}&
	returns the set of valid, invalid, known destinations\\
\multicolumn{2}{|l|}{$\fnqD:\tQUEUES \rightarrow \pow(\tIP)$}&
	returns the set of destinations with unsent packets\\
\multicolumn{2}{|l|}{$\cap,~\cup,~\bigcup\{\ldots\},~\dots$}&
	standard set-theoretic functions\\
\multicolumn{2}{|l|}{$\fnaddprec : \tROUTE\times \pow(\tIP) \to \tROUTE$}&
	adds a set of precursors to a routing table entry\\	
\multicolumn{2}{|l|}{$\fnaddprecrt : \tRT\times\tIP\times \pow(\tIP) \rightharpoonup \tRT$}&
	adds a set of precursors to an entry inside a table\\	
\multicolumn{2}{|l|}{$\fnupd:\tRT \times \tROUTE \rightharpoonup \tRT$}&
	updates a routing table with a route (if fresh enough)\\
\multicolumn{2}{|l|}{$\fninv:\tRT \times (\tIP\rightharpoonup\tSQN) \rightarrow \tRT$}&
	invalidates a set of routes within a routing table\\
\multicolumn{2}{|l|}{$\fnnrreqid: \pow(\tIP\times\tRREQID) \times \tIP \rightarrow \tRREQID$}&
	generates a new route request identifier\\
\multicolumn{2}{|l|}{$\newpktID:\tDATA \times \tIP \rightarrow \tMSG$}&
	generates a message with new application layer data\\
\multicolumn{2}{|l|}{$\pktID:\tDATA \times \tIP \times \tIP \rightarrow \tMSG$}&
	generates a message containing application layer data\\
\multicolumn{2}{|l|}{$\rreqID:\NN \mathord\times \tRREQID \mathord\times \tIP \mathord\times \tSQN\mathord\times\tSQNK \mathord\times \tIP \mathord\times \tSQN \mathord\times \tIP \rightarrow \tMSG$}&
	generates a route request\\
\multicolumn{2}{|l|}{$\rrepID:\NN \times \tIP \times \tSQN \times \tIP \times \tIP \rightarrow \tMSG$}&
	generates a route reply\\
\multicolumn{2}{|l|}{$\rerrID:(\tIP\rightharpoonup\tSQN) \times \tIP \rightarrow \tMSG$}&
	generates a route error message\\
\hline
\end{tabular}
\vspace*{-2pt}
\end{table*}

\section{Modelling AODV}\label{sec:modelling_AODV}
Our formalisation of AODV tries to accurately model the protocol as
defined in the IETF RFC 3561 specification \cite{rfc3561}.  The model
focusses on layer $3$ of the protocol stack, i.e., the routing and
forwarding of messages and packets, and abstracts from lower layer
network protocols and mechanisms such as the Carrier Sense Multiple
Access (CSMA) protocol. 

The presented formalisation includes all core
components of the protocol, but, at the moment, abstracts from timing
issues and optional protocol features. This keeps our specification
manageable. A consequence of not modelling timing issues is that statements such as
\quote{Can a route expire before a data packet is transmitted?}~\cite{CK05} cannot be analysed,
for in our model routes do not expire at all.
Our plan is to extend our model step by step. 
The model allows us to reason about protocol behaviour and to
prove critical protocol characteristics.
A detailed list of abstractions made can be found
  in~\cite[Section~3]{TR13}.

In this section, we present a specification of the AODV protocol
using process algebra.
The model includes a mechanism to describe the delivery of data
packets; though this is not part of the protocol itself it is
necessary to trigger any AODV activity.  Our model consists of $7$
processes, named $\AODV$, $\NEWPKT$, $\PKT$, $\RREQ$, $\RREP$, $\RERR$ and
$\QMSG$:

  \algsetup{linenodelimiter=.,linenosize=\tiny}
  \begin{algorithm*}
    {\footnotesize
      \caption{The basic routine}
      \label{pro:aodv}
      \begin{algorithmic}[1]
\DEFPROCESS{\AODV}{\ip\comma\sn\comma\rt\comma\rreqs\comma\queues}
	\IFempty
		\receiveL{\msg}\ .																															\label{aodv:line2}
		\COMLINE{depending on the message, the node calls different processes}	
		\PAR						\label{aodv:line3}
		\IF[new DATA packet]{$\msg = \newpkt{\data}{\dip}$}																	\label{aodv:line4}
			\newpktP{\data}{\dip}{\ip}{\sn}{\rt}{\rreqs}{\queues}																	\label{aodv:line5}
		\ELSIF[incoming DATA packet]{$\msg = \pkt{\data}{\dip}{\oip}$}  \label{aodv:line6}
			\pktP{\data}{\dip}{\oip}{\ip}{\sn}{\rt}{\rreqs}{\queues}																	\label{aodv:line7}
		\ELSIF[RREQ]{$\msg = \rreq{\hops}{\rreqid}{\dip}{\dsn}{\dsk}{\oip}{\osn}{\sip}$}								\label{aodv:line8}
			\COMLINE{update the route to \sip\ in \rt}																					\label{aodv:line9}
			\UPD{\rt:=\upd{\rt}{(\sip,0,\unkno,\val,1,\sip,\emptyset)}}																\label{aodv:line10}
			\COMMENT{$0$ is used since no sequence number is known}%
			\rreqP{\hops}{\rreqid}{\dip}{\dsn}{\dsk}{\oip}{\osn}{\sip}{\ip}{\sn}{\rt}{\rreqs}{\queues}					\label{aodv:line11}
		\ELSIF[RREP]{$\msg = \rrep{\hops}{\dip}{\dsn}{\oip}{\sip}$}															\label{aodv:line12}
			\COMLINE{update the route to \sip\ in \rt}																					\label{aodv:line13}
			\UPD{\rt:=\upd{\rt}{(\sip,0,\unkno,\val,1,\sip,\emptyset)}}																\label{aodv:line14}
			\rrepP{\hops}{\dip}{\dsn}{\oip}{\sip}{\ip}{\sn}{\rt}{\rreqs}{\queues}												\label{aodv:line15}
		\ELSIF[RERR]{$\msg = \rerr{\dests}{\sip}$}																					\label{aodv:line16}
			\COMLINE{update the route to \sip\ in \rt}																					\label{aodv:line17}
			\UPD{\rt:=\upd{\rt}{(\sip,0,\unkno,\val,1,\sip,\emptyset)}}																\label{aodv:line18}
			\rerrP{\dests}{\sip}{\ip}{\sn}{\rt}{\rreqs}{\queues}																			\label{aodv:line19}
		\ENDIFii
		\ENDPAR																																		\label{aodv:line20}
		\ELSIF[send a queued data packet if a valid route is known]{$\mbox{Let } \dip\in\qD{\queues}\cap\akD{\rt}$}				\label{aodv:line22}
			\UPD{\data:=\head{\selq{\queues}{\dip}}}													\label{aodv:line23}
			\STARTPRIO
			 	\unicast{\nhop{\rt}{\dip}}{\pkt{\data}{\dip}{\ip}}\ . 											\label{aodv:line24}
				\UPD{\queues:=\drop{\dip}{\queues}}\COMMENT{drop {\data} from the {\queues} for {\dip} if the transmission was successful}													\label{aodv:line26}
				\aodvL{\ip}{\sn}{\rt}{\rreqs}{\queues}			\label{aodv:line27}
			 \PRIO
				\COMspec{an error is produced and the routing table is updated}							\label{aodv:line29}
				\UPD{\dests:=\{(\rip,\inc{\sqn{\rt}{\rip}})\,|\,\rip\in\akD{\rt}\wedge \nhop{\rt}{\rip}=\nhop{\rt}{\dip}\}}		\label{aodv:line30}
				\UPD{\rt:=\inv{\rt}{\dests}}															\label{aodv:line32}
				\UPD{\queues:=\setrrf{\queues}{\dests}}\label{aodv:line32a}
				\UPD{\pre:=\bigcup\{\precs{\rt}{\rip}\,|\,(\rip,*)\in\dests\}}									\label{aodv:line31}
				\UPD{\dests:=\{(\rip,\rsn)\,|\,(\rip,\rsn)\in\dests\wedge \precs{\rt}{\rip}\not=\emptyset\}}				\label{aodv:line31a}
				\groupcast{\pre}{\rerr{\dests}{\ip}}\ .
				\aodv{\ip}{\sn}{\rt}{\rreqs}{\queues}													\label{aodv:line33}
 		 	\ENDPRIO
		\ELSIF[a route discovery process is initiated]{$\mbox{Let } \dip\in\qD{\queues}-\akD{\rt}\wedge\fD{\queues}{\dip}=\nonpen$}	\label{aodv:line34}		
			\UPD{\queues:=\unsetrrf{\queues}{\dip}}\COMMENT{set \penFlag to \pen}				\label{aodv:line35}		
			\UPD{\sn:=\inc{\sn}}\COMMENT{increment own sequence number}								\label{aodv:line36}
			\COMLINE{update \rreqs\ by adding $(\ip,\nrreqid{\rreqs}{\ip})$}								\label{aodv:line37}
			\UPD{\rreqid:=\nrreqid{\rreqs}{\ip}}							\label{aodv:line38a}
			\UPD{\rreqs := \rreqs\cup\{(\ip,\rreqid)\}}							\label{aodv:line38b}
			\broadcast{\rreq{$0$}{\rreqid}{\dip}{\sqn{\rt}{\dip}}{\sqnf{\rt}{\dip}}{\ip}{\sn}{\ip}}\ .					\label{aodv:line39}
			\aodv{\ip}{\sn}{\rt}{\rreqs}{\queues}														\label{aodv:line40}			
	\ENDIFii

	\end{algorithmic}
    }
  \end{algorithm*}

\begin{itemize}
	\item The {basic process} $\AODV$ reads a message from the message queue and,
		depending on the type of the message, calls other
		processes. When there is no message handling going on,
		the process initiates the transmission of queued data packets or generates
		a new route request (if packets are stored for a destination, no route 
		to this destination is known and no route request for this destination is pending).
        \item The processes $\NEWPKT$ and $ \PKT$ describe all actions
                performed by a node when a data packet is
                received.  The former process handles a newly injected
		packet. The latter describes all actions performed
		when a node receives data from another node via the protocol. This
		includes accepting the packet (if the
                node is the destination), forwarding the packet (if
                the node is not the destination) and sending an error
                message (if forwarding fails).
	\item The process $\RREQ$ models all events that might occur
                after a route request has been received. This includes
                updating the node's routing table, forwarding the
                route request as well as the initiation of a route
                reply if a route to the destination is known.
	\item Similarly, the $\RREP$ process describes the reaction
		of the protocol to an incoming route reply.
	\item The process $\RERR$ models the part of AODV which handles error messages.
                In particular, it describes the modification and
		forwarding of the AODV error message.
	\item The last process $\QMSG$ concerns message handling. Whenever a message is received,
	  it is first stored in a message queue. If the corresponding node is able to handle a message
	  it pops the oldest message from the queue and handles it. An example where a node is not ready
	  to process an incoming message immediately is when it is already handling a message.
\end{itemize}

In the remainder of the section, we provide a formal specification for
each of these processes and explain them step by step. Our
specification can be split into three parts: the brown lines describe
updates to be performed on the node's data, e.g., its routing table;
the black lines are other process algebra constructs
(cf.\ \Sect{awn}); and the blue lines are
ordinary comments.

\subsection{The Basic Routine}\label{ssec:proc_aodv}

The {basic process} $\AODV$ either reads a message from the
corresponding queue, sends a queued data packet if a route to the destination has been established, 
or initiates a new route discovery process in case of queued
data packets with invalid or unknown routes.
This process maintains five data variables, {\ip}, {\sn}, {\rt}, {\rreqs} and
{\queues}, in which it stores its own identity, its own sequence number, its current routing
table, the list of route requests seen, and its current store
of queued data packets that await transmission (cf.\ \Sect{types}).

The message handling is described in Lines~\ref{aodv:line2}--\ref{aodv:line20}.
First, the message has to be read from the queue of stored messages
(\receive{\msg}). After that, the process $\AODV$ checks the type of
the message and calls a process that can handle the message: in case
of a newly injected data packet, the process $\NEWPKT$ is called; 
in case of  an incoming data packet, the process $\PKT$ is
called; in case that the incoming message is an AODV control message
(route request, route reply or route error), the node updates its routing table.
More precisely, if there is no entry to the message's sender $\sip$, the
receiver-node creates an entry with the unknown sequence number $0$
and hop count $1$; in case  there is already a 
routing table entry $(\sip,\dsn,*,*,*,*,\pre)$, then this entry is updated to
$(\sip,\dsn,\unkno,\val,1,\sip,\pre)$
(cf. Lines~\ref{aodv:line10}, \ref{aodv:line14} and~\ref{aodv:line18}). 
Afterwards, the processes $\RREQ$, $\RREP$ and $\RERR$ are called, respectively.

The second part of $\AODV$ (Lines~\ref{aodv:line22}--\ref{aodv:line33}) initiates the sending of a data packet.
For that, it has to be checked if there is a queued data packet for a destination that has a known and valid route
in the routing table ($\qD{\queues}$ ${}\cap\akD{\rt}\not=\emptyset$). In case that there is more than one destination with
stored data and a known route, an arbitrary destination is chosen and denoted by $\dip$
(Line~\ref{aodv:line22}).\footnote{Although the word ``let'' is not part of the syntax, we add it to stress
the nondeterminism happening here.}\linebreak
Moreover $\data$ is set to the first
queued data packet from the application layer that should be sent
($\data:=\fnhead(\selq{\queues}{\dip})$).\footnote{Following the RFC,
data packets waiting for a route should be buffered ``first-in, first-out'' (FIFO).} 
This data packet is unicast to the next hop on the route to $\dip$.
If the unicast is successful, the data packet $\data$ is removed from $\queues$ (Line~\ref{aodv:line26}).
Finally, the process calls itself---stating that the node is ready for
handling a new message, initiating the sending of another packet towards a destination, etc.
In case the unicast is not successful, the data packet has not been transmitted.
Therefore $\data$ is not removed from $\queues$. Moreover, the node knows that the link
to the next hop on the route to $\dip$ is faulty and, most probably, broken.
An error message is initiated. Generally, route error and link breakage processing requires the
following steps: (a) invalidating existing routing table entries, (b) listing affected destinations,
(c) determining which neighbours may be affected (if any), and
(d) delivering an appropriate AODV error message to such neighbours \cite{rfc3561}.
Therefore, the process determines all valid destinations $\dests$ that have
this unreachable node as next hop (Line~\ref{aodv:line30}) and marks
the routing table entries for these destinations
as invalid (Line~\ref{aodv:line32}), while incrementing their sequence numbers (Line~\ref{aodv:line30}).
In Line~\ref{aodv:line32a}, we set, for all invalidated routing table entries, the \penFlag to \nonpen,
thereby indicating that a new route discovery process may need to be initiated.
In Line~\ref{aodv:line31} the recipients of the error message
are determined. These are the precursors of the invalidated
destinations, i.e., the neighbouring nodes listed as having a route to one of the
affected destinations passing through the broken link.
Finally, an error message is sent to them (Line~\ref{aodv:line33}),
listing only those invalidated destinations with a non-empty set of
precursors (Line~\ref{aodv:line31a}).

  \algsetup{linenodelimiter=.,linenosize=\tiny}
  \begin{algorithm*}
    {\footnotesize
      \caption{Routine for handling a newly injected data packet}
      \label{pro:newpkt}
      \begin{algorithmic}[1]
\DEFPROCESS{\NEWPKT}{\data\comma\dip\,\comma\,\ip\comma\sn\comma\rt\comma\rreqs\comma\queues}
	\IF[the DATA packet is intended for this node]{$\dip=\ip$}															\label{newpkt:line2}
		\deliverL{\data}\ .
		\aodv{\ip}{\sn}{\rt}{\rreqs}{\queues}																						\label{newpkt:line3}
	\ELSIF[the DATA packet is not intended for this node]{$\dip\not=\ip$}										\label{newpkt:line4}
		\UPD{\queues:=\add{\data}{\dip}{\queues}}	\ .
		\aodv{\ip}{\sn}{\rt}{\rreqs}{\queues}																						\label{newpkt:line5}	
	\ENDIFii

	\end{algorithmic}
    }
  \end{algorithm*}

The third and final part of $\AODV$ (Lines~\ref{aodv:line34}--\ref{aodv:line40}) initiates a route
discovery process. 
This is done when there is  at least one queued data
packet for a destination without a valid
routing table entry, that is not 
waiting for a reply in response to a route request process initiated before. 
Following the RFC, the process generates a new route request.
This is achieved in four steps:
First, the \penFlag is set to {\pen} (Line~\ref{aodv:line35}), meaning that no further
route discovery processes for this destination need to be initiated.%
\footnote{The RFC does not describe packet handling in detail; hence
the \penFlag is not part of the RFC's RREQ generation process.}
Second, the node's own sequence number is increased by~$1$
(Line~\ref{aodv:line36}). Third, by determining
$\nrreqid{\rreqs}{\ip}$, a new route request identifier is created and
stored---together with the node's $\ip$---in the set $\rreqs$ of route
requests already seen (Line~\ref{aodv:line38b}). Fourth, the message
itself is sent (Line~\ref{aodv:line39}) using broadcast. In contrast to
\textbf{unicast}, transmissions via \textbf{broadcast} are not checked
on success.  The information inside the message follows strictly the
RFC. In particular, the hop count is set to $0$, the route request identifier previously created is used, etc.
This ends the initiation of the route discovery process.

\subsection{Data Packet Handling}\label{ssec:proc_pkt}

  \algsetup{linenodelimiter=.,linenosize=\tiny}
  \begin{algorithm*}
    {\footnotesize
      \caption{Routine for handling a received data packet}
      \label{pro:pkt}
      \begin{algorithmic}[1]
\DEFPROCESS{\PKT}{\data\comma\dip\comma\oip\,\comma\,\ip\comma\sn\comma\rt\comma\rreqs\comma\queues}
	\IF[the DATA packet is intended for this node]{$\dip=\ip$}																				\label{pkt2:line2}
		\deliverL{\data}\ .
		\aodv{\ip}{\sn}{\rt}{\rreqs}{\queues}																											\label{pkt2:line3}
	\ELSIF[the DATA packet is not intended for this node]{$\dip\not=\ip$}															\label{pkt2:line4}
	 	\PAR
			\IF[valid route to \dip]{$\dip\in\akD{\rt}$}																									\label{pkt2:line5}
				\COMLINE{forward packet}
				\STARTPRIO
					\unicast{\nhop{\rt}{\dip}}{\pkt{\data}{\dip}{\oip}}\ . \aodv{\ip}{\sn}{\rt}{\rreqs}{\queues}						\label{pkt2:line7}
				\PRIO
					\COMspec{If the packet transmission is unsuccessful, a RERR message is generated}		
					\UPD{\dests:=\{(\rip,\inc{\sqn{\rt}{\rip}})\,|\,\rip\in\akD{\rt}\wedge \nhop{\rt}{\rip}=\nhop{\rt}{\dip}\}}			\label{pkt2:line9}
					\UPD{\rt:=\inv{\rt}{\dests}}																												\label{pkt2:line10}
					\UPD{\queues:=\setrrf{\queues}{\dests}}																						\label{pkt2:line11}
					\UPD{\pre:=\bigcup\{\precs{\rt}{\rip}\,|\,(\rip,*)\in\dests\}}																	\label{pkt2:line12}
					\UPD{\dests:=\{(\rip,\rsn)\,|\,(\rip,\rsn)\in\dests\wedge \precs{\rt}{\rip}\not=\emptyset\}}							\label{pkt2:line13}
					\groupcast{\pre}{\rerr{\dests}{\ip}}\ . \aodv{\ip}{\sn}{\rt}{\rreqs}{\queues}											\label{pkt2:line14}
				\ENDPRIO
			\ELSIF[no valid route to \dip]{$\dip\not\in\akD{\rt}$}																				\label{pkt2:line15}
				\COMLINE{no local repair occurs; data is lost}																					\label{pkt2:line16}
				\PAR	
					\IF[invalid route to \dip]{$\dip\in\ikD{\rt}$}																						\label{pkt2:line18}
						\COMLINE{if the route is invalid, a RERR is sent to the precursors}
						\groupcast{\precs{\rt}{\dip}}{\rerr{\{(\dip,\sqn{\rt}{\dip})\}}{\ip}}\ .											
						\aodv{\ip}{\sn}{\rt}{\rreqs}{\queues}																							\label{pkt2:line20}
					\ELSIF[route not in \rt]{$\dip\not\in\ikD{\rt}$}																					\label{pkt2:line21}
						\aodvL{\ip}{\sn}{\rt}{\rreqs}{\queues}																							\label{pkt2:line22}
					\ENDIFii
				\ENDPAR																																			\label{pkt2:line23}									
			\ENDIFii
		\ENDPAR																																					\label{pkt2:line24}
	\ENDIFii

	\end{algorithmic}
    }
  \end{algorithm*}

The processes $\NEWPKT$ and $\PKT$ describe all actions performed by a
node when a data packet is injected by a client hooked up to the
local node or received via the protocol, respectively.  For
the process $\PKT$, this includes the acceptance
(if the node is the destination), the forwarding (if the node is not
the destination), as well as the sending of an error message in case
something went wrong.
The process $\NEWPKT$ does not include the initiation of a new route request; this is part of the process $\AODV$.
Although packet handling itself
is not part of AODV, it is necessary to include it in our
formalisation, since a failure to transmit a data packet triggers AODV
activity.

The process $\NEWPKT$ first checks whether the node is the intended addressee
of the data packet. If this is the
case, it delivers the data and returns to the basic routine $\AODV$.
If the node is not the intended destination ($\dip\not=\ip$, Line~\ref{newpkt:line4}),
the $\data$ is added to the data queue for {\dip} (Line~\ref{newpkt:line5}),\footnote{If no
data for destination $\dip$ was already queued, the function \hyperlink{add}{$\fnadd$} creates a
fresh queue for $\dip$, and set the request-required flag to $\nonpen$; otherwise, the
request-required flag keeps the value it had already.}
which finishes the handling of a newly injected data packet.
The further handling of queued data (forwarding it to the next hop on the way to the destination in
case a valid route to the destination is known,
and otherwise initiating a new route request if still required)
is the responsibility of the main process {\AODV}.

Similar to $\NEWPKT$, the process $\PKT$ first checks if it is the intended addressee
of the data packet. If this is the
case, it delivers the data and returns to the basic routine $\AODV$.
If the node is not the intended destination
($\dip\not=\ip$, Line~\ref{pkt2:line4}) more activity is needed.

In case that the node has a valid route to the $\data$'s destination $\dip$
($\dip\in\akD{\rt}$), it forwards the packet using a unicast to the
next hop $\nhop{\rt}{\dip}$ on the way to $\dip$.
Similar to the unicast of the process $\AODV$, it has to be checked
whether the transmission is successful: no further action is necessary
if the transmission succeeds, and the node returns to the basic
routine $\AODV$. If the transmission fails, the link to the next hop
$\nhop{\rt}{\dip}$ is assumed to be broken. As before, all
destinations $\dests$ that are reached via that broken link are
determined (Line~\ref{pkt2:line9}) and all precursors interested in at
least one of these destinations are informed via an error message
(Line~\ref{pkt2:line14}).  Moreover, all the routing table entries
using the broken link have to be invalidated in the node's routing
table $\rt$ (Line~\ref{pkt2:line10}), and all corresponding \penFlag{}s
are set to \nonpen\ (Line~\ref{pkt2:line11}).

In case that the node has no valid route to the destination $\dip$ ($\dip\not\in\akD{\rt}$),
the data packet is lost and possibly an error message is sent. If 
there is an (invalid) route to the \data's destination {\dip} in
the routing table (Line~\ref{pkt2:line18}), the possibly affected
neighbours can be determined and the error message is sent to these
precursors (Line \ref{pkt2:line20}). If there is no information about a
route towards $\dip$ nothing happens (and the basic process {\AODV} is called again).
\newpage

\subsection{Receiving Route Requests}\label{ssec:proc_rreq}
  \algsetup{linenodelimiter=.,linenosize=\tiny}
  \begin{algorithm*}
    {\footnotesize
      \caption{RREQ handling}
      \label{pro:rreq}
      \begin{algorithmic}[1]
\DEFPROCESS{\RREQ}{\hops\comma\rreqid\comma\dip\comma\dsn\comma\dsk\comma\oip\comma\osn\comma\sip\,\comma\,\ip\comma\sn\comma\rt\comma\rreqs\comma\queues}
	\IF[the RREQ has been received previously]{$(\oip\,\comma\,\rreqid)\in\rreqs$}																							\label{rreq:line2}
		\aodvL{\ip}{\sn}{\rt}{\rreqs}{\queues} \COM{silently ignore RREQ, i.e. do nothing}																			\label{rreq:line3}
	\ELSIF[the RREQ is new to this node]{$(\oip\,\comma\,\rreqid)\not\in\rreqs$}																								\label{rreq:line4}
		\UPD{\rt:=\upd{\rt}{(\oip,\osn,\kno,\val,\hops+1,\sip,\emptyset)}}	\COMMENT{update the route to \oip\ in \rt}									\label{rreq:line6}
		\UPD{\rreqs:=\rreqs\cup\{(\oip,\rreqid)\}}		\COMMENT{update \rreqs\ by adding $(\oip\,\comma\,\rreqid)$}											\label{rreq:line8}
		\PAR																																																\label{rreq:line9}
		\IF[this node is the destination node]{$\dip=\ip$}																															\label{rreq:line10}
			\UPD{\sn:=\max(\sn,\dsn)}	\COMMENT{update the sqn of \ip}																									\label{rreq:line12}
			\COMLINE{unicast a RREP towards \oip\ of the RREQ}																											\label{rreq:line13}
			\STARTPRIO
					\unicast{\nhop{\rt}{\oip}}{{\rrep{$0$}{\dip}{\sn}{\oip}{\ip}}}\ . 																								\label{rreq:line14a}								
					\aodv{\ip}{\sn}{\rt}{\rreqs}{\queues}																																	\label{rreq:line14}
				\PRIO
					\COMspec{If the transmission is unsuccessful, a RERR message is generated}																\label{rreq:line15}
					\UPD{\dests:=\{(\rip,\inc{\sqn{\rt}{\rip}})\,|\,\rip\in\akD{\rt}\wedge \nhop{\rt}{\rip}=\nhop{\rt}{\oip}\}}												\label{rreq:line16}
					\UPD{\rt:=\inv{\rt}{\dests}}																																					\label{rreq:line18}			
					\UPD{\queues:=\setrrf{\queues}{\dests}}																															\label{rreq:line18a}
					\UPD{\pre:=\bigcup\{\precs{\rt}{\rip}\,|\,(\rip,*)\in\dests\}}																										\label{rreq:line17}
					\UPD{\dests:=\{(\rip,\rsn)\,|\,(\rip,\rsn)\in\dests\wedge \precs{\rt}{\rip}\not=\emptyset\}}																\label{rreq:line17a}
					\groupcast{\pre}{\rerr{\dests}{\ip}}\ .																																	
					\aodv{\ip}{\sn}{\rt}{\rreqs}{\queues}																																	\label{rreq:line19}
				\ENDPRIO	
		\ELSIF[this node is not the destination node]{$\dip\not=\ip$}																											\label{rreq:line20}
			\PAR																																															\label{rreq:line21}
			\IF[$\!$valid route to \dip\ that is fresh enough$\!$]{$\!\dip\mathbin\in\akD{\rt} \wedge \dsn \mathbin\leq  \sqn{\rt}{\!\dip} \wedge\sqnf{\rt}{\!\dip}\mathbin=\kno\!$}		\label{rreq:line22}
					\COMLINE{update \rt\ by adding precursors}																														\label{rreq:line23}
					\UPD{\rt := \addprecrt{\rt}{\dip}{\{\sip\}}}																																\label{rreq:line24}
					\UPD{\rt := \addprecrt{\rt}{\oip}{\{\nhop{\rt}{\dip}\}}}																												\label{rreq:line25}
				\COMLINE{unicast a RREP towards the \oip\ of the RREQ}
				\STARTPRIO
					\unicast{\nhop{\rt}{\oip}}{\rrep{\dhops{\rt}{\dip}}{\dip}{\sqn{\rt}{\dip}}{\oip}{\ip}}\ .\\																	\label{rreq:line26}
					\aodvL{\ip}{\sn}{\rt}{\rreqs}{\queues}																																	\label{rreq:line26a}
				\PRIO
					\COMspec{If the transmission is unsuccessful, a RERR message is generated}
					\UPD{\dests:=\{(\rip,\inc{\sqn{\rt}{\rip}})\,|\,\rip\in\akD{\rt}\wedge \nhop{\rt}{\rip}=\nhop{\rt}{\oip}\}}												\label{rreq:line28}
					\UPD{\rt:=\inv{\rt}{\dests}}																																					\label{rreq:line30}			
					\UPD{\queues:=\setrrf{\queues}{\dests}}																															\label{rreq:line30a}		
					\UPD{\pre:=\bigcup\{\precs{\rt}{\rip}\,|\,(\rip,*)\in\dests\}}																										\label{rreq:line29}
					\UPD{\dests:=\{(\rip,\rsn)\,|\,(\rip,\rsn)\in\dests\wedge \precs{\rt}{\rip}\not=\emptyset\}}																\label{rreq:line29a}
					\groupcast{\pre}{\rerr{\dests}{\ip}}\ . 																																	\label{rreq:line31}
					\aodv{\ip}{\sn}{\rt}{\rreqs}{\queues}
				\ENDPRIO
			\ELSIF[$\!$no valid route that is fresh enough$\!$]{$\dip\mathbin{\not\in}\akD{\rt} \vee \sqn{\rt}{\!\dip} <  \dsn \vee\sqnf{\rt}{\!\dip}\mathbin=\unkno$}					\label{rreq:line32}
				\COMLINE{no further update of \rt}
				\broadcast{\rreq{$\hops+1$}{\rreqid}{\dip}{\max(\sqn{\rt}{\dip}\comma\dsn)}{\dsk}{\oip}{\osn}{\ip}}\ .										\label{rreq:line34}
				\aodvL{\ip}{\sn}{\rt}{\rreqs}{\queues}				\label{rreq:line35}
			\ENDIFii
			\ENDPAR																																													\label{rreq:line36}
		\ENDIFii
		\ENDPAR																																														\label{rreq:line37}
	\ENDIFii

	\end{algorithmic}
    }
  \end{algorithm*}

The process $\RREQ$ models all events that may occur after a route
request has been received.

{\RREQ} first reads the unique identifier $(\oip,\rreqid)$ of the route request received.
If this pair is already stored in the node's data $\rreqs$, the route request has been handled
before and the message can silently be ignored (Lines~\ref{rreq:line2}--\ref{rreq:line3}).

If the received message is new to this node, i.e.,
\[
(\oip,\rreqid)\not\in\rreqs\quad \mbox{(Line~\ref{rreq:line4})}\,,
\]
the node
establishes a route of length $\hops\mathord+1$ back to the originator $\oip$
of the message. If this route is ``better'' than the route to $\oip$
in the current routing table, the routing table is updated by this
route (Line~\ref{rreq:line6}).  Moreover the unique identifier has to
be added to the set $\rreqs$ of already seen (and handled) route
requests (Line~\ref{rreq:line8}).

After these updates the process checks if the node is the intended
destination ($\dip=\ip$, Line~\ref{rreq:line10}). In that case, a
route reply must be initiated: first, the node's sequence number
is---according to the RFC---set to the max\-imum of the current sequence
number and the destination sequence number stemming from the RREQ message
(Line~\ref{rreq:line12}).%
Then the reply is unicast to the next hop on
the route back to the originator {\oip} of the route request. The
content of the new route reply is as follows: the hop count is set to
$0$, the destination and originator are copied from the route request
received and the destination's sequence number is the node's own
sequence number \sn; of course the sender's IP of this message has to be set
to the node's $\ip$. As before (cf.\ Sections~\ref{ssec:proc_aodv} and
\ref{ssec:proc_pkt}), the process invalidates the corresponding routing table entries, sets \penFlag{}s and sends an error message to all
relevant precursors if the unicast transmission fails (Lines~\ref{rreq:line16}--\ref{rreq:line19}).

  \algsetup{linenodelimiter=.,linenosize=\tiny}
  \begin{algorithm*}
    {\footnotesize
      \caption{RREP handling}
      \label{pro:rrep}
      \begin{algorithmic}[1]
\DEFPROCESS{\RREP}{\hops\comma\dip\comma\dsn\comma\oip\comma\sip\,\comma\,\ip\comma\sn\comma\rt\comma\rreqs\comma\queues}
	\IF[the routing table has to be updated]{$\rt\not=\upd{\rt}{(\dip\comma\dsn\comma\kno\comma\val\comma\hops+1\comma\sip\comma\emptyset)}$}						\label{rrep:line3}
		\UPD{\rt:=\upd{\rt}{(\dip\comma\dsn\comma\kno\comma\val\comma\hops+1\comma\sip\comma\emptyset)}}			\label{rrep:line5}
		\PAR\label{rrep:line5a}
		\IF[this node is the originator of the corresponding RREQ]{$\oip = \ip$}																	\label{rrep:line6}
			\COMLINE{a packet may now be sent; this is done in the process \AODV}
			\aodvL{\ip}{\sn}{\rt}{\rreqs}{\queues}																													\label{rrep:line8}
		\ELSIF[this node is not the originator; forward RREP]{$\oip \not= \ip$}																	\label{rrep:line9}
			\PAR
				\IF[valid route to \oip]{$\oip\in\akD{\rt}$}																											\label{rrep:line11}
					\COMLINE{add next hop towards $\oip$ as precursor and forward the route reply}									\label{rrep:line12}									
					\UPD{\rt := \addprecrt{\rt}{\dip}{\{\nhop{\rt}{\oip}\}}}																						\label{rrep:line12a}									
					\UPD{\rt := \addprecrt{\rt}{\nhop{\rt}{\dip}}{\{\nhop{\rt}{\oip}\}}}																		\label{rrep:line12b}
					\STARTPRIO
						\unicast{\nhop{\rt}{\oip}}{\rrep{$\hops+1$}{\dip}{\dsn}{\oip}{\ip}}\ .															\label{rrep:line13}
						\aodvL{\ip}{\sn}{\rt}{\rreqs}{\queues}																										\label{rrep:line14}
					\PRIO
						\COMspec{If the transmission is unsuccessful, a RERR message is generated}
						\UPD{\dests:=\{(\rip,\inc{\sqn{\rt}{\rip}})\,|\,\rip\in\akD{\rt}\wedge \nhop{\rt}{\rip}=\nhop{\rt}{\oip}\}}					\label{rrep:line16}
						\UPD{\rt:=\inv{\rt}{\dests}}																														\label{rrep:line18}							
												\UPD{\queues:=\setrrf{\queues}{\dests}}																		\label{rrep:line16a}
						\UPD{\pre:=\bigcup\{\precs{\rt}{\rip}\,|\,(\rip,*)\in\dests\}}																			\label{rrep:line17}
						\UPD{\dests:=\{(\rip,\rsn)\,|\,(\rip,\rsn)\in\dests\wedge \precs{\rt}{\rip}\not=\emptyset\}}									\label{rrep:line17a}
						\groupcast{\pre}{\rerr{\dests}{\ip}}\ .\ \aodv{\ip}{\sn}{\rt}{\rreqs}{\queues} 												\label{rrep:line20}
					\ENDPRIO
				\ELSIF[no valid route to \oip]{$\oip\not\in\akD{\rt}$}																						\label{rrep:line21}
					\aodvL{\ip}{\sn}{\rt}{\rreqs}{\queues}																											\label{rrep:line21a}
				\ENDIFii																																						\label{rrep:line22}
			\ENDPAR																																							\label{rrep:line23}
		\ENDIFii	
		\ENDPAR																																								\label{rrep:line23a}
	\ELSIF[the routing table is not updated]{$\rt=\upd{\rt}{(\dip\comma\dsn\comma\kno\comma\val\comma\hops+1\comma\sip\comma\emptyset)}$}							\label{rrep:line25}
		\aodvL{\ip}{\sn}{\rt}{\rreqs}{\queues}																														\label{rrep:line26}
	\ENDIFii
	\end{algorithmic}
    }
  \end{algorithm*}

If the node is not the destination $\dip$ of the message but an intermediate hop
along the path from the originator to the destination, it is allowed to generate
a route reply only if the information in its own routing table is fresh enough. This means
that
(a) the node has a valid route to the destination,
(b) the destination sequence number in the node's existing routing table entry
for the destination ($\sqn{\rt}{\dip}$) is greater than or equal to
the requested destination sequence number $\dsn$ of the message and
(c) the sequence number $\sqn{\rt}{\dip}$ is known, i.e., $\sqnf{\rt}{\dip}=\kno$.
If these three conditions are satisfied---the check is done in Line~\ref{rreq:line22}---the
node generates a new route reply and sends it to the next hop on the
way back to the originator {\oip} of the received route
request.\footnote{This next hop will often, 
  but not always, be $\sip$; see \cite{TR13}.}. To this end, it copies the sequence number for the
destination $\dip$ from the routing table $\rt$ into the destination sequence number field of the RREP message and
it places its distance in hops from the destination ($\dhops{\rt}{\dip}$) in the corresponding field of the new reply
(Line~\ref{rreq:line26}). The unicast might fail, which
causes the usual error handling (Lines~\ref{rreq:line28}--\ref{rreq:line31}).
Just before transmitting the unicast, the intermediate node updates the forward route entry to
$\dip$ by placing the last hop node ($\sip$)\footnote{This is a mistake in the RFC; it should
be $\nhop{\rt}{\oip}$.} into the precursor list for the forward route entry (Line~\ref{rreq:line24}).
Likewise, it updates the reverse route entry to {\oip} by placing the first hop $\nhop{\rt}{\dip}$
towards $\dip$ in the precursor list for that entry
(Line~\ref{rreq:line25}).\footnote{Unless the \emph{gratuitous RREP flag}
is set, which we do not model in this paper, this update is rather useless,
as the precursor $\nhop{\rt}{\dip}$ in general is not aware that it
has a route to $\oip$.}

If the node is not the destination and there is either no route to the
destination $\dip$ inside the routing table or the route is not fresh enough,
the route request received has to be forwarded. This happens in Line~\ref{rreq:line34}.
The information inside the forwarded request is mostly copied from the request received.
Only the hop count is increased by $1$ and the destination sequence number is set
to the maximum of the destination sequence number in the RREQ packet
and the current sequence number for $\dip$ in the routing table.
In case $\dip$ is an unknown destination, $\sqn{\rt}{\dip}$ returns the
unknown sequence number $0$.

\subsection{Receiving Route Replies}\label{ssec:rrep}

The process $\RREP$ describes the reaction of the protocol to an incoming route reply.
Our model first checks if a forward routing table entry is going to be
created or updated (Line~\ref{rrep:line3}). This is the case if (a) the
node has no known route to the destination, or 
(b) the destination sequence number in the node's existing routing table entry
for the destination ($\sqn{\rt}{\dip}$) is smaller than the destination sequence number $\dsn$ in the RREP message, or
(c) the two destination sequence numbers are equal and, in addition,
either the incremented hop count of the RREP received is strictly smaller than the
one in the routing table, or the entry for $\dip$ in the routing table
is invalid. 
Hence Line~\ref{rrep:line3} could be replaced~by
\begin{algorithmic}[3]%
  {
	\STATE\hspace{-1.6em}\algorithmicif$\dip\not\in\kD{\rt} \vee \sqn{\rt}{\dip}\mathbin<\dsn \vee (\sqn{\rt}{\dip}\mathbin=\dsn
	\newline{}\hspace{-1.2em}\mathop\wedge(\dhops{\rt}{\dip}\mathbin>\hops\mathord+1\mathop\vee\status{\rt}{\dip}\mathbin=\inval))\!$
\algorithmicthen\,.\,\footnotemark}
 \end{algorithmic}
 
 \footnotetext{
In case $\dip\not\in\kD{\rt}$, the terms
  $\dhops{\rt}{\dip}$ and $\status{\rt}{\dip}$ are not defined.
  In such a case, according to 
  the convention of Footnote~\ref{fn:undefvalues} in
  \Sect{awn}, the atomic formulas
  $\dhops{\rt}{\dip}\mathbin>\hops\mathord+1$ and $\status{\rt}{\dip}\mathbin=\inval$
  evaluate to {\tt false}. However, in case one would use lazy evaluation of the outermost disjunction,
  the evaluation of the expression would be  independent of the choice
  of a convention for interpreting undefined terms appearing in formulas.\label{fn:lazy}
  }
 
In case that one of these conditions is true, the routing table is
updated in Line~\ref{rrep:line5}.
If the node is the intended addressee of the route reply ($\oip=\ip$) 
the protocol returns to its basic process $\AODV$.
Otherwise ($\oip\not=\ip$) the message should be forwarded.
Following the RFC~\cite{rfc3561},
``If the current node is not the node indicated by the Originator IP Address
in the RREP message AND a forward route has been created or
updated {[\dots]}, the node consults its route table entry
for the originating node to determine the next hop for the RREP
packet, and then forwards the RREP towards the originator using the
information in that route table entry.''
This action needs a valid route to the originator $\oip$ of the route
request to which the current message is a reply ($\oip\in\akD{\rt}$, Line~\ref{rrep:line11}).
The content of the RREP message to be sent is mostly copied from the RREP received;
only the sender has to be changed (it is now the node's $\ip$) and the
hop count is incremented.
Prior to the unicast, the node $\nhop{\rt}{\oip}$, to which the message is
sent, is added to the list of precursors for the routes to $\dip$
(Line~\ref{rrep:line12a}) and to the next hop on the route to $\dip$ (Line~\ref{rrep:line12b}).
Although not specified in the RFC, it would make sense to also add a precursor to the reverse
route by 
$\assignment{\rt := \addprecrt{\rt}{\oip}{\{\nhop{\rt}{\dip}\}}}$\label{pg:small_precursor_improvement}.
As usual, if the unicast fails, the affected routing table entries are invalidated and the precursors of all
routes using the broken link are determined and an error message is sent (Lines~\ref{rrep:line16}--\ref{rrep:line20}).
In the unlikely situation that a reply should be forwarded but no valid route is known by the node,
nothing happens.
Following the RFC, no precursor has to be notified and no error message
has to be sent---even if there is an invalid route.

If a forward routing table entry is not created nor updated, 
the reply is silently ignored and the basic process is called (Lines~\ref{rrep:line25}--\ref{rrep:line26}).

  \algsetup{linenodelimiter=.,linenosize=\tiny}
  \begin{algorithm*}
    {\footnotesize
      \caption{RERR handling}
      \label{pro:rerr}
      \begin{algorithmic}[1]
\DEFPROCESS{\RERR}{\dests\comma\sip\,\comma\,\ip\comma\sn\comma\rt\comma\rreqs\comma\queues}
		\COMLINE{invalidate broken routes}												\label{rerr:line4}
		\UPD{\dests:=\{(\rip,\rsn)\,|\,(\rip,\rsn)\in\dests\wedge\rip\in\akD{\rt}\wedge \nhop{\rt}{\rip}=\sip\wedge \sqn{\rt}{\rip}<\rsn\}}			\label{rerr:line2}
		\UPD{\rt:=\inv{\rt}{\dests}}															\label{rerr:line5}
		\UPD{\queues:=\setrrf{\queues}{\dests}}\label{rerr:line5a}
		\COMLINE{forward the RERR to all precursors for \rt\ entries for broken connections}			\label{rerr:line1}
		\UPD{\pre:=\bigcup\{\precs{\rt}{\rip}\,|\,(\rip,*)\in\dests\}}									\label{rerr:line3}
		\UPD{\dests:=\{(\rip,\rsn)\,|\,(\rip,\rsn)\in\dests\wedge \precs{\rt}{\rip}\not=\emptyset\}}				\label{rerr:line3a}
		\groupcast{\pre}{\rerr{\dests}{\ip}}\ . \aodv{\ip}{\sn}{\rt}{\rreqs}{\queues}						\label{rerr:line6}

	\end{algorithmic}
    }
  \end{algorithm*}

\newpage

\subsection{Receiving Route Errors}

The process
$\RERR$ models the part of AODV that handles error messages.
An error message consists of a set $\dests$ of pairs of
an unreachable destination IP address $\rip$ and
the corresponding unreachable destination sequence number $\rsn$.

If a node receives an error message from a neighbour for one or
more valid routes, it has---under some conditions---to invalidate the entries for those routes
in its own routing table and forward the error message.  The node
compares the set $\dests$ of unavailable destinations from the
incoming error message with its own entries in the routing table.  If
the routing table lists a valid route with a $(\rip,\rsn)$-combination
from $\dests$ and if the next hop on this route is the sender $\sip$
of the error message, this entry may be affected by the error message. 
In our formalisation, we have added the requirement
$\sqn{\rt}{\rip}<\rsn$, saying that
the entry is affected by the error message only if the ``incoming'' sequence number
is larger than the one stored in the routing table, meaning
that it is based on fresher information.\footnote{This additional
    requirement is in the spirit of Section 6.2 of the RFC~\cite{rfc3561} on
    updating routing table entries, but in contradiction with Section
    6.11 of the RFC on handling $\RERR$ messages. In
    \cite{AODVloop} we show that the reading of Section
    6.11 of the RFC gives rise to routing loops.}
In this case, the entry has to be invalidated and all precursors of this particular route have
to be informed. This has to be done for all affected routes.

  \algsetup{linenodelimiter=.,linenosize=\tiny}
  \begin{algorithm*}
    {\footnotesize
      \caption{Message queue}
      \label{pro:queues}
      \begin{algorithmic}[1]
\DEFPROCESS{\QMSG}{\msgs}
	\IFempty
		\COMLINE{store incoming message at the end of \msgs}			\label{queues:line1}
		\receiveL{\msg}\ . 			
		\Qmsg{\append{\msg}{\msgs}}								\label{queues:line2}
	\ELSIF[the queue is not empty]{$\msgs\not=[\,]$}					\label{queues:line3}
		\PAR
		\COMLINE{pop top message and send it to another sequential process}								\label{queues:line4}
		\sendL{\head{\msgs}}\ .\ \Qmsg{\tail{\msgs}}					\label{queues:line5}
		\COMLINE{or receive and store an incoming message}				\label{queues:line7}
		\STATE $+$\,  \receive{\msg}\ . \Qmsg{\append{\msg}{\msgs}}                     \label{queues:line8}
		\ENDPAR
	\ENDIFii

	\end{algorithmic}
    }
  \end{algorithm*}

In fact, the process first determines all $(\rip,\rsn)$-pairs that have effects on its
own routing table and that may have to be forwarded as content of a new error message
(Line~\ref{rerr:line2}). After that, all entries to unavailable destinations are invalidated
(Line~\ref{rerr:line5}), and 
as usual when routing table entries are invalidated, the \penFlag{}s are set to {\nonpen} (Line~\ref{rerr:line5a}).
In Line~\ref{rerr:line3} the set of all precursors (affected
neighbours) of the unavailable destinations are summarised in the set $\pre$.
Then, the set {\dests} is ``thinned out'' to only those destinations that have at least one precursor---
only these destinations are transmitted in the forwarded error message
(Line~\ref{rerr:line3a}).
Finally, the message is sent (Line~\ref{rerr:line6}).

\subsection{The Message Queue and Synchronisation}\label{ssec:message_queue}

We assume that any message sent by a node \dval{sip} to a node
\dval{ip} that happens to be within transmission range of \dval{sip}
is actually received by \dval{ip}.  For this reason, \dval{ip} should
always be able to perform a receive action, regardless of which state
it is in.
However, the main process {\AODV} that runs on the node \dval{ip} can
reach a state, such as {\PKT}, {\RREQ}, {\RREP} or {\RERR}, in which
it is not ready to perform a receive action. For this reason we
introduce a process $\QMSG$, modelling a message queue,
that runs in parallel with {\AODV} or any other process that might be called.
Every incoming message is first stored in this queue, and piped from
there to the process {\AODV}, whenever {\AODV} is ready to handle a
new message. The process {\QMSG} is always ready to receive a new
message, even when {\AODV} is not.  The whole parallel process running
on a node is then given by an expression of the form
\[
(\xi,\aodv{\ip}{\sn}{\rt}{\rreqs}{\queues})\,\parl\,(\xii,\Qmsg{\msgs})
\,.\]

\subsection{Initial State}\label{ssec:initial}

To finish our specification, we have to define an initial state.  The
initial network expression is an encapsulated parallel composition of
node expressions $\dval{ip}\mathop{:}P\mathop{:}R$, where the (finite) number of nodes
and the range $R$ of each node expression is left unspecified (can be
anything). However, each node in the parallel composition is required
to have a unique IP address \dval{ip}.  The initial process $P$ of
\dval{ip} is given by the expression
$$(\xi,\aodv{\ip}{\sn}{\rt}{\rreqs}{\queues})\ \parl\ (\xii,\Qmsg{\msgs})\,,$$
with
\begin{equation}\label{eq:initialstate_rt}
\begin{array}{l}
\xi(\ip)=\dval{ip}
\wedge
\xi(\sn)=1
\wedge
\xi(\rt)=\emptyset
{}\wedge
\xi(\rreqs)=\emptyset\\
\qquad{}\wedge
\xi(\queues)=\emptyset
\wedge
\xii(\msgs)=[\,]\,.
\end{array}
\end{equation}
This says that initially each node is correctly informed about its own
identity; its own sequence number is initialised with $1$ and its
routing table, the list of RREQs seen, the store of queued data
packets as well as the message queue are empty.

\section{Invariants}\label{sec:invariants}
\newcommand{\rtord}[1][\dval{dip}]{\ensuremath{\sqsubseteq_{#1}}}
\newcommand{\decremented}{\mathbin{\stackrel{\bullet}{\raisebox{0pt}[2pt]{$-$}}}1}
\newcommand{\rtequiv}[1][\dval{dip}]{\ensuremath{\approx_{#1}}}
\newcommand{\rtsord}[1][\dval{dip}]{\ensuremath{\sqsubset_{#1}}}
\newcommand{\hopsc}{\dval{hops}_c}
\newcommand{\dipc}{\dval{dip}_{c}}
\newcommand{\ripc}{\dval{rip}_{c}}
\newcommand{\oipc}{\dval{oip}_{c}}
\newcommand{\rreqidc}{\dval{rreqid}_{c}}
\newcommand{\dsnc}{\dval{dsn}_c}
\newcommand{\rsnc}{\dval{rsn}_c}
\newcommand{\destsc}{\dval{dests}_c}
\newcommand{\osnc}{\dval{osn}_c}
\newcommand{\ipc}{\dval{ip}_{c}}
\newcommand{\xiN}[2][N]{\xi_{#1}^{#2}}
\newcommand{\zetaN}[2][N]{\zeta_{#1}^{#2}}
\newcommand{\RN}[2][N]{R_{#1}^{#2}}
\newcommand{\PN}[2][N]{P_{#1}^{#2}}

Using \awn and the proposed
model of AODV we can now formalise and prove crucial properties of
AODV\@.  In this section we verify properties that can be expressed as
invariants, i.e., statements that hold all the time when the protocol
is executed.

The most important invariant we establish is \emph{loop freedom}; most
prior results can be regarded as stepping stones towards this goal.
Next to that we also formalise and discuss \emph{route correctness}.

\subsection{State and Transition Invariants}\label{ssec:transition invariants}

A \emph{(state) invariant} is a statement that holds for all reachable
states of our model. Here states are network expressions, as
formally defined in \cite{TR13} and described in \Sect{awn}.
An invariant is usually verified by showing that it holds
for all possible initial states, and that, for any transition
$N\ar{\ell}N'$ between network expressions derived by our operational
semantics, if it holds for state $N$ then it also holds for state $N'$.
In this paper we abstain from a formal definition of the
operational semantics, and hence do not define the
labelled transition relation $\ar{}$ between network states.
Instead we verify invariants by checking that they are preserved under
any execution of any line in one of the Processes \ref{pro:aodv}--\ref{pro:queues}.
In \cite{TR13} we formally document that such a check yields the
required result.

Besides (state) invariants, we also establish statements we call
\emph{transition invariants}.  A transition invariant is a statement
that holds for each reachable transition $N\ar{\ell}N'$ between 
network expressions derived by the operational semantics.
Again these transitions correspond with lines in one of the
Processes \ref{pro:aodv}--\ref{pro:queues}; they either describe a
relation between the states $N$ and $N'$ before and after executing
the instruction---e.g.\ that the value of a specific variable
maintained by our processes will never decrease---or they describe a
relation between the instruction being executed (such as a broadcast of a
message involving a certain value) and the state right beforehand
(such as a comparable value maintained by the broadcasting node).
Transition invariants are simply checked by going though all
appropriate lines in Processes \mbox{\ref{pro:aodv}--\ref{pro:queues}}.
In a few cases we use
\hypertarget{induction-on-reachability}{\emph{induction on reachability}};
this amounts to assuming that the same relation holds for
instructions executed earlier. Again we refer to \cite{TR13} for the
soundness of this approach.

\mbox{\!In} our formalisation of transition invariants,
we write $N\ar{R:\starcastP{\dval{m}}}_\dval{ip}N'$ to indicate that
our network moves from state $N$ to state $N'$ by means of a
\textbf{broadcast}, \textbf{unicast} or \textbf{groupcast} of the
message $m$, executed by node \dval{ip}, while the current
transmission range of this node is $R$.

The following observations are crucial in establishing many of our invariants.
\begin{proposition}\rm\label{prop:preliminaries}~
\begin{enumerate}[(a)]
\item\label{it:preliminariesi} With the exception of new packets that
  are submitted to a node by a client of AODV, every message received and handled by the
  main routine of AODV has to be sent by some node before.\label{before}
  More formally, we consider an arbitrary
  path 
  \[ N_0\ar{\ell_1}N_1\ar{\ell_2} \ldots \ar{\ell_k} N_k\]
  \noindent with
  $N_0$ an initial state in our model of AODV\@. If the transition
  $N_{k-1}\ar{\ell_k} N_k$ results from a synchronisation involving the
  action $\receive{\msg}$ from\linebreak Line~\ref{aodv:line2} of
  Pro.~\ref{pro:aodv}---performed by the node \dval{ip}---where the
  variable {\msg} is assigned the value $m$, then either
  $m=\newpkt{\dval{d}}{\dval{dip}}$ or one of the $\ell_i$ with $i<k$
  stems from an action $\starcastP{m}$ of a node $\dval{ip}'$ of the network.
\item\label{it:preliminariesii} No node can receive a message directly from itself.
  Using the formalisation above, we need $\dval{ip}\neq\dval{ip}'$.
\end{enumerate}
\end{proposition}
\prf
The only way Line~\ref{aodv:line2} of Pro.~\ref{pro:aodv} can be
executed, is through a synchronisation of the main process \AODV\ with
the message queue \QMSG\ (Pro.~\ref{pro:rerr}) running on the same
node. This involves the action $\send{m}$ of \QMSG\@.  Here $m$ is
popped from the message queue {\msgs}, which started out empty. So at
some point \QMSG\ must have performed the action $\receive{m}$. However,
this action is blocked by the encapsulation operator $[\_]$, except when $m$ has the form
$\newpkt{\dval{d}}{\dval{dip}}$ or when it synchronises with an action
$\starcastP{m}$ of another node.
\eprf
At first glance Part\eqref{it:preliminariesii} does not seem to reflect reality. Of course, an application running on a local node has to be able to send 
data packets to another application running on the same node. However,
in any practical implementation, when a node sends a message to itself, the message 
will be delivered to the corresponding application on the local node without ever being ``seen'' by AODV 
or any other routing protocol. Therefore,
from AODV's perspective, no node can receive a message (directly) from itself.

\subsection{Notions and Notations}
Before formalising and proving invariants, we introduce some useful notions and notations.

All processes except $\QMSG$ maintain the five data variables {\ip}, {\sn},
{\rt}, {\rreqs} and {\queues}. Next to that $\QMSG$ maintains the variable $\msgs$.
Hence, these $6$ variables can be evaluated at any time.
Moreover, every node expression in the transition system looks like
\[
\dval{ip}:\left(\xi,P\ \parl\ \xii,\Qmsg{\msgs}\right):R
\,,\]
where $P$ is a state in one of the following sequential processes:
\[\begin{array}{l}
\aodv{\ip}{\sn}{\rt}{\rreqs}{\queues}\,,\\[0.5mm]
\newpktPL{\data}{\dip}{\ip}{\sn}{\rt}{\rreqs}{\queues}\,,\\[0.5mm]
\pktPL{\data}{\dip}{\oip}{\ip}{\sn}{\rt}{\rreqs}{\queues}\,,\\[0.5mm]
\RREQ({\hops}\comma{\rreqid}\comma{\dip}\comma
	{\dsn}\comma{\dsk}\comma{\oip}\comma{\osn}\comma{\sip}\;\!\comma\;\!{}\\
\phantom{\RREQ(}{\ip}\comma{\sn}\comma{\rt}\comma{\rreqs}\comma{\queues})\\[0.5mm]
\RREP({\hops}\comma{\dip}\comma{\dsn}\comma{\oip}\comma{\sip}\;\!\comma\;\!{}\\
\phantom{\RREQ(}{\ip}\comma{\sn}\comma{\rt}\comma{\rreqs}\comma{\queues})\\[0.5mm]
\rerrPL{\dests}{\sip}{\ip}{\sn}{\rt}{\rreqs}{\queues}\,.\\[0.5mm]
\end{array}\]

\noindent Hence the state of the transition system for a node $\dval{ip}$
is determined by
the process $P$,
the range $R$, and
the two valuations $\xi$ and $\xii$.
If a network consists of a (finite) set $\IP\subseteq\tIP$ of nodes, a
reachable network expression $N$ is an encapsulated parallel composition
of node expressions---one for each $\dval{ip}\in\IP$.
In this section, we assume $N$ and $N'$ to be reachable
network expressions in our model of AODV\@.
To distill information about a node from $N$,
we define the following projections:
\[
\begin{array}{@{}l@{\,\mathbin{:=}\,}l@{\;\mbox{where}\;}l@{{}\mathop:}c@{\mathop:{}}l@{\ \mbox{is a node expr.\ of $N\!$}}@{}l@{}}
\PN{\dval{ip}}       & P,        & \dval{ip} & (*,P\parl *,*)       & *   &\,,\\[0.5mm]
\RN{\dval{ip}}       &R,         & \dval{ip} & (*,*\parl *,*)        & \!R  &\,,\\[0.5mm]
\xiN{\dval{ip}}       & \xi,      & \dval{ip}  & (\xi,*\parl *,*)     & *   &\,,\\[0.5mm]
\zetaN{\dval{ip}}   &\zeta,   & \dval{ip}   & (*,*\parl \zeta,*) & *   &\,.
\end{array}
\]
\noindent
For example,
$\PN{\dval{ip}}$ determines the sequential process the node is currently working in,
$\RN{\dval{ip}}$ denotes the set of all nodes currently within transmission range of $\dval{ip}$, and 
$\xiN{\dval{ip}}(\rt)$ evaluates the current routing table maintained by node \dval{ip} in the network expression $N$.
In the forthcoming proofs, when discussing the effects of an action,
identified by a line number in one of the processes of our model, $\xi$ denotes the current valuation
\plat{$\xiN{\dval{ip}}$}, where \dval{ip} is the address of
the local node, executing the action under consideration, and
$N$ is the network expression obtained right before this action occurs, corresponding
with the line number under consideration. When considering the effects of
several actions, corresponding to several line numbers, $\xi$ is
always interpreted most locally. For instance, in the proof of \Prop{msgsending}\eqref{prop:msgsendingRREQ},
case \hyperlink{forinstance}{\textbf{Pro.~\ref*{pro:rreq}, Line~\ref*{rreq:line34}}}, we write
\begin{myquote}
  Hence \ldots $\ipc:=\xi(\ip)=\dval{ip}$ and $\xiN{\ipc}=\xi$ (by \Eq{uniqueidwithxi}).
  At Line~\ref{rreq:line6} we update the routing table using
  $\dval{r}:=\xi(\oip,\osn,\kno,\val,\hops\mathord+1,\sip,\emptyset)$
  as new entry.  The routing table does not change between
  Lines~\ref{rreq:line6} and~\ref{rreq:line34}; nor do the values of
  {\hops}, {\oip} and {\osn}.
\end{myquote}
Writing $N_k$ for a network expression in which the local node \dval{ip} is about to
execute Line~$k$, this passage can be reworded as
\begin{myquote}
  Hence \ldots $\ipc:=\xiN[N_{\ref*{rreq:line34}}]{\dval{ip}}(\ip)=\dval{ip}$ and
  $\xiN[N_{\ref*{rreq:line34}}]{\ipc}=\xiN[N_{\ref*{rreq:line34}}]{\dval{ip}}$
  (by \Eq{uniqueidwithxi}). 
  $\xiN[N_{\ref*{rreq:line8}}]{\dval{ip}}(\rt)$~:=\\
  {
  \renewcommand{\comma}{\mathbin{\text{\hskip-0.17em$,$\hskip-0.17em}}} 
  $\xiN[N_{\ref*{rreq:line6}}]{\dval{ip}}\hspace{-1pt}(\upd{\rt}{(\oip\comma\osn\comma\kno\comma\val\comma\hops\mathord+1\comma\sip\comma\emptyset)})$\\
  }%
   $:=~\upd{\xiN[N_{\ref*{rreq:line6}}]{\dval{ip}}(\rt)}{(\xiN[N_{\ref*{rreq:line6}}]{\dval{ip}}(\oip)\comma\xiN[N_{\ref*{rreq:line6}}]{\dval{ip}}(\osn)\comma\ldots)}.$\\
$\begin{array}{@{}r@{~}l@{\,\wedge\,}r@{~}l@{}}
\xiN[N_{\ref*{rreq:line8}}]{\dval{ip}}(\rt)
=&\xiN[N_{\ref*{rreq:line34}}]{\dval{ip}}(\rt) & \xiN[N_{\ref*{rreq:line6}}]{\dval{ip}}(\hops)
=&\xiN[N_{\ref*{rreq:line34}}]{\dval{ip}}(\hops)\, \wedge\\ \xiN[N_{\ref*{rreq:line6}}]{\dval{ip}}(\oip)
=&\xiN[N_{\ref*{rreq:line34}}]{\dval{ip}}(\oip) & \xiN[N_{\ref*{rreq:line6}}]{\dval{ip}}(\osn)
=&\xiN[N_{\ref*{rreq:line34}}]{\dval{ip}}(\osn).
\end{array}$
\end{myquote}
In all of case \hyperlink{forinstance}{\textbf{Pro.~\ref*{pro:rreq}, Line~\ref*{rreq:line34}}}, through the
statement of the proposition, $N$ is bound to $N_{\ref*{rreq:line34}}$,
so that $\xiN{\dval{ip}}=\xiN[N_{\ref*{rreq:line34}}]{\dval{ip}}$.

In \SSect{rt} we have defined functions that work on evaluated
routing tables $\xiN{\dval{ip}}(\rt)$, such as $\fnnhop$.
To ease readability, we abbreviate
\plat{$\nhop{\xiN{\dval{ip}}(\rt)}{\dval{dip}}$} by \plat{$\nhp{\dval{ip}}$}.
Similarly, we 
use 
\plat{$\sq{\dval{ip}}$}, \plat{$\dhp{\dval{ip}}$}, \plat{$\sta{\dval{ip}}$}, \plat{$\sr{\dval{ip}}$}, 
\plat{$\kd{\dval{ip}}$}, \plat{$\akd{\dval{ip}}$}
and \plat{$\ikd{\dval{ip}}$} for
\linebreak
\plat{$\sqn{\xiN{\dval{ip}}(\rt)\!}{\!\dval{dip}}$},
\plat{$\dhops{\xiN{\dval{ip}}(\rt)\!}{\!\dval{dip}}$},
\plat{$\status{\xiN{\dval{ip}}(\rt)\!}{\!\dval{dip}}$,}\linebreak
 \plat{$\selr{\xiN{\dval{ip}}(\rt)}{\dval{ip}}$},
\plat{$\kD{\xiN{\dval{ip}}(\rt)}$}, 
\plat{$\akD{\xiN{\dval{ip}}(\rt)}$}
and\linebreak
 \plat{$\ikD{\xiN{\dval{ip}}(\rt)}$}, respectively.

\subsection{Basic Properties}
In this section we show some of the most fundamental invariants for AODV. 
The first one is already stated in the RFC~\cite[Sect. 3]{rfc3561}.
\begin{proposition}\rm
\label{prop:invarianti_itemiii}
Any sequence number of a given node $\dval{ip}$ increases monotonically, i.e., never decreases, 
and is never unknown.
  That is, for $\dval{ip}\mathbin\in\IP$, if $N \ar{\ell} N'$
  then $1\leq\xiN{\dval{ip}}(\sn)\leq\xiN[N']{\dval{ip}}(\sn)$.
\end{proposition}

\prf
In all initial states the invariant is satisfied, as 
all sequence numbers of all nodes are set to $1$
(see~\eqref{eq:initialstate_rt} in Section~\ref{ssec:initial}).
The Processes \ref{pro:aodv}--\ref{pro:queues} of
\Sect{modelling_AODV} change a node's sequence number
only through the functions
{\fninc} and $\max$.
This occurs at two places only:
\begin{description}
  \item[Pro.~\ref{pro:aodv}, Line~\ref{aodv:line36}:]
    Here $\xiN{\dval{ip}}(\sn)\mathbin\leq
    \inc{\xiN{\dval{ip}}(\sn)} \mathbin= \xiN[N']{\dval{ip}}(\sn)$.
  \item[Pro.~\ref{pro:rreq}, Line~\ref{rreq:line12}:]
    {}\hspace{-0.05pt}Here $\xiN{\dval{ip}}(\sn)\mathbin\leq
    \max({\xiN{\dval{ip}}(\sn)}{,}*) \mathbin= \xiN[N']{\dval{ip}}(\sn)$.
\end{description}
From this and the fact that all sequence numbers are initialised with
$1$ we get $1\leq\xiN{\dval{ip}}(\sn)$.
\eprf

The proof strategy used above can be generalised.

\begin{remark}\label{rem:remark}
Most of the forthcoming proofs can be done by showing the statement
for each initial state and then checking all locations in the
processes where the validity of the invariant is possibly changed.
Note that routing table entries are only changed by the functions
\hyperlink{update}{$\fnupd$}, \hyperlink{invalidate}{$\fninv$}
or \hyperlink{addprert}{$\fnaddprecrt$}.
Thus we have to show that an invariant dealing with routing tables is
satisfied after the execution of these functions if it was valid
before.  In our proofs, we go through all occurrences of these functions.
In case the invariant does not make statements about precursors, 
the function $\fnaddprecrt$ need not be considered.
\end{remark}
To ease readability we defer most of the proofs to the appendix;
and show only the most important ones in the main text.

\begin{proposition}\rm\label{prop:destinations maintained}
  The set of known destinations of any node increases monotonically.
  That is, for $\dval{ip}\mathbin\in\IP$,\linebreak[2] if $N \ar{\ell} N'$
  then $\kd{\dval{ip}}\subseteq\kd[N']{\dval{ip}}$.
\end{proposition}
\prfatend
None of the functions used to change routing tables removes an entry altogether.
\eprf

\begin{proposition}\rm\label{prop:dsn increase}
In each node's routing table, the
  sequence number for a given destination increases monotonically, i.e., never decreases.
  That is, for $\dval{ip},\dval{dip}\mathbin\in\IP$, if $N \ar{\ell} N'$
  then $\sq{\dval{ip}}\leq\fnsqn_{N'}^\dval{ip}(\dval{dip})$.
\end{proposition}
\prfnoboxatend
The only function that can decrease a destination sequence number is 
\hyperlink{invalidate}{$\fninv$}. 
When invalidating routing table entries using the function $\inv{\rt}{\dests}$, sequence numbers are copied from {\dests} to the corresponding entry in \rt. 
It is sufficient to show that for all \plat{$(\dval{rip},\dval{rsn})\in{}$}
\plat{$\xiN{\dval{ip}}(\dests)$} we have
$\sq[\dval{rip}]{\dval{ip}}\leq\dval{rsn}$, as all other sequence numbers in routing table
entries remain unchanged.
\begin{description}
 \myitem[Pro.~\ref{pro:aodv}, Line~\ref{aodv:line32}; Pro.~\ref{pro:pkt}, Line~\ref{pkt2:line10};Pro.~\ref{pro:rreq}, Lines~\ref{rreq:line18}, \ref{rreq:line30}; Pro.~\ref{pro:rrep},
Line~\ref{rrep:line18}:] The set {\dests} is constructed immediately before the invalidation procedure. For $(\dval{rip},\dval{rsn})\in\xiN{\dval{ip}}(\dests)$, we have
$
   \sq[\dval{rip}]{\dval{ip}} \leq \inc{\sq[\dval{rip}]{\dval{ip}}} = \dval{rsn}.
$
\item[Pro.~\ref{pro:rerr}, Line~\ref{rerr:line5}:] When constructing {\dests} in Line~\ref{rerr:line2}, the side condition $\xiN[N_{\ref*{rerr:line2}}]{\dval{ip}}(\sqn{\rt}{\rip})<\xiN[N_{\ref*{rerr:line2}}]{\dval{ip}}(\rsn)$ is taken into account, which immediately yields the claim for $(\dval{rip},\dval{rsn})\in\xiN{\dval{ip}}(\dests)$.
\qed
\end{description}
\eprf

Our next invariant tells that each node is correctly informed about
its own identity.
\begin{proposition}\rm\label{prop:self-identification}
For each $\dval{ip}\in\IP$ and each reachable state $N$ we have
$\xiN{\dval{ip}}(\ip)=\dval{ip}$.
\end{proposition}
\prfatend
According to \SSect{initial} the claim holds for each initial state, and none of our processes has an
assignment changing the value of the variable $\ip$.
\eprf
This proposition will be used implicitly in many of the proofs to follow.
In particular, for all $\dval{ip}',\dval{ip}''\mathbin{\in}\IP$%
\begin{equation}\label{eq:uniqueidwithxi}
\xiN{\dval{ip}'}(\ip)=\dval{ip}'' \Rightarrow \dval{ip}'=\dval{ip}''\wedge \xiN{\dval{ip}'}=\xiN{\dval{ip}''}\,.
\end{equation}

Next, we show that every AODV control message contains the IP address of the sender.
\begin{proposition}\rm\label{prop:ip=ipc} If an AODV control message is
  sent by node $\dval{ip}\in\IP$, the node sending this message
  identifies itself correctly:
$	N\ar{R:\starcastP{m}}_{\dval{ip}}N' \Rightarrow \dval{ip}=\ipc$,
where the message $m$ is either
$\rreq{*}{*}{*}{*}{*}{*}{*}{\ipc}$,
$\rrep{*}{*}{*}{*}{\ipc}$, or
$\rerr{*}{\ipc}$.
\end{proposition}
The proof is straightforward: whenever such a message is sent in
one of the processes of \Sect{modelling_AODV}, $\xi(\ip)$ is set as the last argument.
\qed

\begin{proposition}\rm\label{prop:positive hopcount}
All routing table entries have a hop count greater than or equal to $1$.
\begin{equation}\label{eq:inv_length}
(*,*,*,*,\dval{hops},*,*)\in\xiN{\dval{ip}}(\rt) \Rightarrow \dval{hops}\geq1
\end{equation}
\end{proposition}

\prfnoboxatend
All initial states trivially satisfy the invariant since all routing tables are empty.
The functions \hyperlink{invalidate}{\fninv} and \hyperlink{addprert}{\fnaddprecrt} do not affect
the invariant, since they do not change the hop count of a routing table entry.
Therefore, we only have to look at the application calls of \hyperlink{update}{\fnupd}.
In each case, if the update does not change the routing table entry beyond its precursors
({the last clause} of \hyperlink{update}{\fnupd}), the invariant is trivially
preserved; hence we examine the cases that an update actually occurs.
	\begin{description}
		\item[Pro.~\ref{pro:aodv}, Lines~\ref{aodv:line10}, \ref{aodv:line14}, \ref{aodv:line18}:]
			All these updates have a hop count equal to $1$; hence the invariant is preserved.
		\item[Pro.~\ref{pro:rreq}, Line~\ref{rreq:line6}; Pro.~\ref{pro:rrep}, Line~\ref{rrep:line5}:]
			Here, $\xi(\hops)+1$ is used for the update. Since $\xi(\hops)\in\NN$, the invariant is maintained.
			\qed
	\end{description}
\eprf

\begin{proposition}\rm\label{prop:starcastNew}~
\begin{enumerate}[(a)]
\item If a route request with hop count $0$ is sent by a node
  $\ipc\in\IP$, the sender must be the originator.
	\begin{equation}\label{inv:starcast_i}
	N\ar{R:\starcastP{\rreq{0}{*}{*}{*}{*}{\oipc}{*}{\ipc}}}_{\dval{ip}}N' \Rightarrow \oipc\mathbin=\ipc
	\end{equation}
\item If a route reply with hop count $0$ is sent by a node
  $\ipc\in\IP$, the sender must be the destination.
	\begin{equation}\label{inv:starcast_i_rrep}
        N\ar{R:\starcastP{\rrep{0}{\dipc}{*}{*}{\ipc}}}_{\dval{ip}}N' \Rightarrow \dipc\mathbin=\ipc
	\end{equation}
\end{enumerate}
\end{proposition}
\prfnoboxatend
\begin{enumerate}[(a)]
\item We have to check that the consequent holds whenever a route request is sent. In all the processes there
are only two locations where this happens.
\begin{description}
	\item[\Pro{aodv}, Line~\ref{aodv:line39}:]
		A request with content 
		$
		\xi(0\comma*\comma*\comma*\comma*\comma\ip\comma*\comma\ip)
		$
		 is sent. Since the sixth and the eighth component are
                 the same ($\xi(\ip)$), the claim holds.
	\item[Pro.~\ref{pro:rreq}, Line~\ref{rreq:line34}:]
	The message has the form 
	$\rreqID(\nosp{\xi(\hops)}$\newline $\nosp{+1}\comma\nosp{*}\comma\nosp{*}\comma\nosp{*}\comma\nosp{*}\comma\nosp{*}\comma\nosp{*}\comma\nosp{*})$
	Since $\xi(\hops)\in\NN$, $\xi(\hops)+1\not=0$ and hence the
        antecedent does not hold.
\end{description}
\item We have to check that the consequent holds whenever a route reply is sent. In all the processes there
are only three locations where this happens.
\begin{description}
	\item[\Pro{rreq}, Line~\ref{rreq:line14a}:]
		A reply with content 
		$
		\xi(0\comma\dip\comma*\comma*\comma\ip)
		$
		is sent. By Line~\ref{rreq:line10} we have
                $\xi(\dip)=\xi(\ip)$, so the claim holds.
	\item[Pro.~\ref{pro:rreq}, Line~\ref{rreq:line26}:]
	$\rrep{\dhops{\rt}{\dip}}{*}{*}{*}{*}$
	is the form of the message.
        By \Prop{positive hopcount}, $\dhops{\rt}{\dip}>0$, so the
        antecedent does not hold.
	\item[Pro.~\ref{pro:rrep}, Line~\ref{rrep:line13}:]
	$\rrep{\xi(\hops)\mathord+1}{*}{*}{*}{*}$
	is the form of the message.
	Since $\xi(\hops)\in\NN$, $\xi(\hops)+1\not=0$ and hence the
        antecedent does not hold.
\qed
\end{description}
\end{enumerate}
\eprf

\begin{proposition}\rm\label{prop:rte}~
\hfuzz .3pt
\begin{enumerate}[(a)]
\item\label{it_a} Each routing table entry
with $0$ as its destination sequence number has a sequence-number-status flag valued unknown.
\begin{equation}\label{eq:unk_sqn}
(\dval{dip},0,\dval{f},*,*,*,*)\in\xiN{\dval{ip}}(\rt) \Rightarrow \dval{f}=\unkno
\end{equation}
\item\label{it_d} Unknown sequence numbers can only occur at $1$-hop connections.
\begin{equation}\label{eq:inv_viia}
(*,*,\unkno,*,\dval{hops},*,*)\in\xiN{\dval{ip}}(\rt) \Rightarrow \dval{hops}=1
\end{equation}
\item\label{it_c} $1$-hop connections must contain the destination as next hop.
\begin{equation}\label{eq:inv_vii}
(\dval{dip},*,*,*,1,\dval{nhip},*)\in\xiN{\dval{ip}}(\rt)\Rightarrow \dval{dip}=\dval{nhip}
\end{equation}
\item\label{it_e}
If the sequence number $0$ occurs within a routing table entry, the
hop count as well as the next hop can be determined.
\begin{equation}\label{eq:inv_viib}
\begin{array}{ll}
&(\dval{dip},0,\dval{f},*,\dval{hops},\dval{nhip},*)\in\xiN{\dval{ip}}(\rt)\\
\Rightarrow& \dval{f}=\unkno \wedge \dval{hops}=1\wedge \dval{dip}=\dval{nhip}
\end{array}
\end{equation}
\end{enumerate}
\end{proposition}
\prfnoboxatend 
At the initial states all routing tables are empty. Since \hyperlink{invalidate}{$\fninv$} and \hyperlink{addprert}{$\fnaddprecrt$}
  change neither the sequence-number-status flag, nor the next hop or the hop count of a routing
  table entry, and---by \Prop{dsn increase}---cannot decrease the sequence number of a destination,
  we only have to look at the application calls of \hyperlink{update}{\fnupd}.
  As before, we only examine the cases that an update actually occurs.
\begin{enumerate}[(a)]
\item
 Function calls of the form \hyperlink{update}{\upd{\dval{rt}}{\dval{r}}} always preserve the invariant:
in case {\fnupd} is
  given an argument for which it is not defined,\footnote{In \SSect{well-definedness}
  we will show that this cannot occur.} the process algebra blocks and no change
  of the routing table is performed \cite[Section~4]{TR13};
in case one of the first four clauses in the definition of \hyperlink{update}{\fnupd} is used, this
follows because $\upd{\dval{rt}}{\dval{r}}$ is defined only when $\pi_{2}(\dval{r})\mathbin=0\Leftrightarrow\pi_{3}(\dval{r})\mathbin=\unkno$;
 in
case the fifth clause is used it follows because $\pi_{3}(\dval{r})=\unkno$; and in case the last
clause is used, it follows by induction, since the invariant was already valid before the update.
\item\begin{description}
		\item[Pro.~\ref{pro:aodv}, Lines~\ref{aodv:line10}, \ref{aodv:line14}, \ref{aodv:line18}:]
			All these updates have an unknown sequence number and hop count equal to $1$.
                        By Clause 5 of
			\hyperlink{update}{\fnupd}, these sequence-number-status flag and hop count
			are transferred literally into the routing table; hence the invariant is preserved.
		\item[Pro.~\ref{pro:rreq}, Line~\ref{rreq:line6} and Pro.~\ref{pro:rrep}, Line~\ref{rrep:line5}:]
		  In these updates the sequence-number-status flag is set to {\kno}. By the definition of
		  \hyperlink{update}{\fnupd}, this value ends up in the routing table.
		  Hence the assumption of the invariant to be proven
		  is not satisfied. 
	\end{description}
\item 	\begin{description}
		\item[Pro.~\ref{pro:aodv}, Lines~\ref{aodv:line10}, \ref{aodv:line14}, \ref{aodv:line18}:]
			The new entries, which have the form $\xi(\sip,0,\unkno,\val,1,\sip,\emptyset)$, satisfy the invariant; even if the routing table is actually
			updated with one of the new routes, the invariant holds afterwards.
		\item[Pro.~\ref{pro:rreq}, Line~\ref{rreq:line6}; Pro.~\ref{pro:rrep}, Line~\ref{rrep:line5}:]
			The route that might be inserted into the routing table
			has hop count $\hops\mathord+1$, $\hops\in\NN$.
		 	It can only be equal to $1$ if the received message had hop count $\hops=0$.
			In that case
                        Invariant~\eqref{inv:starcast_i}, resp.~\eqref{inv:starcast_i_rrep}, guarantees that the invariant remains unchanged.
	\end{description}
\item Immediate from Parts~\eqref{it_a} to \eqref{it_c}. \qed
\end{enumerate}
\pagebreak[3]
\eprf

\pagebreak[3]
\begin{proposition}\rm\label{prop:msgsendingii}~
\begin{enumerate}[(a)]
\item Whenever an originator sequence number is sent as part of a route request message, it is known, i.e.,
it is greater than or equal to $1$.
	\begin{equation}\label{inv:starcast_sqni}	
		N\ar{R:\starcastP{\rreq{*}{*}{*}{*}{*}{*}{\osnc}{*}}}_{\dval{ip}}N' \Rightarrow \osnc \geq1
	\end{equation}
\item Whenever a destination sequence number is sent as part of a route reply message, it is known, i.e.,
it is greater than or equal to $1$.
 	\begin{equation}\label{inv:starcast_sqnii}
		N\ar{R:\starcastP{\rrep{*}{*}{\dsnc}{*}{*}}}_{\dval{ip}}N'\Rightarrow\dsnc\geq1
 	\end{equation}
\end{enumerate}
\end{proposition}
\prfnoboxatend~
\begin{enumerate}[(a)]
\item
We have to check that the consequent holds whenever a route request is sent.
\begin{description}
	\item[Pro.~\ref{pro:aodv}, Line~\ref{aodv:line39}:] A route request is initiated. 
		The originator sequence number is a copy of the node's own sequence 
		number, i.e., $\osnc=\xi(\sn)$. By \Prop{invarianti_itemiii}, 
		we get $\osnc\geq 1$.
	\item[Pro.~\ref{pro:rreq}, Line~\ref{rreq:line34}:] $\osnc:=\xi(\osn)$
		is not changed within Pro.~\ref{pro:rreq}; it stems, through Line~\ref{aodv:line8}
		of Pro.~\ref{pro:aodv}, from an incoming RREQ message (Pro.~\ref{pro:aodv}, Line~\ref{aodv:line2}).
		For this incoming RREQ message, using
		\Prop{preliminaries}(\ref{it:preliminariesi}) and
		\hyperlink{induction-on-reachability}{induction on reachability}, the
		invariant holds; hence the claim follows immediately. 
\end{description}
\item We have to check that the consequent holds whenever a route reply is sent.
\begin{description}
	\item[Pro.~\ref{pro:rreq}, Line~\ref {rreq:line14}:] The destination initiates a route reply. 
		The sequence number is a copy of the node's own sequence number, i.e., 
		$\dsnc=\xi(\sn)$. By \Prop{invarianti_itemiii}, we get $\dsnc\geq 1$.
	\item[Pro.~\ref{pro:rreq}, Line~\ref{rreq:line26}:] 
		The sequence number used for the message is copied from the routing table;
		its value is $\dsnc:=\sqn{\xi(\rt)}{\xi(\dip)}$. 
		By Line~\ref{rreq:line22}, we know that $\status{\xi(\rt)}{\xi(\dip)}=\kno$ and hence,
                by Invariant~\Eq{unk_sqn}, $\dsnc\geq 1$. Thus the invariant is maintained.
	\item[Pro.~\ref{pro:rrep}, Line~\ref{rrep:line13}:] $\dsnc:=\xi(\dsn)$
		is not changed within Pro.~\ref{pro:rrep}; it stems, through Line~\ref{aodv:line12}
		of Pro.~\ref{pro:aodv}, from an incoming RREP message (Pro.~\ref{pro:aodv}, Line~\ref{aodv:line2}).
		For this incoming RREP message the invariant holds and hence the claim follows immediately.
\qed
\end{description}
\end{enumerate}
\eprf
\pagebreak[3]
\begin{proposition}\rm\label{prop:msgsending}~
\begin{enumerate}[(a)]
\item\label{prop:msgsendingRREQ}
If a route request is sent (forwarded) by a node $\ipc$ different from
the originator of the request 
then the content of $\ipc$'s routing table
must be fresher or at least as good as the information inside the message.\hspace{-1mm}
	\begin{equation}\label{inv:starcast_ii}	
	\begin{array}{ll}
	  &N\ar{R:\starcastP{\rreq{\hopsc}{*}{*}{*}{*}{\oipc}{\osnc}{\ipc}}}_{\dval{ip}}N'
          \\&{}\wedge\ipc\neq\oipc\\
	  \Rightarrow&
	  \oipc\in\kd{\ipc}
	  \wedge\big(\sq[\oipc]{\ipc}>\osnc
	  \\
	  &{}\vee (\sq[\oipc]{\ipc}=\osnc \wedge \dhp[\oipc]{\ipc}\leq\hopsc\\
	  &\phantom{{}\vee (}{}\wedge \sta[\oipc]{\ipc}=\val)\big)
	\end{array}
	\end{equation}
\item
If a route reply is sent by a node $\ipc$, different from
the destination of the route, then the content of $\ipc$'s routing table
must be consistent with the information inside the message.
 	\begin{equation}\label{inv:starcast_iv}
 	\begin{array}{ll}
 	  &N\ar{R:\starcastP{\rrep{\hopsc}{\dipc}{\dsnc}{*}{\ipc}}}_{\dval{ip}}N'
          \\&{}\wedge\ipc\neq\dipc\\
 	  \Rightarrow&
	  \dipc\in\kd{\ipc}
 	  \wedge \sq[\dipc]{\ipc} = \dsnc\\
	  &{}\wedge  \dhp[\dipc]{\ipc}=\hopsc\wedge \sta[\dipc]{\ipc}=\val
 	\end{array}
 	\end{equation}
\end{enumerate}
\end{proposition}
\prfnoboxatend~
\begin{enumerate}[(a)]
\item
We have to check all cases where a route request is sent:
\begin{description}
	\item[Pro.~\ref{pro:aodv}, Line~\ref{aodv:line39}:]
		A new route request is initiated with $\ipc=\oipc:=\xi(\ip)=\dval{ip}$.
		Here the antecedent of \eqref{inv:starcast_ii} is not satisfied.
	\item[Pro.~\ref{pro:rreq}, Line~\ref{rreq:line34}:]
                \hypertarget{forinstance}{The broadcast message has the form\linebreak
                $\xi(\rreq{\hops\mathord+1}{\rreqid}{\dip}{\max(\sqn{\rt}{\dip},\dsn)}{\dsk}{\oip}{\osn}{\ip}).$
                Hence $\hopsc:=\xi(\hops)\mathord+1$,
                $\oipc:=\xi(\oip)$, $\osnc:=\xi(\osn)$,
                $\ipc:=\xi(\ip)=\dval{ip}$ and $\xiN{\ipc}=\xi$
                (by \Eq{uniqueidwithxi}).
		At Line~\ref{rreq:line6} we update the routing table using
		$\dval{r}\mathbin{:=}\xi(\oip,\osn,\kno,\val,\hops\mathord+1,\sip,\emptyset)$ 
		as new entry.
		The routing table does not change between
                	Lines~\ref{rreq:line6}
                	and~\ref{rreq:line34}; nor do the values of the variables
                	{\hops}, {\oip} and {\osn}.}
		If the new (valid) entry is inserted into the routing table, then
                	one of the first four cases in the definition of \hyperlink{update}{$\fnupd$}
                	must have applied---the fifth case cannot apply, since $\pi_3(\dval{r})=\kno$.
                	Thus, using that $\oipc\neq\ipc$,
		\[
		\begin{array}{@{}r@{{}={}}l@{}}
		\sq[\oipc]{\ipc} & \sqn{\xi(\rt)}{\xi(\oip)}=\xi(\osn) = \osnc\\
		\dhp[\oipc]{\ipc} & \dhops{\xi(\rt)}{\xi(\oip)} = \xi(\hops)\mathop+1 \\& \hopsc\\
		\sta[\oipc]{\ipc}& \status{\xi(\rt)}{\xi(\oip)} = \xi(\val) = \val\,.
		\end{array}
		\]	
		In case the new entry is not inserted into the routing
		table (the sixth case of \hyperlink{update}{$\fnupd$}), we have
		$\sq[\oipc]{\ipc}$ ${}=\sqn{\xi(\rt)}{\xi(\oip)}\geq\xi(\osn)=\osnc$,
		and in case that		
		$\sq[\oipc]{\ipc}=\osnc$ we see that
		$\dhp[\oipc]{\ipc} = \dhops{\xi(\rt)}{\xi(\oip)}\leq \xi(\hops)\mathop{+}1 = \hopsc$
		and moreover $\sta[\oipc]{\ipc}=\val$.
		Hence the invariant holds.
\end{description}
\item We have to check all cases where a route reply is sent.
\begin{description}
	\item[Pro.~\ref{pro:rreq}, Line~\ref {rreq:line14}:]
		A new route reply with
		$\ipc:=\xi(\ip)=\dval{ip}$ is initiated.
		Moreover, by Line~\ref{rreq:line10}, $\dipc:=\xi(\dip)=\xi(\ip)=\dval{ip}$ and 
		thus $\ipc=\dipc$.
		Hence, the antecedent of \eqref{inv:starcast_iv} is not satisfied.
	\item[Pro.~\ref{pro:rreq}, Line~\ref{rreq:line26}:]
	        We have $\ipc:=\xi(\ip)=\dval{ip}$, so	$\xiN{\ipc}=\xi$.
		This time, by Line~\ref{rreq:line20}, 
		$\dipc:= \xi(\dip)\neq\xi(\ip)=\ipc$.
		By Line~\ref{rreq:line22} there is a valid routing table entry for $\dipc:=\xi(\dip)$.
		\[\begin{array}{r@{{}:={}}c@{{}={}}l}
		\dsnc&\sqn{\xi(\rt)}{\xi(\dip)}&\sq[\dipc]{\ipc}\,,
		\\
		\hopsc&\dhops{\xi(\rt)}{\xi(\dip)}&\dhp[\dipc]{\ipc}\,.
		\end{array}\]
	\item[Pro.~\ref{pro:rrep}, Line~\ref{rrep:line13}:]
		The RREP message has the form
		\[\xi(\rrep{\hops\mathop{+}1}{\dip}{\dsn}{\oip}{\ip})\,.\]
		Hence $\hopsc:=\xi(\hops)\mathord+1$,
                $\dipc:=\xi(\dip)$, $\dsnc:=\xi(\dsn)$,
                $\ipc:=\xi(\ip)=\dval{ip}$ and \mbox{$\xiN{\ipc}=\xi$}.
		Using $(\xi(\dip),$ $\xi(\dsn),\kno,\val,\xi(\hops)\mathord{+}1,\xi(\sip),\emptyset)$ as
		new entry,\linebreak the routing table is updated at Line~\ref{rrep:line5}.
		With exception of its precursors, which are irrelevant here, the routing table
		does not change between Lines~\ref{rrep:line5} and \ref{rrep:line13};
		nor do the values of the variables {\hops}, {\dip} and {\dsn}.
		Line~\ref{rrep:line3} guarantees that
		during the update in Line~\ref{rrep:line5},
		the new entry is inserted into the routing table,
		so
		\[
		\begin{array}{r@{{}={}}l}
		\sq[\dipc]{\ipc}& \sqn{\xi(\rt)}{\xi(\dip)} = \xi(\dsn) = \dsnc\\
		\dhp[\dipc]{\ipc} & \dhops{\xi(\rt)}{\xi(\dip)} = \xi(\hops)\mathop+1\\& \hopsc\\
		\sta[\dipc]{\ipc} & \status{\xi(\rt)}{\xi(\dip)} = \xi(\val) \\& \val\,.\hfill\qed
		\end{array}\]
\end{description}
\end{enumerate}
\eprf

\begin{proposition}\rm\label{prop:starcastrerr} Any sequence number appearing in a route error message
 stems from an invalid destination
 and is equal to the sequence number for that destination in
 the sender's routing table at the time of sending.
	\begin{equation}\label{inv:starcast_rerr}
	 \begin{array}{ll}
	  &N\ar{R:\starcastP{\rerr{\destsc}{\ipc}}}_{\dval{ip}}N'
	  \\&{}\wedge (\ripc,\rsnc)\in\destsc\\
	  \Rightarrow&\ripc\in\ikd{\dval{ip}} \wedge \rsnc = \sq[\ripc]{\dval{ip}}
	\end{array}
	\end{equation}
\end{proposition}

\prfnoboxatend
We have to check that the consequent holds whenever a route error message is sent.
In all the processes there are only seven locations where this happens.
\begin{description}
\item[Pro.~\ref{pro:aodv}, Line~\ref{aodv:line33}:]
The set $\destsc$ is constructed in Line~\ref{aodv:line31a} as a subset of 
$\xiN[N_{\ref*{aodv:line31a}}]{\dval{ip}}(\dests)=\xiN[N_{\ref*{aodv:line32}}]{\dval{ip}}(\dests)$.
For each pair $(\ripc,\rsnc)\in\xiN[N_{\ref*{aodv:line32}}]{\dval{ip}}(\dests)$
one has $\ripc=\xiN[N_{\ref*{aodv:line30}}]{\dval{ip}}(\rip)\in\fnakD_{N_{\ref*{aodv:line30}}}^{\dval{ip}}$.
Then in Line~\ref{aodv:line32}, using the function \hyperlink{invalidate}{$\fninv$},
$\status{\xi(\rt)}{\ripc}$ is set to $\inval$ and 
$\sqn{\xi(\rt)}{\ripc}$ to $\rsnc$.
Thus we obtain $\ripc\in\ikd{\dval{ip}}$ and
$\sq[\ripc]{\dval{ip}} = \rsnc$.
\myitem[Pro.~\ref{pro:pkt}, Line~\ref{pkt2:line14};
      Pro.~\ref{pro:rreq}, Lines~\ref{rreq:line19},~\ref{rreq:line31};
      Pro.~\ref{pro:rrep}, Line~\ref{rrep:line20};
      Pro.~\ref{pro:rerr}, Line~\ref{rerr:line6}:]
Exactly as above.
\item[Pro.~\ref{pro:pkt}, Line~\ref{pkt2:line20}:]
The set $\destsc$ contains only one single element. Hence $\ripc\mathbin{:=}\xiN{\dval{ip}}(\dip)$ and $\rsnc\mathbin{:=}\xiN{\dval{ip}}(\sqn{\rt}{\dip})$.
By Line~\ref{pkt2:line18}, we have $\ripc=\xiN{\dval{ip}}(\dip)\in\ikd{\dval{ip}}$. The remaining claim follows by 
$
\rsnc=\xiN{\dval{ip}}(\sqn{\rt}{\dip})=\sqn{\xiN{\dval{ip}}(\rt)}{\xiN{\dval{ip}}(\dip)} = \sq[\ripc]{\dval{ip}}.
$\qed
\end{description}
\eprf

\subsection{Well-Definedness}\label{ssec:well-definedness}
We have to ensure that our specification of AODV is 
actually well defined. Since many functions introduced in \Sect{types} are only partial,
it has to be checked that these functions are either defined when they are used, 
or are subterms of atomic formulas. In the latter case, those formula would evaluate 
to {\tt false} (cf.\ Footnote~\ref{fn:undefvalues}).

The first proposition shows that the functions defined in \Sect{types} respect the data structure.
In fact, these properties are required (or implied) by our data structure.
\begin{proposition}\rm\label{prop:invarianti}~
\begin{enumerate}[(a)]
\item\label{prop:invarianti_itemii} In each routing table
 there is at most one entry for each destination.
\item\label{prop:invarianti_itemiv} In each store of queued data
  packets there is at most one data queue for each destination.
\item\label{prop:invarianti_dests} Whenever a set of pairs $(\dval{rip},\dval{rsn})$ is assigned
  to the variable {\dests} of type $\tIP\rightharpoonup\tSQN$,
or to the first argument of the function $\rerrID$, this
  set is a partial function, i.e., there is at most one entry
  $(\dval{rip},\dval{rsn})$ for each destination $\dval{rip}$.
\end{enumerate}
\end{proposition}

\prfnoboxatend~
\begin{enumerate}[(a)]
\item In all initial states the invariant is satisfied, as a routing
table starts out empty
(see \eqref{eq:initialstate_rt} in Section~\ref{ssec:initial}).
None of the Processes \ref{pro:aodv}--\ref{pro:queues} of
\Sect{modelling_AODV} changes a routing table directly;
the only way a routing table can be changed is through the functions
\hyperlink{update}{$\fnupd$}, \hyperlink{invalidate}{$\fninv$} and
\hyperlink{addprert}{$\fnaddprecrt$}. The latter two
only change the sequence number, the validity status and the precursors of an existing route.
This kind of update has no effect on the invariant.
The first function inserts a new entry into a routing table only if the
destination is unknown, that is, if no entry for this destination
already exists in the routing table; otherwise the existing entry is replaced.
Therefore the invariant is maintained.
\item In any initial state the invariant is satisfied, as each store of
  queued data packets starts out empty. In Processes
  \ref{pro:aodv}--\ref{pro:queues} of \Sect{modelling_AODV}
  a store is updated only through the functions \hyperlink{add}{\fnadd} and \hyperlink{drop}{\fndrop}.
  These functions respect the invariant.
\item This is checked by inspecting all assignments to $\dests$ in
  Processes \ref{pro:aodv}--\ref{pro:queues}.
\begin{description}
\item[Pro.~\ref{pro:aodv}, Line~\ref{aodv:line16}:]
  {}\hspace{-2.4pt}The message $\xi(\msg)$ is received in Line~\ref{aodv:line2}, and
  hence, by \Prop{preliminaries}(\ref{it:preliminariesi}), sent by
  some node before. The content of the message does not change during
  transmission, and we assume there is only one way to read a message
  $\xi(\msg)$ as $\rerr{\xi(\dests)}{\xi(\sip)}$. By induction, we may
  assume that when the other node composed the message, a partial
  function was assigned to the first argument $\xi(\dests)$ of $\rerrID$.
\myitem[Pro.~\ref{pro:aodv}, Line~\ref{aodv:line30}; 
	Pro.~\ref{pro:pkt}, Line~\ref{pkt2:line9}; 
	Pro.~\ref{pro:rreq}, Lines~\ref{rreq:line16},~\ref{rreq:line28}; 
	Pro.~\ref{pro:rrep}, Line~\ref{rrep:line16}:]
  The assigned sets have the form
  $\{(\xi(\rip),\inc{\sqn{\xi(\rt)}{\xi(\rip)}})\mid \dots\})$.
  Since $\fninc$ and $\fnsqn$ are functions, for each $\xi(\rip)$ there is only
  one pair $(\xi(\rip),\inc{\sqn{\xi(\rt)}{\xi(\rip)}})$.
\myitem[Pro.~\ref{pro:aodv}, Line~\ref{aodv:line31a}; 
	Pro.~\ref{pro:pkt}, Line~\ref{pkt2:line13}; 
	Pro.~\ref{pro:rreq}, Lines~\ref{rreq:line17a},~\ref{rreq:line29a}; 
	Pro.~\ref{pro:rrep}, Line~\ref{rrep:line17a}; 
	Pro.~\ref{pro:rerr}, Line~\ref{rerr:line3a}:]
In each of these cases a set $\xi(\dests)$ constructed four lines before is used to construct a new set.
By the invariant to be proven, these sets are already partial functions. 
From these sets some values are removed.
Since subsets of partial functions are again partial functions, the claim follows immediately.
\item[Pro.~\ref{pro:rerr}, Line~\ref{rerr:line2}:] Similar to the previous case except that the set $\xi(\dests)$ to be thinned 
out is not constructed before but stems from an incoming RERR message.
\item[Pro.~\ref{pro:pkt}, Lines~\ref{pkt2:line20}:]
The set is explicitly given and consists of only one element; thus the claim is trivial.
\qed
\end{description}
\end{enumerate}
\eprf

\noindent
Property~\eqref{prop:invarianti_itemii}
is stated in the RFC~\cite{rfc3561}.

Next, we show that a function is  used in the specification of AODV only when it is defined,
with $\fnnhop$ and $\fnfD$ as possible exceptions.
In this paper, we only give the proof for $\fnupd$;
for the remaining functions $\fnselroute$, $\fnstatus$, $\fndhops$, $\fnprecs$, $\fnaddprecrt$, $\fnhead$, $\fntail$ and $\fndrop$
the proofs are straightforward, inspecting the locations of
function calls; detailed proofs can be found in \cite[Section 7.4]{TR13}.

\begin{proposition}\rm\label{prop:upd_well_defined}
In our specification of AODV, the\linebreak function \hyperlink{update}{$\fnupd$} is used only when it is defined. 
\end{proposition}
\prfnoboxatend
$\upd{\dval{rt}}{\dval{r}}$ is defined only under the assumptions $\pi_{4}(\dval{r})\mathbin=\val$,
\mbox{$\pi_{2}(\dval{r})\mathbin=0 \Leftrightarrow \pi_{3}(\dval{r})\mathbin=\unkno$} and $\pi_{3}(r)=\unkno\Rightarrow\pi_{5}(r)=1$.
In \Pro{aodv}, Lines~\ref{aodv:line10}, \ref{aodv:line14} and \ref{aodv:line18}, the entry
$\xi(\sip,0,\unkno,\val,1,\sip, \emptyset)$ is used as second argument, which obviously satisfies the 
assumptions. The function is used at four other locations: 
\begin{description}
\item[\Pro{rreq}, Line~\ref{rreq:line6}:]
Here, the entry $\xi(\oip, \osn, \kno, \val, \hops + 1, \sip, \emptyset)$ is
used as $\dval{r}$ to update the routing table. 
This entry fulfils $\pi_{4}(\dval{r})=\val$. 
Since $\pi_{3}(\dval{r})=\kno$, it remains to show that $\pi_{2}(\dval{r})=\xi(\osn)\geq1$.
The sequence number $\xi(\osn)$ stems, through Line~\ref{aodv:line8}
of Pro.~\ref{pro:aodv}, from an incoming RREQ message and is not changed within Pro.~\ref{pro:rreq}.
Hence, by Invariant~\eqref{inv:starcast_sqni}, $\xi(\osn)\geq 1$.
\item[\Pro{rrep}, Lines~\ref{rrep:line3},~\ref{rrep:line5},~\ref{rrep:line25}:]
                  The update is similar to the one of Pro.~\ref{pro:rreq}, Line~\ref{rreq:line6}. 
                  The only difference is that the information stems from an incoming RREP message and 
                  that a routing table entry to $\xi(\dip)$ (instead of  $\xi(\oip)$) is established. 
                  Therefore, the proof is similar to the one of Pro.~\ref{pro:rreq}, Line~\ref{rreq:line6}; instead 
                  of Invariant~\eqref{inv:starcast_sqni} we use Invariant~\eqref{inv:starcast_sqnii}. \qed
\end{description}
\eprf

\noindent
The functions $\fnnhop$ and $\fnfD$ need a closer inspection. 

\begin{proposition}\rm\label{prop:nhop_well_defined}
In our specification of AODV, the\linebreak function \hyperlink{nhop}{$\fnnhop$} is either used within formulas or if it is defined; hence it is only used in a meaningful way. 
\end{proposition}
\prfnoboxatend
The function $\nhop{\dval{rt}}{\dval{dip}}$ is defined iff $\dval{dip}\in\kD{\dval{rt}}$. 
\begin{description}
\myitem[\Pro{aodv}, Line~\ref{aodv:line30};
    \Pro{pkt}, Line~\ref{pkt2:line9}; 
    \Pro{rreq}, Lines~\ref{rreq:line16}, \ref{rreq:line28}; 
    \Pro{rrep}, Line~\ref{rrep:line16}; 
    \Pro{rerr}, Line~\ref{rerr:line2}:] The function is used within a formula.
\item[\Pro{aodv}, Line~\ref{aodv:line24}:] Line~\ref{aodv:line22} states $\xi(\dip)\in\akD{\xi(\rt)}$; hence $\nhop{\xi(\rt)}{\xi(\dip)}$ is defined.
\item[\Pro{pkt}, Line~\ref{pkt2:line7}:] By Line~\ref{pkt2:line5}, $\xi(\dip)\in\akD{\xi(\rt)}$.
\item[\Pro{rreq}, Lines~\ref{rreq:line14a}, \ref{rreq:line26}:] In Line~\ref{rreq:line6} the entry for destination\linebreak $\xi(\oip)$ is updated;
by this $\xi(\oip)\mathbin\in\kD{\xi(\rt)}$.
\item[\Pro{rreq}, Line~\ref{rreq:line25}:] By Line~\ref{rreq:line22} $\xi(\dip)\in\akD{\xi(\rt)}$.
\item[\Pro{rrep}, Lines~\ref{rrep:line12a},  \ref{rrep:line13}:]  By Line~\ref{rrep:line11} $\xi(\oip)\in\akD{\xi(\rt)}$.
\item[\Pro{rrep}, Line~\ref{rrep:line12b}:] In Line~\ref{rrep:line5} the entry for destination $\xi(\dip)$ is updated;
by this $\xi(\dip)\mathbin\in\kD{\xi(\rt)}$. By Line~\ref{rrep:line11} $\xi(\oip)\in\akD{\xi(\rt)}$.
\end{description}
  If $\fnnhop$ is used within a formula, then
  $\nhop{\dval{rt}}{\dval{rip}}$ may not be defined, namely
  if $\dval{rip}\not\in\kD{\dval{rt}}$. In such a case, according to the convention of
  Footnote~\ref{fn:undefvalues} in \Sect{awn},  the atomic formula
  in which this term occurs evaluates to {\tt false}, and thereby is
  defined properly.\qed
\eprf

\noindent
If one chooses to use lazy evaluation for conjunction, then \hyperlink{nhop}{$\fnnhop$} is only used where it is defined.
Lastly, the function $\fnfD$ is called only in Pro.~\ref{pro:aodv} in Line~\ref{aodv:line34}, within a formula. 
Again, if one uses lazy evaluation for conjunction, then \hyperlink{nhop}{$\fnfD$} is used only where it is defined.

\subsection{The Quality of Routing Table Entries}\label{ssec:quality}

In this section we define a total preorder $\rtord$ on routing table
entries for a given destination \dval{dip}. Entries are ordered
by the \emph{quality} of the information they provide. This order will be
defined in such a way that 
(a) the quality of a node's routing table entry for \dval{dip}
will only increase over time, and 
(b)  the quality of valid routing table entries along a route to \dval{dip} strictly increases every hop
(at least prior to reaching \dval{dip}).
This order allows us 
to prove \emph{loop freedom} of AODV in the next section.

A main ingredient in the definition of the quality preorder is the
sequence number of a routing table entry. A higher sequence number
denotes fresher information. However, it generally is not the
case that along a route to \dval{dip} found by AODV the sequence
numbers are only increasing. This is since AODV increases the
sequence number of an entry at an intermediate node when invalidating
it.  To ``compensate'' for that we introduce the concept of a \emph{net
sequence number}. It is defined by a function $\fnnsqn:\tROUTE\to\tSQN$
\[\begin{array}{@{}r@{\hspace{0.5em}}c@{\hspace{0.5em}}l@{}}
\fnnsqn(\dval{r})&:=&\left\{
\begin{array}{ll}
\ifs{\pi_2(\dval{r})}{\pi_4(\dval{r})=\val
                             \vee\pi_2(\dval{r})=0}\\
\ow{\pi_2(\dval{r})-1}
\end{array}\right.
\end{array}\]
For $n\in\NN$ define $n\decremented:=\max(n\mathord-1,0)$; hence $\hyperlink{inc}{\fninc}(n)\mbox{$\decremented$}$ ${}=n$.
Then $\fnnsqn(r)\mathbin=\pi_2(r)\decremented$ if $\pi_4(r)\mathbin=\inval$.

To model increase in quality, we define $\rtord$ by first comparing the net sequence numbers of
two entries---a larger net sequence number denotes fresher
and higher quality information. In case the net sequence numbers are
equal, we decide on their hop counts---the entry with the least hop
count is the best. This yields the following lexicographical order:

Assume two routing table entries $\dval{r},\dval{r}'\in\tROUTE$ with
$\pi_1(\dval{r})$ ${}=\pi_1(\dval{r}')=\dval{dip}$. Then
\[
\begin{array}{rl}
\dval{r} \rtord \dval{r}':\Leftrightarrow&
\fnnsqn(\dval{r}) < \fnnsqn(\dval{r}')\\
&{}\vee (\fnnsqn(\dval{r})=\fnnsqn(\dval{r}')\wedge
   \pi_5(\dval{r}) \geq \pi_5(\dval{r}'))\,.
\end{array}
\]

To reason about AODV, net sequence numbers and the quality preorder are
lifted to routing tables. As for \hyperlink{sqn}{$\fnsqn$} we define a total function 
to determine net sequence numbers.
\hypertarget{nsqn}{
\[\begin{array}{l}
  \fnnsqn : \tRT\times\tIP\to \tSQN\\
  \nsqn{\dval{rt}}{\dval{dip}}{}:={}
    \left\{
      \begin{array}{ll}
        \ifsnewline	{\fnnsqn(\selr{\dval{rt}}{\dval{dip}})}	
        		{\selr{\dval{rt}}{\dval{dip}}\mbox{ is defined}}\\
        \ow{0}
      \end{array}
    \right.\\[2ex]
    \phantom{\nsqn{\dval{rt}}{\dval{dip}}}{}\textcolor{white}{:}={}
    \left\{
      \begin{array}{ll}
        \ifsnewline {\sqn{\dval{rt}}{\dval{dip}}}
        			{\status{\dval{rt}}{\dval{dip}}=\val}\\
        \ow{\sqn{\dval{rt}}{\dval{dip}}\decremented}
      \end{array}
    \right.
  \end{array}\]
}
If two routing tables $\dval{rt}$ and $\dval{rt}'$ have a routing table entry to $\dval{dip}$, i.e., $\dval{dip}\in\kD{\dval{rt}}\cap\kD{\dval{rt}'}$,
 the preorder can be lifted as well.
\[\begin{array}{r@{\hspace{0.5em}}r@{\hspace{0.5em}}l}
\dval{rt} \rtord \dval{rt}' &:\Leftrightarrow&
\selr{\dval{rt}}{\dval{dip}} \rtord \selr{\dval{rt}'}{\dval{dip}}\\
&\Leftrightarrow&\nsqn{\dval{rt}}{\dval{dip}} < \nsqn{\dval{rt}'}{\dval{dip}} \vee\\
&&\big(\nsqn{\dval{rt}}{\dval{dip}} = \nsqn{\dval{rt}'}{\dval{dip}}\\
&&\phantom{\big(} \wedge
\dhops{\dval{rt}}{\dval{dip}} \geq \dhops{\dval{rt}'}{\dval{dip}}\big)
\end{array}
\]
\noindent For all destinations $\dval{dip}\in\IP$, 
the relation $\rtord$ on routing tables with an entry for \dval{dip} is total preorder.
The equivalence relation induced by $\rtord$ is denoted by $\rtequiv$.

As with \fnsqn, we abbreviate $\fnnsqn$:
$
\nsq{\dval{ip}}\ :=$\linebreak $ \nsqn{\xiN{\dval{ip}}(\rt)}{\dval{dip}}.
$
Note that
\begin{equation}\label{eq:sqn_vs_nsqn}
\sq{\dval{ip}}\decremented\leq \nsq{\dval{ip}}\leq \sq{\dval{ip}}\,.
\end{equation}
After setting up this notion of quality, we now show that routing tables, when modified by AODV, 
never decrease their quality.

\begin{proposition}\rm\label{prop:qual}~
Assume a routing table $\dval{rt}\in\tRT$ with $\dval{dip}\in\kD{\dval{rt}}$.
\begin{enumerate}[(a)]
\item\label{it:qual_upd}
An \hyperlink{update}{\fnupd} of $\dval{rt}$ can only increase the quality of the routing table.
That is, for all routes \dval{r} such that \upd{\dval{rt}}{\dval{r}} is defined
($\pi_{4}(\dval{r})\mathbin=\val$,
 $\pi_{2}(\dval{r})\mathbin=0$ $\Leftrightarrow$ $\pi_{3}(\dval{r})\mathbin=\unkno$ and
 $\pi_{3}(\dval{r})\mathbin=\unkno\Rightarrow\pi_{5}(\dval{r})\mathbin=1$)
we have
  \begin{equation}\label{eq:qual_upd}
    \dval{rt}\rtord\upd{\dval{rt}}{\dval{r}}\,.
\vspace{-2pt}
  \end{equation}
\item\label{it:qual_inv}
An \hyperlink{invalidate}{$\fninv$} on $\dval{rt}$ does not change the
quality of the routing table if, for each $(\dval{rip},\dval{rsn})\in\dval{dests}$,
\dval{rt} has a valid entry for \dval{rip}, and
  \begin{itemize}
  \item \dval{rsn} is the by one incremented sequence number from the routing table, or 
  \item both \dval{rsn} and the sequence number in the routing table are $0$.
  \end{itemize}
That is, for all partial functions $\dval{dests}$ (subsets of $\tIP\times\tSQN$)
\vspace{-2pt}
  \begin{equation}\label{eq:qual_inv}
  \begin{array}{cl}
  &\big((\dval{rip},\dval{rsn})\in\dval{dests} \Rightarrow \dval{rip}\in\akD{\dval{rt}}\\
  &\phantom{\big(}{}\wedge
  \dval{rsn}=\inc{\sqn{\dval{rt}}{\dval{rip}}}\big) \\
  \Rightarrow&
    \dval{rt}\rtequiv\inv{\dval{rt}}{\dval{dests}}\,.
  \end{array}
\vspace{-2pt}
  \end{equation}
\item\label{it:qual_addpre}
If precursors are added to an entry of $\dval{rt}$, the quality of the routing table
does not change.
 That is, for all $\dval{dip}\in\IP$ and sets of precursors $\dval{npre}\in\pow(\IP)$,
\vspace{-2pt}
  \begin{equation}\label{eq:qual_addpre}
    \dval{rt}\rtequiv\addprecrt{\dval{rt}}{\dval{dip}}{\dval{npre}}\,.
\vspace{-2pt}
  \end{equation}
\end{enumerate}
\end{proposition}

\prfnoboxatend
For the proof we denote the routing table after the update by $\dval{rt}'$.
  \begin{enumerate}[(a)]
 \item By assumption, there is an entry $(\dval{dip},\dval{dsn}_{\dval{rt}},*,\dval{f}_{\dval{rt}},\dval{hops}_{\dval{rt}},$ $*,*)$ 
 for $\dval{dip}$ in $\dval{rt}$. In case $\pi_{1}(r) \not=\dval{dip}$ the quality of the routing table w.r.t.\ \dval{dip}
stays the same, since the entry for \dval{dip} is not changed.
 
We first assume that $\dval{r} := (\dval{dip},0,\unkno,\val,1,*,*)$. This 
means that the Clause 5 in the definition of \hyperlink{update}{\fnupd} 
is used. The updated routing table entry to $\dval{dip}$ has the form $(\dval{dip},\dval{dsn}_{\dval{rt}},\unkno,\val,1,*,*)$.
So
 \begin{eqnarray*}
 &\nsqn{\dval{rt}}{\dval{dip}} \leq \sqn{\dval{rt}}{\dval{dip}}=\dval{dsn}_{\dval{rt}} =  \nsqn{\dval{rt}'}{\dval{dip}}\,,\\
  \mbox{and} &
 \dhops{\dval{rt}}{\dval{dip}} = \dval{hops}_{\dval{rt}} \geq 1 =  \dhops{\dval{rt}'}{\dval{dip}}\,.
 \end{eqnarray*}
The first inequality holds by~\eqref{eq:sqn_vs_nsqn}; the penultimate step by Invariant~\eqref{eq:inv_length}.

Next, we assume that the sequence number is known and therefore the 
route used for the update has the form $\dval{r} = (\dval{dip},\dval{dsn},\kno,\val,\dval{hops},*,*)$ with $\dval{dsn}\geq1$.
After the performed update the routing entry for $\dval{dip}$ either has the form 
$(\dval{dip},\dval{dsn}_{\dval{rt}},*,\dval{f}_{\dval{rt}},\dval{hops}_{\dval{rt}},*,*)$ or 
$(\dval{dip},\dval{dsn},\kno,\val,$\linebreak $\dval{hops},*,*)$.
In the former case the invariant is trivially preserved;
in the latter, we know, by definition of \fnupd, that either
(i) $\dval{dsn}_{\dval{rt}}<\dval{dsn}$, 
(ii) $\dval{dsn}_{\dval{rt}}=\dval{dsn} \wedge \dval{hops}_{\dval{rt}}>\dval{hops}$, or
(iii) $\dval{dsn}_{\dval{rt}}=\dval{dsn} \wedge \dval{f}_{\dval{rt}}=\inval$ holds. 
We complete the proof of the invariant by a case distinction.
\begin{description}\hfuzz 50pt
\item[(i) holds:]
First, $\nsqn{\dval{rt}}{\dval{dip}}\leq\dval{dsn}_{\dval{rt}}<\dval{dsn}=\sqn{\dval{rt}'}{\dval{dip}}=\nsqn{\dval{rt}'}{\dval{dip}}$. Since $\dval{dsn}_{\dval{rt}}$ is strictly smaller than $\nsqn{\dval{rt}'}{\dval{dip}}$, there is nothing more to prove.
\item[(iii) holds:]
We have $\nsqn{\dval{rt}}{\dval{dip}}=\dval{dsn}_{\dval{rt}}\decremented<\dval{dsn}=\sqn{\dval{rt}'}{\dval{dip}}=\nsqn{\dval{rt}'}{\dval{dip}}$.
The inequality holds since either $\dval{dsn}_{\dval{rt}}\decremented=0<1\leq\dval{dsn}$ or 
$\dval{dsn}_{\dval{rt}}\decremented=\dval{dsn}_{\dval{rt}}-1<\dval{dsn}_{\dval{rt}}=\dval{dsn}$.
\item[(ii) holds but (iii) does not:] Then $\dval{f}_{\dval{rt}}=\val$.
In this case the update does not change the net sequence number for $\dval{dip}$:
$\nsqn{\dval{rt}}{\dval{dip}}=\dval{dsn}_{\dval{rt}}=\dval{dsn}=\nsqn{\dval{rt}'}{\dval{dip}}$.\linebreak 
By (ii), the hop count decreases:\\
\mbox{}\quad $\dhops{\dval{rt}}{\dval{dip}} = \dval{hops}_{\dval{rt}}>\dval{hops} = \dhops{\dval{rt}'}{\dval{dip}}$\,.
\end{description}
\item
Assume that \hyperlink{invalidate}{$\fninv$} modifies an
entry of the form $(\dval{rip},\dval{dsn},*,\dval{flag},*,*,*)$.
Let $(\dval{rip},\dval{rsn})\mathbin\in\dval{dests}$; then $\dval{flag}\mathbin=\val$ and
the update results in the
entry $(\dval{rip},\inc{\dval{dsn}},*,$ $\inval,*,*,*)$.
By definition of net sequence numbers,\vspace{-2pt}
\[
\begin{array}{rl}
\nsqn{\dval{rt}}{\dval{rip}} =& \sqn{\dval{rt}}{\dval{rip}} =
  \dval{dsn}
  =
  \inc{\dval{dsn}}\decremented \\
   =& \nsqn{\dval{rt}'}{\dval{rip}}\,.
\end{array}
\]
Since the hop count is not changed by $\fninv$, we also have
$\dhops{\dval{rt}}{\dval{rip}} =\dhops{\dval{rt}'}{\dval{rip}}$,
and hence $\dval{rt}\rtequiv\inv{\dval{rt}}{\dval{dests}}$.
 \item The function \hyperlink{addprert}{$\fnaddprecrt$} only modifies a set of precursors; 
 it does not change the sequence number, the validity, the flag, nor the hop count 
 of any entry of the routing table $\dval{rt}$.
\qed
\end{enumerate}
\eprf

\noindent
We can apply this result to obtain the following theorem.

\begin{theorem}\rm\label{thm:state_quality}
In AODV, the quality of routing tables can only be increased, never decreased.
Assume $N \mathbin{\ar{\ell}}N'$ and $\dval{ip},\dval{dip}\mathbin\in\IP$.
If $\dval{dip}\in\kd{\dval{ip}}$, then $\dval{dip}\in\kd[N']{\dval{ip}}$ and
$\xiN{\dval{ip}}(\rt)\rtord\xiN[N']{\dval{ip}}(\rt)$.
\end{theorem}
\prfnobox
If $\dval{dip}\in\kd{\dval{ip}}$, then
$\dval{dip}\in\kd[N']{\dval{ip}}$ follows by \Prop{destinations maintained}.
To show $\xiN{\dval{ip}}(\rt)\rtord\xiN[N']{\dval{ip}}(\rt)$,
by Remark~\ref{rem:remark} and \Prop{qual}(\ref{it:qual_upd}) and~(\ref{it:qual_addpre})
it suffices to check all calls of \hyperlink{invalidate}{$\fninv$}.
\begin{description}
\myitem[Pro.~\ref{pro:aodv}, Line~\ref{aodv:line32}; 
	\Pro{pkt}, Line~\ref{pkt2:line10}; 
	Pro.~\ref{pro:rreq}, Lines~\ref{rreq:line18}, \ref{rreq:line30}; 
	Pro.~\ref{pro:rrep}, Line~\ref{rrep:line18}:] 
	By construction of {\dests} (immediately before the invalidation call) 
	\[\begin{array}{ll}
		&(\dval{rip},\dval{rsn})\in\xiN{\dval{ip}}(\dests)\\
		\Rightarrow&  
		\dval{rip}\in\akD{\xiN{\dval{ip}}(\rt)} \wedge \dval{rsn}=\inc{\sqn{\xiN{\dval{ip}}(\rt)}{\dval{rip}}}
	\end{array}\]
and hence, by \Prop{qual}\eqref{it:qual_inv},
$\xiN{\dval{ip}}(\rt)\rtequiv{}$\linebreak $\inv{\xiN{\dval{ip}}(\rt)}{\xiN{\dval{ip}}(\dests)} =\xiN[N']{\dval{ip}}(\rt)$.
\item[Pro.~\ref{pro:rerr}, Line~\ref{rerr:line5}:]
Assume that \hyperlink{invalidate}{$\fninv$} modifies an
entry having the form $(\dval{rip},\dval{dsn},*,\dval{flag},*,*,*)$.
Let $(\dval{rip},\dval{rsn})\mathbin\in\dval{dests}$; then
the update results in the entry $(\dval{rip},\dval{rsn},*,\inval,*,*,*)$.
Moreover, by Line~\ref{rerr:line2} of Pro.~\ref{pro:rerr}, $\dval{flag}\mathbin=\val$.
By definition of net sequence numbers,
\[\begin{array}{rcl}
\nsqn{\xiN{\dval{ip}}(\rt)}{\dval{rip}} &=& \sqn{\xiN{\dval{ip}}(\rt)}{\dval{rip}}\\
  &\leq& \dval{rsn}\decremented 
  = \nsqn{\xiN[N']{\dval{ip}}(\rt)}{\dval{rip}}\,.
  \end{array}
\]
The second step holds, since
$\sqn{\xiN[N_{\ref*{rerr:line2}}]{\dval{ip}}(\rt)}{\dval{rip}} <\dval{rsn}$, using Line~\ref{rerr:line2}.
Since the hop count is not changed by $\fninv$, we also have
$\dhops{\xiN{\dval{ip}}(\rt)}{\dval{rip}} =\dhops{\xiN[N']{\dval{ip}}(\rt)}{\dval{rip}}$,
and hence $\xiN{\dval{ip}}(\rt)\rtord\xiN[N']{\dval{ip}}(\rt)$.
\qed
\end{description}
\eprf
\Thm{state_quality} states in particular that if $N \ar{\ell}N'$ then
$\nsq{\dval{ip}} \leq \fnnsqn_{N'}^{\dval{ip}}(\dval{dip})$.

\begin{proposition}\rm\label{prop:inv_nsqn}
If, in a reachable network expression $N$, a node $\dval{ip}\mathop\in\IP$ has a
routing table entry to $\dval{dip}$, then also the next hop
\dval{nhip} towards \dval{dip}, if not \dval{dip} itself, has a
routing table entry to $\dval{dip}$, and the net sequence number of
the latter entry is at least as large as that of the former.
\begin{equation}\label{eq:inv_ix}
\begin{array}{ll}
&\dval{dip}\in\kd{\dval{ip}}\wedge\dval{nhip}\not=\dval{dip}\\
\Rightarrow&	\dval{dip}\in\kd{\dval{nhip}} \wedge \nsq{\dval{ip}}\leq \nsq{\dval{nhip}}\,,
\end{array}
\end{equation}
where $\dval{nhip}:=\nhp{\dval{ip}}$ is the IP address of the next hop.
\end{proposition}

\prfnoboxatend As before, we first check the initial states
  of our transition system and then check all locations in
  Processes~\ref{pro:aodv}--\ref{pro:queues} where a routing table might
  be changed. For an initial network expression, the invariant holds
  since all routing tables are empty.
	
A modification of \plat{$\xiN{\dval{nhip}}(\rt)$} is harmless, as it
can only increase $\kd{\dval{nhip}}$ (cf.\ \Prop{destinations maintained})
as well as $\nsq{\dval{nhip}}$ (cf.\ \Thm{state_quality}).

Adding precursors to $\xiN{\dval{ip}}(\rt)$ does not harm since the 
invariant does not depend on precursors.
It remains to examine all calls of \hyperlink{update}{$\fnupd$} and
\hyperlink{invalidate}{$\fninv$} to \plat{$\xiN{\dval{ip}}(\rt)$}.
Without loss of generality we restrict attention to those applications of $\fnupd$
or $\fninv$ that actually modify the entry for \dval{dip}, beyond its
precursors; if $\fnupd$ only adds some precursors in the routing
table, the invariant---which is assumed to hold before---is maintained.
If $\fninv$ occurs,
the next hop \dval{nhip} is not changed.
Since the invariant has to hold before the execution, it follows that 
$\dval{dip}\in\kd{\dval{nhip}}$ also holds
after execution.

\begin{description}
\item[Pro.~\ref{pro:aodv}, Lines~\ref{aodv:line10}, \ref{aodv:line14}, \ref{aodv:line18}:]
	The entry $\xi(\sip\comma0\comma\unkno\comma\val\comma1\comma\sip\comma\emptyset)$ is used for the update; 
	its destination is $\dval{dip}:=\xi(\sip)$.
	Since $\dval{dip}=\xi(\sip)=\nhp[\xi(\sip)]{\dval{ip}}=\nhp[\dval{dip}]{\dval{ip}}=\dval{nhip}$, the antecedent of the invariant to be proven is not satisfied.
\myitem[Pro.~\ref{pro:aodv}, Line~\ref{aodv:line32}; 
	Pro.~\ref{pro:pkt}, Line~\ref{pkt2:line10}; 
	Pro.~\ref{pro:rreq}, Lines~\ref{rreq:line18}, \ref{rreq:line30}; 
	Pro.~\ref{pro:rrep}, Line~\ref{rrep:line18}:]
        In each of these cases, the precondition of  \eqref{eq:qual_inv} is satisfied by 
        the executions of the line immediately before the call of {\fninv} 
        (Pro.~\ref{pro:aodv}, Line~\ref{aodv:line30},
        Pro.~\ref{pro:pkt}, Line~\ref{pkt2:line9}; Pro.~\ref{pro:rreq}, Lines~\ref{rreq:line16}, \ref{rreq:line28}; Pro.~\ref{pro:rrep}, Line~\ref{rrep:line16}).
        Thus, the quality of the routing table w.r.t.\ \dval{dip}, and thereby the net
        sequence number of the routing table entry for \dval{dip},
	remains unchanged. Therefore the invariant is maintained.
\item[Pro.~\ref{pro:rreq}, Line~\ref{rreq:line6}:] 
        Let us assume that the routing table entry $\xi(\oip,\osn,\kno,\val,\hops+1,\sip,*)$ is inserted into $\xi(\rt)$.
	So $\dval{dip}:=\xi(\oip)$, $\dval{nhip}:=\xi(\sip)$,
        $\nsq{\dval{ip}}:=\xi(\osn)$ and $\dhp{\dval{ip}}:=\xi(\hops)+1$.\linebreak[2]
        This information is distilled from a received
	route request message (cf.\ Lines~\ref{aodv:line2} and~\ref{aodv:line8}
	of Pro.~\ref{pro:aodv}).
	By \Prop{preliminaries} this message was sent before, say in state $N^\dagger$;
        by \Prop{ip=ipc} the sender of this message is $\xi(\sip)$.

	By Invariant~\eqref{inv:starcast_ii}, with $\ipc:=\xi(\sip)=\dval{nhip}$,
	~$\oipc:=\xi(\oip)=\dval{dip}$, ~$\osnc:=\xi(\osn)$~ and ~$\hopsc:=\xi(\hops)$,
        and using that $\ipc = \dval{nhip} \neq \dval{dip} = \oipc$, we get that
	$\dval{dip}\in \kd[N^\dagger]{\dval{nhip}}$ and
\begin{eqnarray*}
&\fnsqn_{N^\dagger}^{\dval{nhip}}(\dval{dip})~=~\fnsqn_{N^\dagger}^{\ipc}(\oipc) ~>~ \osnc ~=~ \xi(\osn)\,, \mbox{ or}\\
&\fnsqn_{N^\dagger}^{\dval{nhip}}(\dval{dip})~=~\xi(\osn) \wedge
 \fnstatus_{N^\dagger}^{\dval{nhip}}(\dval{dip})~=~\val\,.
\end{eqnarray*}
We first assume that the first line holds.
Then, by \Thm{state_quality} and \Eq{sqn_vs_nsqn},
\begin{eqnarray*}
\nsq{\dval{nhip}}
&\geq&
\fnnsqn_{N^\dagger}^{\dval{nhip}}(\dval{dip})
\geq
\fnsqn_{N^\dagger}^{\dval{nhip}}(\dval{dip})\decremented\\
&\geq& \xi(\osn)=\nsq{\dval{ip}}\,.
\end{eqnarray*}

We now assume the second line to be valid. 
From this we conclude
\begin{eqnarray*}
\nsq{\dval{nhip}}&\geq&
\fnnsqn_{N^\dagger}^{\dval{nhip}}(\dval{dip})
=\fnsqn_{N^\dagger}^{\dval{nhip}}(\dval{dip})\\
&=&\xi(\osn)=\nsq{\dval{ip}}\,.
\end{eqnarray*}

\item[Pro.~\ref{pro:rrep}, Line~\ref{rrep:line5}:]
                  The update is similar to the one of Pro.~\ref{pro:rreq}, Line~\ref{rreq:line6}. 
                  The only difference is that the information stems from an incoming RREP message and 
                  that a routing table entry to $\xi(\dip)$ (instead of  $\xi(\oip)$) is established. 
                  Therefore, the proof is similar to the one of Pro.~\ref{pro:rreq}, Line~\ref{rreq:line6}; instead 
                  of Invariant~\eqref{inv:starcast_ii} we use Invariant~\eqref{inv:starcast_iv}.
\item[Pro.~\ref{pro:rerr}, Line~\ref{rerr:line5}:]
        Let $N_{\ref*{rerr:line5}}$ and $N$ be the network expressions right before
        and right after executing Pro.~\ref{pro:rerr}, Line~\ref{rerr:line5}.
	The entry for destination $\dval{dip}$ can be affected 
	only if $(\dval{dip},\dval{dsn})$ ${}\in\xiN[N_{\ref*{rerr:line2}}]{\dval{ip}}(\dests)$ for some $\dval{dsn}\in\tSQN$.
	In that case, by Line~\ref{rerr:line2},
        \plat{$(\dval{dip},\dval{dsn})\in\xiN[N_{\ref*{rerr:line2}}]{\dval{ip}}(\dests)$},
	\plat{$\dval{dip}\in\akd[N_{\ref*{rerr:line2}}]{\dval{ip}}$}, and
        \plat{$\fnnhop_{N_{\ref*{rerr:line2}}}^{\dval{ip}}(\dval{dip})={}$}\linebreak \plat{$\xiN[N_{\ref*{rerr:line2}}]{\dval{ip}}(\sip)$}.
By definition of \hyperlink{invalidate}{\fninv},
$\sq{\dval{ip}} = \dval{dsn}$
and $\sta{\dval{ip}}=\inval$, so
$$\nsq{\dval{ip}}=\sq{\dval{ip}}\decremented=\dval{dsn}\decremented\,.$$
Hence we need to show that $\dval{dsn}\decremented \leq \nsq{\dval{nhip}}$.

        The values $\xiN[N_{\ref*{rerr:line2}}]{\dval{ip}}(\dests)$ and
        $\xiN[N_{\ref*{rerr:line2}}]{\dval{ip}}(\sip)$ stem from a received route 
	error message (cf.\ Lines~\ref{aodv:line2} and~\ref{aodv:line16} of Pro.~\ref{pro:aodv}).
	By \Prop{preliminaries}\eqref{it:preliminariesi}, a transition
	labelled 
	\begin{equation*}\colonact{R}{\starcastP{\rerr{\destsc}{\ipc}}}
	\end{equation*} with \plat{$\destsc:=\xiN[N_{\ref*{rerr:line2}}]{\dval{ip}}(\dests)$}
	and \plat{$\ipc:=\xiN[N_{\ref*{rerr:line2}}]{\dval{ip}}(\sip)$} must have occurred before, say in state $N^\dagger$.  
	By \Prop{ip=ipc}, the node casting this message is
	\plat{$\ipc\mathop=
	\xiN[N_{\ref*{rerr:line2}}]{\dval{ip}}(\sip)\mathop=
	\fnnhop_{N_{\ref*{rerr:line2}}}^{\dval{ip}}(\dval{dip})$}
	\plat{${}=
	\nhp{\dval{ip}}=\dval{nhip}$}.
	The penultimate equation holds since the next hop to $\dval{dip}$
	is not changed during the execution of Pro.~\ref{pro:rerr}.
By \Prop{starcastrerr} we have 
\plat{$\dval{dip}\in\fnikD_{N^\dagger}^{\dval{nhip}}$} and 
\plat{$\dval{dsn} \leq \sqn{\xiN[N^\dagger]{\dval{nhip}}(\rt)}{\dval{dip}}$}. 
Hence
\begin{eqnarray*}
\nsq{\dval{nhip}}&\geq&
\fnnsqn_{N^\dagger}^{\dval{nhip}}(\dval{dip}) =
\nsqn{\xiN[N^\dagger]{\dval{nhip}}(\rt)}{\dval{dip}}\\ &=&
\sqn{\xiN[N^\dagger]{\dval{nhip}}(\rt)}{\dval{dip}}\decremented \geq
\dval{dsn}\decremented\,,
\end{eqnarray*} where the first inequality follows by \Thm{state_quality}.
\qed
\end{description}
\eprf

\noindent
To prove loop freedom we will show that on any route established by AODV the quality 
of routing tables increases when going from one node to the next hop. Here, the 
preorder is not sufficient, since we need a strict increase in quality. 
Therefore, on routing tables $\dval{rt}$ and $\dval{rt}'$ that both have an entry to $\dval{dip}$, i.e., $\dval{dip}\in\kD{\dval{rt}}\cap\kD{\dval{rt}'}$,
 we define a relation $\rtsord$ by
\[
\dval{rt} \rtsord \dval{rt}'\ :\Leftrightarrow\  \dval{rt} \rtord \dval{rt}' \wedge \dval{rt} \not\rtequiv \dval{rt}'\,.
\]

\begin{corollary}\label{cor:strictord}
The relation $\rtsord$ is irreflexive and\linebreak transitive. 
\end{corollary}

\begin{theorem}\rm
\label{thm:inv_a}
The quality of the routing table entries for a destination \dval{dip} is strictly increasing
along a route towards \dval{dip},
 until it reaches either \dval{dip} or a node with an invalid routing table entry to \dval{dip}.
\begin{equation}\label{eq:inv_x}
\begin{array}{ll}
&\dval{dip}\in\akd{\dval{ip}}\cap \akd{\dval{nhip}} \wedge\dval{nhip}\not=\dval{dip}\\
\Rightarrow& \xiN{\dval{ip}}(\rt)\rtsord \xiN{\dval{nhip}}(\rt)\,,
\end{array}
\end{equation}
where $N$ is a reachable network expression and $\dval{nhip}:=\nhp{\dval{ip}}$ is the IP address of the next hop.
\end{theorem}

\prfnobox
As before, we first check the initial states of our transition system
and then check all locations in Processes~\ref{pro:aodv}--\ref{pro:queues}
where a routing table might be changed. For an initial network
expression, the invariant holds since all routing tables are empty.
Adding precursors to $\xiN{\dval{ip}}(\rt)$ or $\xiN{\dval{nhip}}(\rt)$
does not affect the invariant, since the invariant does not depend on
precursors, so it suffices to examine all modifications to $\xiN{\dval{ip}}(\rt)$
or $\xiN{\dval{nhip}}(\rt)$ using \hyperlink{update}{$\fnupd$} or
\hyperlink{invalidate}{$\fninv$}. Moreover, without loss of generality we restrict
attention to those applications of $\fnupd$ or $\fninv$ that actually
modify the entry for \dval{dip}, beyond its precursors; if $\fnupd$
only adds some precursors in the routing table, the invariant---which
is assumed to hold before---is maintained. 

Applications of {\fninv} to $\xiN{\dval{ip}}(\rt)$ or
$\xiN{\dval{nhip}}(\rt)$ lead to a network state in which the
antecedent of \Eq{inv_x} is not satisfied.
Now consider an application of $\fnupd$ to \plat{$\xiN{\dval{nhip}}(\rt)$.}
We restrict attention to the case that the antecedent of \Eq{inv_x} is satisfied right after the
update, so that right before the update we have 
$\dval{dip}\in\akd{\dval{ip}} \wedge \dval{nhip}\not=\dval{dip}$.
In the special case that $\sq{\dval{nhip}}\mathbin=0$ right before the update, we have
$\nsq{\dval{nhip}}\mathbin=0$ and thus $\nsq{\dval{ip}}=0$ by Invariant~\Eq{inv_ix}.
Considering that \plat{$\sta{\dval{ip}}=\val$}, this implies $\sq{\dval{ip}}=0$.
By \Prop{rte}(\ref{it_e}) we have $\dval{nhip}=\dval{dip}$,
contradicting our assumptions. It follows that right before the update $\sq{\dval{nhip}}>0$;
so in particular $\dval{dip}\in\kd{\dval{nhip}}$.

An application of $\fnupd$ to \plat{$\xiN{\dval{nhip}}(\rt)$}
that changes \plat{$\sta{\dval{nhip}}$} from {\inval} to {\val} cannot
decrease the sequence number of the entry to \dval{dip} and hence
strictly increases its net sequence number.
Before the $\fnupd$ we had
$\nsq{\dval{ip}}\leq \nsq{\dval{nhip}}$ by Invariant~\eqref{eq:inv_ix},
so afterwards we must have
\mbox{$\nsq{\dval{ip}} < \nsq{\dval{nhip}}$}, and therefore
$\xiN{\dval{ip}}(\rt)\rtsord \xiN{\dval{nhip}}(\rt)$.
An $\fnupd$ to $\xiN{\dval{nhip}}(\rt)$ that maintains
$\sta{\dval{nhip}}=\val$ can only increase the quality of the entry to
\dval{dip} (cf.\ \Thm{state_quality}), and hence maintains Invariant \Eq{inv_x}.

It remains to examine the $\fnupd$s to $\xiN{\dval{ip}}(\rt)$.
\begin{description}
\item[Pro.~\ref{pro:aodv}, Lines~\ref{aodv:line10}, \ref{aodv:line14}, \ref{aodv:line18}:]
	The entry $\xi(\sip\comma0\comma\unkno\comma\val\comma1\comma\sip\comma\emptyset)$ is used for the update; 
	its destination is $\dval{dip}:=\xi(\sip)$.
	Since $\dval{dip}=\nhp[\dval{dip}]{\dval{ip}}=\dval{nhip}$,
	the antecedent of the invariant to be proven is not satisfied.
\item[Pro.~\ref{pro:rreq}, Line~\ref{rreq:line6}:]\hypertarget{729Pro3Line4}{ }\label{pg:729Pro3Line4}
        We assume that  the entry $\xi(\oip,\osn,$ $\kno,\val,\hops+1,\sip,*)$ is inserted into $\xi(\rt)$.
	So $\dval{dip}\mathop{:=}\xi(\oip)$, $\dval{nhip}\mathop{:=}\xi(\sip)$,
        $\nsq{\dval{ip}}\mathop{:=}\xi(\osn)$ and $\dhp{\dval{ip}}:=\xi(\hops)+1$.\linebreak[2]
        This information is distilled from a received
	route request message (cf.\ Lines~\ref{aodv:line2} and~\ref{aodv:line8}
	of Pro.~\ref{pro:aodv}).
	By \Prop{preliminaries} this message was sent before, say in state $N^\dagger$;
        by \Prop{ip=ipc} the sender of this message is $\xi(\sip)$.

	By Invariant~\eqref{inv:starcast_ii}, with $\ipc\mathbin{:=}\xi(\sip)\mathbin{=}\dval{nhip}$,
	~$\oipc\mathbin{:=}\xi(\oip)=\dval{dip}$, ~$\osnc\mathbin{:=}\xi(\osn)$~ and ~$\hopsc\mathbin{:=}\xi(\hops)$,
        and using that $\ipc \mathbin{=} \dval{nhip} \neq \dval{dip} \mathbin{=} \oipc$, we get that
\begin{eqnarray*}
\fnsqn_{N^\dagger}^{\dval{nhip}}(\dval{dip})&=&\fnsqn_{N^\dagger}^{\ipc}(\oipc) > \osnc ~=~ \xi(\osn)\,, \mbox{ or}
\\
\fnsqn_{N^\dagger}^{\dval{nhip}}(\dval{dip})&=& \xi(\osn) \wedge
\fndhops_{N^\dagger}^{\dval{nhip}}(\dval{dip})\leq\xi(\hops)\\
&&\phantom{\xi(\osn)}\wedge\fnstatus_{N^\dagger}^{\dval{nhip}}(\dval{dip})~=~\val\,.
\end{eqnarray*}
We first assume that the first line holds.
Then, by the assumption $\dval{dip}\in\akD{\xiN{\dval{nhip}}(\rt)}$,
the definition of net sequence numbers, and \Prop{dsn increase},
\begin{eqnarray*}
\nsq{\dval{nhip}}&=&\sq{\dval{nhip}} \geq
\fnsqn_{N^\dagger}^{\dval{nhip}}(\dval{dip})\\
&>&\xi(\osn)=\nsq{\dval{ip}}\,.
\end{eqnarray*}
and hence $\xiN{\dval{ip}}(\rt)\rtsord \xiN{\dval{nhip}}(\rt)$.

We now assume the second line to be valid. 
From this we conclude
\begin{eqnarray*} \fnnsqn_{N^\dagger}^{\dval{nhip}}(\dval{dip})
  &=&\fnsqn_{N^\dagger}^{\dval{nhip}}(\dval{dip})
  =\xi(\osn)
  \\&=&\nsq{\dval{ip}}\,.
\end{eqnarray*}

Moreover,\hfill $\fndhops_{N^\dagger}^{\dval{nhip}}(\dval{dip})\leq\xi(\hops) < \xi(\hops)+1=\dhp{\dval{ip}}$.
Hence $\xiN{\dval{ip}}(\rt)\rtsord \xiN[N^\dagger]{\dval{nhip}}(\rt)$.
Together with \Thm{state_quality} and the transitivity of $\rtord$\vspace{-2pt}
this yields $\xiN{\dval{ip}}(\rt)\rtsord \xiN{\dval{nhip}}(\rt)$.
	
\item[Pro.~\ref{pro:rrep}, Line~\ref{rrep:line5}:]
                  This update is similar to the one of Pro.~\ref{pro:rreq}, Line~\ref{rreq:line6}. 
                  The only difference is that the information stems from an incoming RREP message and 
                  that a routing table entry to $\xi(\dip)$ (instead of  $\xi(\oip)$) is established. 
                  Therefore, the proof is similar to the one of Pro.~\ref{pro:rreq}, Line~\ref{rreq:line6}; instead 
                  of Invariant~\eqref{inv:starcast_ii} we use Invariant~\eqref{inv:starcast_iv}.
\qed
\end{description}
\eprf

\subsection{Loop Freedom}\label{ssec:loop freedom}
The ``na\"ive'' notion of loop freedom is a term that informally means
that ``a packet never goes round in cycles without (at some point)
being delivered".  This dynamic definition is not only hard to formalise, 
it is also too restrictive a requirement for AODV\@. There are situations where 
packets are sent in cycles, but which are not considered harmful. 
This can happen when the topology keeps changing. 
We refer to~\cite[Sect. 7.6]{TR13} for an \mbox{example}.

Due to this dynamic behaviour, the sense of loop freedom is much
better captured by a static invariant,
saying that at any given
time the collective routing tables of the nodes do not admit a loop.
Such a requirement does not rule out the dynamic loop alluded to
above. However, in situations where the topology remains stable sufficiently long
it does guarantee that packets will not keep going around in cycles.

\newcommand{\RG}[2]{\mathcal{R}_{#1}(#2)}
To this end we define the \emph{routing graph} of a network expression $N$ with respect to
destination~$\dval{dip}$ by $\RG{N}{\dval{dip}}$ $\mathop{:=}(\IP,E)$, where
all nodes of the network form the set of vertices and there is an
arc $({\dval{ip}},{\dval{ip}}')\in E$ iff $\dval{ip}\mathop{\not=}\dval{dip}$ and
$
(\dval{dip}\comma\nosp{*}\comma\nosp{*}\comma\nosp{\val}\comma\nosp{*}\comma\nosp{\dval{ip}'}\comma\nosp{*})\mathop{\in}\xiN{\dval{ip}}(\rt).
$

An arc in a routing graph states that $\dval{ip}'$ is the next hop on
a valid route to $\dval{dip}$ known by $\dval{ip}$; a path in a routing
graph describes a route towards $\dval{dip}$ discovered by AODV\@.
We say that a network expression $N$ is \emph{loop free} if the
corresponding routing graphs $\RG{N}{\dval{dip}}$ are loop free, for
all $\dval{dip}\mathop{\in}\IP$. A routing protocol, such as AODV, is
\emph{loop free} iff all reachable network expressions are loop free.

Using this definition of a routing graph, \Thm{inv_a} states that 
along a path towards a destination \dval{dip} in the routing
graph of a reachable network expression $N$, until it reaches either
\dval{dip} or a node with an invalided routing table entry to dip,
the quality of the routing table entries for \dval{dip} is strictly increasing.
From this, we can immediately conclude
\begin{theorem}\rm\label{thm:loop free}
The specification of AODV given in \Sect{modelling_AODV} is loop free.
\end{theorem}
\prf
If there were a loop in a routing graph $\RG{N}{\dval{dip}}$, then for
any edge $(\dval{ip},\dval{nhip})$ on that loop one has, by \Thm{inv_a},
$
\xiN{\dval{ip}}(\rt)\rtsord\xiN{\dval{nhip}}(\rt).$
Thus, by transitivity of $\rtsord$, one has
$\xiN{\dval{ip}}(\rt)\rtsord\xiN{\dval{ip}}(\rt)$, which
contradicts the irreflexivity of {\rtsord} (cf.\ \Cor{strictord}).
\eprf

\hypertarget{end}{According to \Thm{loop free} any route to a destination \dval{dip}
established by AODV---i.e.\ a path in $\RG{N}{\dval{dip}}$---ends after finitely many
hops. There are three possible ways in which it could end:
\begin{enumerate}[(1)]
\item by reaching the destination,\label{eq:success}
\item by reaching a node with an invalid entry to \dval{dip}, or\label{eq:invalid}
\item by reaching a node without any entry to \dval{dip}.\label{eq:failure}
\end{enumerate}
\Eq{success} is what AODV attempts to accomplish, whereas
\Eq{invalid} is an unavoidable
due to link breaks in a dynamic topology. It follows directly from \Prop{inv_nsqn} that \Eq{failure} can never occur.%
}

\subsection{Route Correctness}\label{ssec:route correctness}

\newcommand{\CG}[1]{\mathcal{C}_{#1}}
The creation of a routing table entry at node \dval{ip} for destination \dval{dip} is no guarantee
that a route from \dval{ip} to \dval{dip} actually exists. The entry is created based on information
gathered from messages received in the past, and at any time link breaks may occur. The best one could require
of a protocol like AODV is that routing table entries are based on information that was valid at some
point in the past. This is the essence of what we call \emph{route correctness}.

We define a \emph{history} of an AODV-like protocol as a sequence $H=N_0 N_1 \ldots N_k$ of network expressions,
where $N_0$ is an initial state of the protocol, and for $1\leq i\leq k$ there is a transition $N_{i-1}\ar{\ell}N_i$;
we call $H$ a history \emph{of} the state $N_k$.
The \emph{connectivity graph} of a history $H$ is $\CG{H}\mathop{:=}(\IP,E)$, where
the nodes of the network form the set of vertices and there is an
arc $({\dval{ip}},{\dval{ip}}')\in E$ iff ${\dval{ip}}'\in \RN[N_i]{\dval{ip}}$ for some $0\leq i
\leq k$, i.e.\ if at some point during that history node $\dval{ip}'$ was in transmission range of \dval{ip}.
A protocol satisfies the property \emph{route correctness} if 
for every history $H$ of a reachable state $N$
and for every routing table entry $(\dval{dip}\comma\nosp{*}\comma\nosp{*}\comma\nosp{*}\comma\nosp{\dval{hops}}\comma\nosp{\dval{nhip}}\comma\nosp{*})\mathop{\in}\xiN[N]{\dval{ip}}(\rt)$
there is a path $\dval{ip}\rightarrow\dval{nhip}\rightarrow\cdots\rightarrow\dval{dip}$
in $\CG{H}$ from $\dval{ip}$ to $\dval{dip}$ with \dval{hops} hops and (if $\dval{hops}>0$) next hop \dval{nhip}.%
\footnote{A path with $0$ hops consists of a single node only.}

\begin{theorem}\rm\label{thm:route correctness}
Let $H$ be a history of a network state $N$.
 \begin{enumerate}[(a)]
 \item
For each entry $(\dval{dip},*,*,*,\dval{hops},\dval{nhip},*)\mathop{\in}\xiN[N]{\dval{ip}}(\rt)$\linebreak[4]
there is a path $\dval{ip}\rightarrow\dval{nhip}\rightarrow\cdots\rightarrow\dval{dip}$
in $\CG{H}$ from $\dval{ip}$ to $\dval{dip}$ with \dval{hops} hops and (if $\dval{hops}>0$) next hop \dval{nhip}.
 \item For each route request sent in state $N$ there is a corresponding path in the connectivity graph of $H$.\hspace{-1mm}
 	\begin{equation}\label{eq:rcreq}
 	\begin{array}{rcl}
 	  &&N\ar{R:\starcastP{\rreq{\hopsc}{*}{*}{*}{*}{\oipc}{*}{\ipc}}}_{\dval{ip}}N'\\
 	  &\Rightarrow&
           \mbox{there is a path $\ipc\rightarrow\cdots\rightarrow\oipc$ in $\CG{H}$}\\
           &&\mbox{from $\ipc$ to $\oipc$ with $\hopsc$ hops}
 	\end{array}
 	\end{equation}
 \item For each route reply sent in state $N$ there is a corresponding path in the connectivity graph of $H$.
  	\begin{equation}\label{eq:rcrep}
  	\begin{array}{@{}rcl@{}}
  	  &&N\ar{R:\starcastP{\rrep{\hopsc}{\dipc}{*}{*}{\ipc}}}_{\dval{ip}}N'\\
  	  &\Rightarrow&
           \mbox{there is a path $\ipc\rightarrow\cdots\rightarrow\dipc$ in $\CG{H}$}\\
           &&\mbox{from $\ipc$ to $\dipc$ with $\hopsc$ hops}
  	\end{array}
  	\end{equation}
 
 \end{enumerate}
\end{theorem}

\prfnoboxatend
In the course of running the protocol, the set of edges $E$ in the connectivity graph $\CG{H}$ only increases,
so the properties are invariants. We prove
them by simultaneous induction.
\begin{enumerate}[(a)]
\item In an initial state the invariant is satisfied because the routing tables are empty.
  Since entries can never be removed, and the functions $\fnaddprecrt$ and $\fninv$ do
  not affect $\dval{hops}$ and $\dval{nhip}$, it suffices to check all application calls of \hyperlink{update}{$\fnupd$}.
  In each case, if the update does not change the routing table entry beyond its precursors
  (the last clause of \hyperlink{update}{\fnupd}), the invariant is trivially
  preserved; hence we examine the cases that an update actually occurs.
\begin{description}
\item[Pro.~\ref{pro:aodv}, Lines~\ref{aodv:line10}, \ref{aodv:line14}, \ref{aodv:line18}:]
The update changes the entry into $\xi(\sip,*,\unkno,\val,1,\sip,*)$; hence \mbox{$\dval{hops}\mathbin=1$} and
$\dval{nhip}=\dval{dip}:=\xi(\sip)$. The value $\xi(\sip)$ stems through
Lines~\ref{aodv:line8},~\ref{aodv:line12} or~\ref{aodv:line16} of Pro.~\ref{pro:aodv} from an
incoming AODV control message.	By \Prop{preliminaries} this message was sent before, say in state $N^\dagger$;
by \Prop{ip=ipc} the sender of this message is $\xi(\sip)=\dval{nhip}$. Since in state $N^\dagger$
the message must have reached the queue of incoming messages of node \dval{ip}, it must be that
\plat{${\dval{ip}}\mathbin\in \RN[N^\dagger]{\dval{nhip}}$}.
In our formalisation of \awn the connectivity graph is always symmetric~\cite{TR13}: 
${\dval{nhip}}\mathbin\in
\RN[N^\dagger]{\dval{ip}}$ iff \plat{${\dval{ip}}\mathbin\in \RN[N^\dagger]{\dval{nhip}}$}.
It follows that $(\dval{ip},\dval{nhip})\in E$, so there is a 1-hop path in $\CG{H}$ from $\dval{ip}$ to $\dval{dip}$.
\item[Pro.~\ref{pro:rreq}, Line~\ref{rreq:line6}:] 
Here $\dval{dip}:=\xi(\oip)$, $\dval{hops}:=\xi(\hops)\mathord+1$ and $\dval{nhip}:=\xi(\sip)$. 
These values stem from an incoming RREQ message, which must have been sent beforehand, say in state $N^\dagger$.
As in the previous case we obtain $(\dval{ip},\dval{nhip})\in E$.
By Invariant~\Eq{rcreq}, with $\oipc:=\xi(\oip)=\dval{dip}$, $\hopsc:=\xi(\hops)$ and $\ipc:=\xi(\sip)=\dval{nhip}$, 
there is a path $\dval{nhip}\rightarrow\cdots\rightarrow\dval{dip}$ in $\CG{H}$ from $\ipc$ to $\oipc$ with $\hopsc$ hops.
It follows that there is a path $\dval{ip}\rightarrow\dval{nhip}\rightarrow\cdots\rightarrow\dval{dip}$
in $\CG{H}$ from $\dval{ip}$ to $\dval{dip}$ with \dval{hops} hops and next hop \dval{nhip}.
\item[Pro.~\ref{pro:rrep}, Line~\ref{rrep:line5}:]
Here $\dval{dip}:=\xi(\dip)$, $\dval{hops}:=\xi(\hops)\mathord+1$ and $\dval{nhip}:=\xi(\sip)$. 
The reasoning is exactly as in the previous case, except that we deal with an incoming RREP message
and use Invariant~\Eq{rcrep}.
\end{description}

\item We check all occasions where a route request is sent.
\begin{description}
	\item[Pro.~\ref{pro:aodv}, Line~\ref{aodv:line39}:]
		A new route request is initiated with $\ipc=\oipc:=\xi(\ip)=\dval{ip}$
                and $\hopsc:=0$.
                Indeed there is a path in $\CG{H}$ from $\ipc$ to $\oipc$ with $0$ hops.
	\item[Pro.~\ref{pro:rreq}, Line~\ref{rreq:line34}:]
                The broadcast message has the form
			  \[\begin{array}{l@{}l}
                	\xi(\rreqID(&\nosp{\hops\mathord+1}\comma\nosp{\rreqid}\comma\nosp{\dip}\comma\nosp{\max(\sqn{\rt}{\dip},\dsn)}\comma\\
	&\nosp{\dsk}\comma\nosp{\oip}\comma\nosp{\osn}\comma\nosp{\ip}))\,.
			  \end{array}\]
                So $\hopsc\mathord{:=}\xi(\hops)\mathord+1$, $\oipc\mathord{:=}\xi(\oip)$ and $\ipc\mathord{:=}{}\xi(\ip)$ $\mathop{=}\dval{ip}$.
                The values $\xi(\hops)$ and $\xi(\oip)$ stem through Line~\ref{aodv:line8} of Pro.~\ref{pro:aodv} from an
                incoming RREQ message of the form
                \[\xi(\rreq{\hops}{\rreqid}{\dip}{\dsn}{\dsk}{\oip}{\osn}{\sip})\,.\]
                By \Prop{preliminaries} this message was sent before, say in state $N^\dagger$;
                by \Prop{ip=ipc} the sender of this message is $\dval{sip}:=\xi(\sip)$. 
                By induction, using Invariant~\eqref{eq:rcreq}, there is a path $\dval{sip}\rightarrow\cdots\rightarrow\oipc$ in
                $\CG{H^{\dagger}} \subseteq \CG{H}$ from $\dval{sip}$ to $\oipc$ with $\xi(\hops)$ hops.
                It remains to show that there is a $1$-hop path from $\dval{ip}$ to
                $\dval{sip}$. In state $N^\dagger$ the message sent by $\dval{sip}$ must have
                reached the queue of incoming messages of node \dval{ip}, and therefore $\dval{ip}$
                was in transmission range of $\dval{sip}$, i.e., \plat{${\dval{ip}}\mathbin\in \RN[N^\dagger]{\dval{sip}}$}.
                Since the connectivity graph of \awn is always symmetric,
                \plat{${\dval{ip}}\mathbin\in \RN[N^\dagger]{\dval{sip}}$} holds as well. Hence it follows that $(\dval{ip},\dval{sip})\in E$.
\end{description}
\item We check all occasions where a route reply is sent.
\begin{description}
	\item[Pro.~\ref{pro:rreq}, Line~\ref {rreq:line14}:]
		A new route reply with $\hopsc:=0$ and
		$\ipc:=\xi(\ip)=\dval{ip}$ is initiated.
		Moreover, by Line~\ref{rreq:line10}, $\dipc:=\xi(\dip)=\xi(\ip)=\dval{ip}$.
                Thus there is a path in $\CG{H}$ from $\ipc$ to $\dipc$ with $0$ hops.
	\item[Pro.~\ref{pro:rreq}, Line~\ref{rreq:line26}:]
                We have $\ipc\mathbin{:=}\xi(\ip)\mathbin=\dval{ip}$, $\dipc\mathbin{:=}\xi(\dip)$ and $\hopsc\mathbin{:=}\dhp[\dipc]{\dval{ip}}$.
		By Line~\ref{rreq:line22} there is a routing table entry
                \plat{$(\dipc,*,*,*,\hopsc,*,*)\mathop{\in}\xiN[N]{\dval{ip}}(\rt)$}.
                Hence by Invariant~(a), which we may assume to hold when using simultaneous
                induction, there is a path $\dval{ip}\rightarrow\cdots\rightarrow\dipc$
                in $\CG{H}$ from $\dval{ip}=\ipc$ to $\dipc$ with $\hopsc$ hops.
	\item[Pro.~\ref{pro:rrep}, Line~\ref{rrep:line13}:]
		The RREP message has the form\linebreak
		$\xi(\rrep{\hops\mathop{+}1}{\dip}{\dsn}{\oip}{\ip})$
                and the proof goes exactly as for Pro.~\ref{pro:rreq}, Line~\ref{rreq:line34} of Part (b),
                by using $\dipc:=\xi(\dip)$ instead of $\oipc:=\xi(\oip)$, and an incoming RREP
                message instead of an incoming RREQ message.
\qed
\end{description}
\end{enumerate}
\eprf
\noindent\Thm{route correctness}(a) says that the AODV protocol is route correct.
For the proof it is essential that we use the version of \awn where a node $\dval{ip}'$ is in
the range of node $\dval{ip}$, meaning that $\dval{ip}'$ can receive
messages sent by $\dval{ip}$, if and only if $\dval{ip}$ is in the range of $\dval{ip}'$.
If \awn is modified so as to allow asymmetric connectivity graphs, 
as contemplated in \cite{ESOP12,TR13},
it is trivial to construct a 2-node counterexample to route correctness.

A stronger concept of route correctness could require that for every history $H$ of a state $N$ and for each
$(\dval{dip},*,*,*,\dval{hops},\dval{nhip},*)\mathop{\in}\xiN{\dval{ip}}(\rt)$
\begin{itemize}
\item either $\dval{hops}=0$ and $\dval{dip}=\dval{ip}$,
\item or $\dval{hops}=1$ and $\dval{dip}=\dval{nhip}$ and there is a $N^\dagger$ in $H$ such that
  ${\dval{nhip}}\mathbin\in \RN[N^\dagger]{\dval{ip}}$,
\item or $\dval{hops}\mathbin>1$ and there is a $N^\dagger$ in $H$ with \plat{${\dval{nhip}}\mathbin\in \RN[N^\dagger]{\dval{ip}}$}
and \plat{$(\dval{dip},*,*,\val,\dval{hops}\mathord-1,*,*)\mathbin\in\xiN[N^\dagger]{\dval{nhip}}(\rt)$}.
\end{itemize}
It turns out that this stronger form of route correctness does not hold for AODV\@.
It can be violated when a node forwards a route request without
  updating its own (fresher) routing table entry for the originator of the
  route request.

\section{Related Work}\label{sec:relatedwork}
Several process algebras modelling broadcast communication have been
proposed before:
the Calculus of Broadcasting Systems (CBS) \cite{CBS}, 
the $b\pi$-calculus \cite{bpi},
CBS${}^{\#}\!\!$ \cite{NH06},
the Calculus of Wireless Systems (CWS) \!\cite{CWS},
the Calculus of Mobile Ad Hoc Networks (CMAN) \cite{CMAN},
the Calculus for Mobile Ad Hoc Networks (CMN) \cite{CMN},
the $\omega$-calculus \hspace{-2.3pt}\cite{SRS10},
\hspace{-2.2pt}restricted branching process theory (RBPT) \cite{RBPT},
$bA\pi$ \cite{bApi}
and the broadcast psi-calculi~\cite{BHJRVPP11}.
The latter eight of these were specifically designed\linebreak[4] to model MANETs.
However, none of these process calculi provides all  features needed to fully model
routing protocols such as AODV\@, namely data handling, (conditional) unicast and (local) broadcast. 
For example,  all above-mentioned process algebras lack
the feature of guaranteed receipt of messages by destinations within transmission range.
Due to this, it is not possible to analyse properties such as route
discovery and packet delivery \cite{TR13}.
A more detailed discussion of these process algebras can be found in~\cite{TR13}.

Our complete formalisation of AODV has grown from
elaborating a partial and simplified formalisation of AODV in~\cite[Fig.\ 8]{SRS10}.
The features of our process algebra were
largely determined by what we needed to enable a complete and accurate
formalisation of this protocol. 
The same formalism has been used to model the Dynamic MANET On-demand (DYMO) Routing Protocol (also known as AODVv2)~\cite{EHWripe12}.
We conjecture that \awn is also applicable to a wide range of other wireless protocols, such as 
the Dynamic Source Routing (DSR) protocol~\cite{rfc4728},
the Lightweight Underlay Network Ad-hoc Routing (LUNAR)
protocol~\cite{lunar,Tschudin04},
the Optimized Link State Routing (OSLR) protocol~\cite{rfc3626}
or the Better Approach To Mobile Adhoc Networking (B.A.T.M.A.N.)~\cite{batman}.
The specification and the correctness of  the latter three, however, rely heavily on timing aspects; hence
an {\awn}-extension with time \cite{tawn} appears necessary (see also \Sect{conclusion}).

While process algebras such as \awn can be used to formally model and verify the correctness of network routing protocols, test-bed experiments and simulations are complementary tools that can be used to quantitatively evaluate the performance of the protocols. While test-bed experiments are able to capture the full complex characteristics of the wireless medium and its effect on the network routing protocols \cite{MaltzEtAl01,PirzadaEtAl09}, network simulators~\cite{NS2,QUALNET} offer the ease and flexibility of evaluating and comparing the performance of different routing protocols in a large-scale network of hundreds of nodes, coupled with the added advantage of being able to repeat and reproduce the experiments \cite{DasEtAl00,PerkinsBDM01,JacquetEtAl02}.

Loop freedom is a crucial property of network protocols, commonly claimed to hold for AODV \cite{rfc3561}.
Merlin and Segall \cite{MS79} were amongst the first to use sequence numbers to guarantee loop freedom of a routing protocol.
In a companion paper~\cite{AODVloop}
we have shown that several \emph{interpretations} of AODV---consistent ways to revolve
the ambiguities in the RFC---fail to be loop free, while in \cite{TR13} we establish
loop freedom of others by adaptation of the proof presented here.

A preliminary draft of AODV has been shown to be not loop free by Bhargavan et al.\
in~\cite{BOG02}. Their counterexamples to loop freedom have to do with timing issues:
the premature deletion of invalid routes, and a too quick restart of a node after a reboot.
Since then, AODV has changed to such a degree that these examples do not apply to
the current version \cite{rfc3561}. However, similar examples, claimed to apply to the current
version, are reported in \cite{Garcia04,Rangarajan05}; we discuss them in \cite{tawn}.
All these papers propose repairs that avoid these loops through better timing policies.
In contrast, the routing loops documented in \cite{AODVloop} are time-independent.

Previous attempts to prove loop freedom of AODV have been reported
in \cite{AODV99,BOG02,ZYZW09}, but none of these proofs are complete and valid for the current version of AODV \cite{rfc3561}:
\begin{itemize}
\item 
The proof sketch given in~\cite{AODV99} uses the fact that when a loop in a route to a destination
$Z$ is created, all nodes $X_i$ on that loop must have route entries for destination $Z$
with the same destination sequence number. \quote{Furthermore, because the destination sequence
numbers are all the same, the next hop information must have been derived at every node
$X_i$ from the same RREP transmitted by the destination $Z$}~\cite[Page 11]{AODV99}. The latter is not true at all:
some of the information could have been derived from RREQ messages, or from a RREP message transmitted
by an intermediate node that has a route to $Z$. More importantly, the nodes on the loop
may have acquired their information on a route to $Z$ from different RREP or RREQ
messages, that all carried the same sequence number. This is illustrated by 
the routing loop created in \cite[Figure~1]{AODVloop}.

\item Based on an analysis of an early draft of AODV\footnote{Draft version 2 is analysed, dated November 1998; 
the RFC can be seen as version 14, dated July 2001.}~\cite{BOG02} suggests three improvements. The
modified version is then proved to be loop free, using the following invariant
(written in our notation):
 \[\begin{array}{cl}
\multicolumn{2}{l}{\mbox{if $\dval{nhip} = \nhp{\dval{ip}}$, then }}\\
\mbox{(1)}& \sq{\dval{ip}} \leq \sq{\dval{nhip}}, \mbox{ and}\\ 
\mbox{(2)}& \sq{\dval{ip}} = \sq{\dval{nhip}}\\
&\Rightarrow \dhp{\dval{ip}} < \dhp{\dval{nhip}}\,. 
\end{array}
\]
This invariant does not hold for this  modified version of AODV, nor for the
  current version, documented in the RFC\@. It can happen that in a state $N$ where
  $\sq{\dval{ip}} = \sq{\dval{nhip}}$, node $\dval{ip}$ notices that the link to $\dval{nhip}$ is broken.
  Consequently, \dval{ip} invalidates its route to \dval{dip}, which has \dval{nhip} as its next hop.
  According to recommendation (\textbf{A1}) of \cite[Page 561]{BOG02}), node $\dval{ip}$
  increments its sequence number for the (invalid) route to \dval{dip}, resulting in a state $N'$
  for which $\keyw{sqn}_{N'}^{\dval{ip}}(\dval{dip}) > \keyw{sqn}_{N'}^{\dval{nhip}}(\dval{dip})$,
  thereby violating the invariant.

  Note that the invariant of \cite{BOG02} does not restrict itself to the case that the routing
  table entry for \dval{dip} maintained by \dval{ip} is \emph{valid}. Adapting the invariant with such
  a requirement would give rise to a valid invariant, but one whose verification poses
  problems, at least for the current version of AODV\@. These problems led us, in this paper, to use
  \emph{net sequence numbers\/} instead (cf.\ \SSect{quality}).

Recommendation (\textbf{A1}) is assumed to be in
  effect for the (improved) version of AODV analysed in \cite{BOG02}, although it was not in effect for
  the draft of AODV existing at the time. Since then, recommendation (\textbf{A1}) has been
  incorporated in the RFC\@. Looking at the proofs in \cite{BOG02}, it turns out that Lemma~20(1) of
  \cite{BOG02} is invalid. This failure is surprising, given that according to \cite{BOG02} Lemma~20
  is automatically verified by SPIN\@. A possible explanation might be that this lemma \emph{is} obviously
  valid for the version of AODV prior to the recommendations of \cite{BOG02}.

\item Zhou, Yang, Zhang, and Wang \cite{ZYZW09} establish loop freedom of AODV using an adaptation
  of the invariant from \cite{BOG02} with a validity requirement. However, they do not model route
  replies by intermediate nodes. This is a core feature of AODV, and a potential source of routing
  loops \cite{AODVloop}.
\end{itemize}

\section{Conclusion and Future Work}\label{sec:conclusion}
In this paper, we have presented a complete and accurate model of 
the core functionality of the Ad hoc On-Demand Distance Vector routing protocol,
a widely used protocol of practical relevance, using the process algebra {\awn}.
We currently do not model optional
features such as local route repair, expanding ring search, gratuitous route reply and multicast.
We also abstract from all timing issues.
In addition to modelling the complete set of core functionalities of the
AODV protocol, our model also covers the interface to higher protocol layers via the injection
and delivery of application layer data,
as well as the forwarding of data packets at intermediate nodes.
Although this is not part of the AODV protocol specification, it is necessary
for a practical model of any reactive routing protocol, where protocol activity is triggered via the
sending and forwarding of data packets.
The completeness of our model is in contrast to some prior related work, which either 
modelled only very simple protocols, or modelled only a subset of the functionality of relevant WMN
or MANET routing protocols.

The used modelling language \awn is tailored for WMNs and MANETs
and hence covers major aspects of WMN routing
protocols, for example the crucial aspect of data handling, such as maintaining
routing table information.
\awn allows not only  the creation of accurate and concise models  of relatively complex and practically relevant protocols, 
but also supports readability.

The currently predominant practice of informally 
specifying WMN and MANET protocols via English prose has a potential 
for ambiguity and inconsistent interpretation. The ability to provide 
a formal and unambiguous specification of such protocols via {\awn} 
is a significant benefit in its own right.
Through a 
careful analysis of AODV, in particular with respect to the loop-freedom property,
we have demonstrated how  {\awn} 
can be used as a basis for reasoning about
critical protocol correctness properties. 
By establishing invariants that remain valid in a network running AODV, 
we have shown that our model is loop free. 
In contrast to protocol evaluation using simulation, test-bed experiments or model checking, 
where only a finite number of specific network scenarios can be considered, our reasoning with {\awn} is generic and the proofs hold for any possible network scenario in terms of topology and traffic pattern. None of the experimental protocol evaluation approaches can deliver this high degree of assurance about protocol behaviour.
As a ``side product'' we have also shown that, in contrast to common belief, sequence numbers do not 
guarantee loop freedom, even if they are increased monotonically over time and 
incremented whenever a new route request is generated; this result is presented elsewhere~\cite{AODVloop}.

During creation of our model of AODV we 
uncovered several ambiguities in the AODV RFC~\cite{rfc3561}.
In~\cite{TR13} we have analysed {\em all\/}  interpretations of the RFC\@ that stem from the 
ambiguities revealed. 
It turned out that several interpretations can yield unwanted behaviour such as routing loops. 
We also found that implementations of 
AODV behave differently in crucial aspects of protocol behaviour, although they all follow the lines of the RFC\@. 
Of course a specification \quote{needs to be reasonably implementation independent}%
\footnote{\url{http://www.ietf.org/iesg/statement/pseudocode-guidelines.html}}
and can leave some decisions to the software engineer; however it is our belief that any specification should be clear and 
unambiguous enough to guarantee the same behaviour when given to different developers. 
As demonstrated, this is not the case for AODV, and likely
not for many other RFCs provided by the IETF.

To increase the level of confidence of our analysis even further, we 
mechanised \awn\ as well as the presented pen-and-paper proof of loop freedom of AODV in the interactive theorem prover Isabelle/HOL.~\cite{ITP14,ATVA14}
When verifying our (pen-and-paper) proof we did not find any major errors: (1) type checking found a minor typo in the model, (2) one proof invoked an incorrect invariant requiring the addition and proof of a new invariant based on an existing one, (3) a minor flaw in another proof required the addition of a new invariant. 
All these ``flaws'' have been repaired in the
present proof.

There are two directions of future work with regards to AODV:
(a) A further analysis of AODV will require an extension of {\awn} with time and probability:
the former to cover aspects such as AODV's handling (deletion) of
stale routing table entries and the latter to model the probability
associated with lossy links. The loop freedom result presented
here is based on a model in which routing table entries never expire.
Hence it does not rule out AODV routing loops due to premature deletion of
routing table entries. We expect that the resulting algebra will
be also applicable to a wide range of other wireless protocols.
(b) Since AODV was designed without security features in mind, 
it is vulnerable to malicious attacks such as 
routing attacks and forwarding attacks~\cite{YangEtAlSecurityinMANET}.
It may be worthwhile to formally prove that extensions of AODV such 
as SAODV~\cite{Zapata02}
actually protect the route discovery mechanism by 
providing security features like integrity and authentication.

Next to this on-going work, we also aim
to complement {\awn} by model checking.%
\footnote{See \cite{TACAS12} for the first work in that direction.}
Having the ability of automatically deriving a model for model checkers such as {\sc Uppaal} from an \awn specification allows the confirmation and detailed
diagnostics of suspected errors in an early phase of protocol development. 
Model checking is limited to networks of small size---due to state space explosion---
whereas our analysis covers all (static and dynamic) topologies.
However, finding shortcomings in some topologies is  useful to identify problematic behaviour. 
These shortcomings  can be eliminated, even before a more thorough and general analysis using {\awn}.


\begin{acknowledgements}
We thank Annabelle\ McIver and Ansgar\ Fehnker for fruitful discussions.
Further we thank the anonymous referees for the careful evaluation of this 
paper and their feedback.
\end{acknowledgements}

\begin{thebibliography}{10}
\providecommand{\url}[1]{{#1}}
\providecommand{\urlprefix}{URL }
\expandafter\ifx\csname urlstyle\endcsname\relax
  \providecommand{\doi}[1]{DOI~\discretionary{}{}{}#1}\else
  \providecommand{\doi}{DOI~\discretionary{}{}{}\begingroup
  \urlstyle{rm}\Url}\fi

\bibitem{BK85}
Bergstra, J.A., Klop, J.W.: Algebra of communicating processes with
  abstraction.
\newblock Theoretical Computer Science \textbf{37}(1), 77--121 (1985)

\bibitem{Verisim}
Bhargavan, K., Gunter, C.A., Kim, M., Lee, I., Obradovic, D., Sokolsky, O.,
  Viswanathan, M.: {Verisim}: Formal analysis of network simulations.
\newblock {IEEE} Transactions on Software Engineering \textbf{28}(2), 129--145
  (2002).
\newblock \doi{10.1109/32.988495}

\bibitem{BOG02}
Bhargavan, K., Obradovic, D., Gunter, C.A.: Formal verification of standards
  for distance vector routing protocols.
\newblock Journal of the ACM \textbf{49}(4), 538--576 (2002).\newline
\newblock \doi{10.1145/581771.581775}

\bibitem{BB87}
Bolognesi, T., Brinksma, E.: Introduction to the {ISO} specification language
  {LOTOS}.
\newblock Computer Networks \textbf{14}, 25--59 (1987).
\newblock \doi{10.1016/0169-7552(87)90085-7}

\bibitem{BHJRVPP11}
Borgstr{\"o}m, J., Huang, S., Johansson, M., Raabjerg, P., Victor, B., Pohjola,
  J.{\AA}., Parrow, J.: Broadcast psi-calculi with an application to wireless
  protocols.
\newblock In: G.~Barthe, A.~Pardo, G.~Schneider (eds.) Software Engineering and
  Formal Methods {(SEFM'11)}, \emph{Lecture Notes in Computer Science}, vol.
  7041, pp. 74--89. Springer (2011).
\newblock \doi{10.1007/978-3-642-24690-6\_7}

\bibitem{ATVA14}
Bourke, T., van Glabbeek, R.J., H\"ofner, P.: A mechanized proof of loop
  freedom of the (untimed) {AODV} routing protocol.
\newblock In: F.~Cassez, J.F. Raskin (eds.) Automated Technology for
  Verification and Analysis (ATVA'14), \emph{Lecture Notes in Computer
  Science}, vol. 8837, pp. 47--63. Springer (2014).
\newblock \doi{10.1007/978-3-319-11936-6\_5}

\bibitem{ITP14}
Bourke, T., van Glabbeek, R.J., H{\"o}fner, P.: Showing invariance
  compositionally for a process algebra for network protocols.
\newblock In: G.~Klein, R.~Gamboa (eds.) Interactive Theorem Proving (ITP'14),
  \emph{Lecture Notes in Computer Science}, vol. 8558, pp. 144--159. Springer
  (2014).
\newblock \doi{10.1007/978-3-319-08970-6\_10}

\bibitem{tawn}
Bres, E., van Glabbeek, R.J., H\"ofner, P.: {T-AWN}: A timed process alebra for
  wireless networks.
\newblock To appear (2016)

\bibitem{CK05}
Chiyangwa, S., Kwiatkowska, M.: A timing analysis of {AODV}.
\newblock In: Formal Methods for Open Object-based Distributed Systems
  {(FMOODS'05)}, \emph{Lecture Notes in Computer Science}, vol. 3535, pp.
  306--322. Springer (2005).
\newblock \doi{10.1007/11494881\_20}

\bibitem{rfc3626}
Clausen, T., Jacquet, P.: Optimized link state routing protocol {(OLSR)}.
\newblock RFC 3626 (Experimental), Network Working Group (2003).
\newblock \urlprefix\url{http://www.ietf.org/rfc/rfc3626.txt}

\bibitem{DasEtAl00}
Das, S.R., Casta\~{n}eda, R., Yan, J.: Simulation-based performance evaluation
  of routing protocols for mobile ad hoc networks.
\newblock Mobile Networks and Applications \textbf{5}(3), 179--189 (2000).
\newblock \doi{10.1023/A:1019108612308}

\bibitem{EHWripe12}
Edenhofer, S., H{\"o}fner, P.: Towards a rigorous analysis of {AODVv2}
  {(DYMO)}.
\newblock In: Rigorous Protocol Engineering {(WRiPE '12)}. IEEE (2012).\newline
\newblock \doi{10.1109/ICNP.2012.6459942}

\bibitem{bpi}
Ene, C., Muntean, T.: A broadcast-based calculus for communicating systems.
\newblock In: Parallel {\&} Distributed Processing Symposium {(IPDPS '01)}, pp.
  1516--1525. IEEE Computer Society (2001).\newline
\newblock \doi{10.1109/IPDPS.2001.925136}

\bibitem{TACAS12}
Fehnker, A., van Glabbeek, R.J., H{\"o}fner, P., McIver, A.K., Portmann, M.,
  Tan, W.L.: Automated analysis of {AODV} using {UPPAAL}.
\newblock In: C.~Flanagan, B.~K{\"onig} (eds.) Tools and Algorithms for the
  Construction and Analysis of Systems {(TACAS '12)}, \emph{Lecture Notes in
  Computer Science}, vol. 7214, pp. 173--187. Springer (2012).
\newblock \doi{10.1007/978-3-642-28756-5\_13}

\bibitem{ESOP12}
Fehnker, A., van Glabbeek, R.J., H{\"o}fner, P., McIver, A.K., Portmann, M.,
  Tan, W.L.: A process algebra for wireless mesh networks.
\newblock In: H.~Seidl (ed.) European Symposium on Programming {(ESOP '12)},
  \emph{Lecture Notes in Computer Science}, vol. 7211, pp. 295--315. Springer
  (2012).
\newblock \doi{10.1007/978-3-642-28869-2\_15}

\bibitem{TR13}
Fehnker, A., van Glabbeek, R.J., H{\"{o}}fner, P., McIver, A.K., Portmann, M.,
  Tan, W.L.: A process algebra for wireless mesh networks used for modelling,
  verifying and analysing {AODV}.
\newblock Technical Report 5513, NICTA (2013).
\newblock \urlprefix\url{http://arxiv.org/abs/1312.7645}

\bibitem{Garcia-Luna-Aceves89}
Garcia-Luna-Aceves, J.J.: A unified approach to loop-free routing using
  distance vectors or link states.
\newblock In: Symposium Proceedings on Communications, Architectures \&
  Protocols {(SIGCOMM '89)}, \emph{ACM SIGCOMM Computer Communication Review},
  vol. 19(4), pp. 212--223. ACM Press (1989).
\newblock \doi{10.1145/75246.75268}

\bibitem{Garcia04}
Garcia-Luna-Aceves, J.J., Rangarajan, H.: A new framework for loop-free
  on-demand routing using destination sequence numbers.
\newblock In: Mobile Ad-hoc and Sensor Systems {(MASS' 04)}, pp. 426--435. IEEE
  (2004).
\newblock \doi{10.1109/MAHSS.2004.1392182}

\bibitem{RBPT}
Ghassemi, F., Fokkink, W., Movaghar, A.: Restricted broadcast process theory.
\newblock In: A.~Cerone, S.~Gruner (eds.) Software Engineering and Formal
  Methods {(SEFM '08)}, pp. 345--354. IEEE Computer Society (2008).\newline
\newblock \doi{10.1109/SEFM.2008.25}

\bibitem{AODVloop}
van Glabbeek, R.J., H{\"o}fner, P., Tan, W.L., Portmann, M.: Sequence numbers
  do not guarantee loop freedom ---{AODV} can yield routing loops---.
\newblock In: Modeling, Analysis and Simulation of Wireless and Mobile Systems
  {(MSWiM '13)}, pp. 91--100. ACM Press (2013).
\newblock \doi{10.1145/2507924.2507943}

\bibitem{CMAN}
Godskesen, J.C.: A calculus for mobile ad hoc networks.
\newblock In: A.L. Murphy, J.~Vitek (eds.) Coordination Models and Languages
  {(COORDINATION '07)}, \emph{Lecture Notes in Computer Science}, vol. 4467,
  pp. 132--150. Springer (2007).
\newblock \doi{10.1007/978-3-540-72794-1\_8}

\bibitem{bApi}
Godskesen, J.C.: Observables for mobile and wireless broadcasting systems.
\newblock In: D.~Clarke, G.A. Agha (eds.) Coordination Models and Languages
  {(COORDINATION '10)}, \emph{Lecture Notes in Computer Science}, vol. 6116,
  pp. 1--15. Springer (2010).
\newblock \doi{10.1007/978-3-642-13414-2\_1}

\bibitem{GS05}
Griffin, T.G., Sobrinho, J.: Metarouting.
\newblock {SIGCOMM} Computer Communication Review \textbf{35}(4), 1--12 (2005).
\newblock \doi{10.1145/1090191.1080094}

\bibitem{Zapata02}
Guerrero-Zapata, M., Asokan, N.: {Securing Ad Hoc Routing Protocols}.
\newblock In: Proceedings of the 2002 ACM Workshop on Wireless Security (WiSe
  2002), pp. 1--10. ACM Press (2002).
\newblock \doi{10.1145/570681.570682}

\bibitem{Ho85}
Hoare, C.A.R.: Communicating Sequential Processes.
\newblock Prentice Hall, Englewood Cliffs (1985)

\bibitem{MSWIM12}
H{\"o}fner, P., van Glabbeek, R.J., Tan, W.L., Portmann, M., McIver, A.K.,
  Fehnker, A.: A rigorous analysis of {AODV} and its variants.
\newblock In: Modeling, Analysis and Simulation of Wireless and Mobile Systems
  {(MSWiM '12)}, pp. 203--212. ACM Press (2012).\newline
\newblock \doi{10.1145/2387238.2387274}

\bibitem{IEEE80211s}
IEEE: {IEEE Standard for Information Technology---Telecommunications and
  information exchange between systems---Local and metropolitan area
  networks---Specific requirements Part 11: Wireless LAN Medium Access Control
  (MAC) and Physical Layer (PHY) specifications Amendment 10: Mesh Networking}
  (2011).
\newblock
  \urlprefix\url{http://ieeexplore.ieee.org/xpl/articleDetails.jsp?arnumber=6018236}

\bibitem{JacquetEtAl02}
Jacquet, P., Laouiti, A., Minet, P., Viennot, L.: Performance of multipoint
  relaying in ad hoc mobile routing protocols.
\newblock In: E.~Gregori, M.~Conti, A.T. Campbell, G.~Omidyar, M.~Zukerman
  (eds.) Networking Technologies, Services, and Protocols; Performance of
  Computer and Communication Networks; Mobile and Wireless Communications
  {(NETWORKING '02)}, Lecture Notes in Computer Science, pp. 387--398. Springer
  (2002).\newline
\newblock \doi{10.1007/3-540-47906-6\_31}

\bibitem{rfc4728}
Johnson, D., Hu, Y., Maltz, D.: The dynamic source routing protocol {(DSR)} for
  mobile ad hoc networks for {IPv4}.
\newblock RFC 4728 (Experimental), Network Working Group (Errata Exist) (2007).
\newblock \urlprefix\url{http://www.ietf.org/rfc/rfc4728.txt}

\bibitem{MaltzEtAl01}
Maltz, D., Broch, J., Johnson, D.B.: Lessons from a full-scale multihop
  wireless ad hoc network testbed.
\newblock IEEE Personal Communications \textbf{8}(1), 8--15 (2001).\newline
\newblock \doi{10.1109/98.904894}

\bibitem{MS79}
Merlin, P.M., Segall, A.: A failsafe distributed routing protocol.
\newblock IEEE Transactions on Communications \textbf{27}(9), 1280--1287
  (1979).
\newblock \doi{10.1109/TCOM.1979.1094552}

\bibitem{CMN}
Merro, M.: An observational theory for mobile ad hoc networks (full version).
\newblock Information and Computation \textbf{207}(2), 194--208 (2009).
\newblock \doi{10.1016/j.ic.2007.11.010}

\bibitem{CWS}
Mezzetti, N., Sangiorgi, D.: Towards a calculus for wireless systems.
\newblock Electronic Notes in Theoretical Computer Science \textbf{158},
  331--353 (2006).\newline
\newblock \doi{10.1016/j.entcs.2006.04.017}

\bibitem{Mi89}
Milner, R.: Communication and Concurrency.
\newblock Prentice Hall (1989)

\bibitem{MK10}
Miskovic, S., Knightly, E.W.: Routing primitives for wireless mesh networks:
  Design, analysis and experiments.
\newblock In: Conference on Information Communications {(INFOCOM '10)}, pp.
  2793--2801. IEEE (2010).
\newblock \doi{10.1109/INFCOM.2010.5462111}

\bibitem{NH06}
Nanz, S., Hankin, C.: A framework for security analysis of mobile wireless
  networks.
\newblock Theoretical Computer Science \textbf{367}, 203--227 (2006).
\newblock \doi{10.1016/j.tcs.2006.08.036}

\bibitem{batman}
Neumann, A., Aichele C.~Lindner, M., Wunderlich, S.: Better approach to mobile
  ad-hoc networking {(B.A.T.M.A.N.)}.
\newblock Internet-Draft (Experimental), Network Working Group (2008).
\newblock
  \urlprefix\url{http://tools.ietf.org/html/draft-openmesh-b-a-t-m-a-n-00}

\bibitem{NS2}
The network simulator {ns-2}.
\newblock \urlprefix\url{http://nsnam.isi.edu/nsnam/index.php/Main_Page} 

\bibitem{rfc3561}
Perkins, C.E., Belding-Royer, E.M., Das, S.: {Ad hoc On-Demand Distance Vector
  {(AODV)} Routing}.
\newblock RFC 3561 (Experimental), Network Working Group (2003).
\newblock \urlprefix\url{http://www.ietf.org/rfc/rfc3561.txt}

\bibitem{PerkinsBDM01}
Perkins, C.E., Belding-Royer, E.M., Das, S.R., Marina, M.K.: Performance
  comparison of two on-demand routing protocols for ad hoc networks.
\newblock IEEE Personal Communications \textbf{8}(1), 16--28 (2001).
\newblock \doi{10.1109/98.904895}

\bibitem{AODVv2}
Perkins, C.E., Ratliff, S., Dowdell, J.: Dynamic {MANET} on-demand {(AODVv2)}
  routing.
\newblock Internet Draft (Standards Track), Mobile Ad hoc Networks Working
  Group (2013).
\newblock \urlprefix\url{http://tools.ietf.org/html/draft-ietf-manet-aodvv2-02}

\bibitem{AODV99}
Perkins, C.E., Royer, E.M.: {Ad-hoc On-Demand Distance Vector Routing}.
\newblock In: Mobile Computing Systems and Applications {(WMCSA '99)}, pp.
  90--100. IEEE (1999).
\newblock \doi{10.1109/MCSA.1999.749281}

\bibitem{PPI08}
Pirzada, A.A., Portmann, M., Indulska, J.: Performance analysis of multi-radio
  {AODV} in hybrid wireless mesh networks.
\newblock Computer Communications \textbf{31}(5), 885--895 (2008).
\newblock \doi{10.1016/j.comcom.2007.12.012}

\bibitem{PirzadaEtAl09}
Pirzada, A.A., Portmann, M., Wishart, R., Indulska, J.: {SafeMesh:} a wireless
  mesh network routing protocol for incident area communications.
\newblock Pervasive and Mobile Computing \textbf{5}(2), 201--221 (2009).\newline
\newblock \doi{10.1016/j.pmcj.2008.11.005}

\bibitem{CBS}
Prasad, K.V.S.: A calculus of broadcasting systems.
\newblock Science of Computer Programming \textbf{25}(2-3), 285--327 (1995).
\newblock \doi{10.1016/0167-6423(95)00017-8}

\bibitem{AODV-ST}
Ramachandran, K., Buddhikot, M.M., Chandranmenon, G., Miller, S.,
  Belding-Royer, E.M., Almeroth, K.: On the design and implementation of
  infrastructure mesh networks.
\newblock In: {IEEE} Workshop on Wireless Mesh Networks {(WiMesh'05)}. IEEE
  (2005)

\bibitem{Rangarajan05}
Rangarajan, H., Garcia-Luna-Aceves, J.J.: Making on-demand routing protocols
  based on destination sequence numbers robust.
\newblock In: Communications {(ICC '05)}, vol.~5, pp. 3068--3072 (2005).
\newblock \doi{10.1109/ICC.2005.1494958}

\bibitem{QUALNET}
{SCALABLE Network Technologies}: {QualNet} communications simulation platform.
\newblock \urlprefix\url{http://web.scalable-networks.com/content/qualnet} 

\bibitem{SRS10}
Singh, A., Ramakrishnan, C.R., Smolka, S.A.: A process calculus for mobile ad
  hoc networks.
\newblock Science of Computer Programming \textbf{75}, 440--469 (2010).
\newblock \doi{10.1016/j.scico.2009.07.008}

\bibitem{SBM06}
Subramanian, A.P., Buddhikot, M.M., Miller, S.: Interference aware routing in
  multi-radio wireless mesh networks.
\newblock In: {IEEE} Workshop on Wireless Mesh Networks {(WiMesh '06)}. IEEE
  (2006)

\bibitem{lunar}
Tschudin, C.F.: Lightweight underlay network ad hoc routing {(LUNAR)} protocol.
\newblock Internet Draft (Expired), Mobile Ad Hoc Networking Working Group
  (2004).
\newblock
  \urlprefix\url{http://user.it.uu.se/~rmg/pub/draft-tschudin-manet-lunar-00.txt}

\bibitem{Tschudin04}
Tschudin, C.F., Gold, R., Rensfelt, O., Wibling, O.: {LUNAR:} a lightweight
  underlay network ad-hoc routing protocol and implementation.
\newblock In: Y.~Koucheryavy, J.~Harju, A.~Koucheryavy (eds.) Next Generation
  Teletraffic and Wired/Wireless Advanced Networking {(NEW2AN '04)} (2004)

\bibitem{YangEtAlSecurityinMANET}
Yang, H., Luo, H., Ye, F., Lu, S., Zhang, L.: {Security in Mobile Ad Hoc
  Networks: Challenges and Solutions}.
\newblock IEEE Wireless Communications, \textbf{11}(1), 38--47 (2004).
\newblock \doi{10.1109/MWC.2004.1269716}

\bibitem{Zave12}
Zave, P.: Using lightweight modeling to understand {Chord}.
\newblock SIGCOMM Comput. Commun. Rev. \textbf{42}(2), 49--57 (2012).
\newblock \doi{10.1145/2185376.2185383}

\bibitem{ZYZW09}
Zhou, M., Yang, H., Zhang, X., Wang, J.: The proof of {AODV} loop freedom.
\newblock In: Wireless Communications \& Signal Processing (WCSP '09). IEEE
  (2009).
\newblock \doi{10.1109/WCSP.2009.5371479}

\end{thebibliography}

\vspace{1cm}
\appendix
\section{Omitted Proofs}
\printproofs
\end{document}